\documentclass[10pt,english]{bourbaki}
\usepackage{amsmath}
\usepackage{amsthm}
\usepackage{amsfonts}
\usepackage{amssymb}
\usepackage{epic}
\usepackage{eepic}
\usepackage{times}

\theoremstyle{plain}
\newtheorem{thm}{Theorem}
\newtheorem{defn}{Definition}

\theoremstyle{remark}
\newtheorem{rem}{Remark}

\newcommand{\sst}{\scriptscriptstyle}

\newcommand{\beq}{\begin{equation}}
\newcommand{\eeq}{\end{equation}}

\newcommand{\id}{\mbox{id}}

\newcommand{\pa}{\partial}
\newcommand{\ot}{\otimes}
\newcommand{\ra}{\rightarrow}

\newcommand{\ti}{\times}

\newcommand{\fr}[2]{{\textstyle \frac{#1}{#2} }}

\newcommand{\fsl}{{\mathfrak s}{\mathfrak l}}

\newcommand{\usl}{{\mathcal U}_{q}(\fsl_2)}
\newcommand{\bra}{\langle}
\newcommand{\ket}{\rangle}
\newcommand{\al}{\alpha}
\newcommand{\bal}{\bar{\alpha}}

\newcommand{\be}{\beta}
\newcommand{\ga}{\gamma}
\newcommand{\Ga}{\Gamma}
\newcommand{\de}{\delta}
\newcommand{\De}{\Delta}
\newcommand{\ep}{\epsilon}
\newcommand{\la}{\lambda}
\newcommand{\om}{\omega}
\newcommand{\Om}{\Omega}
\newcommand{\si}{\sigma}
\newcommand{\up}{\Upsilon}
\newcommand{\vf}{\varphi}

\newcommand{\bT}{\bar{T}}
\newcommand{\ba}{\bar{a}}

\newcommand{\bw}{\bar{w}}

\newcommand{\bz}{\bar{z}}

\newcommand{\bbe}{\bar{\beta}}

\newcommand{\CB}{{\mathcal B}}
\newcommand{\CC}{{\mathcal C}}
\newcommand{\CD}{{\mathcal D}}

\newcommand{\CF}{{\mathcal F}}
\newcommand{\CG}{{\mathcal G}}
\newcommand{\CH}{{\mathcal H}}
\newcommand{\CI}{{\mathcal I}}

\newcommand{\CO}{{\mathcal O}}  
\newcommand{\CP}{{\mathcal P}}  
  
\newcommand{\CR}{{\mathcal R}}
\newcommand{\CS}{{\mathcal S}}
\newcommand{\CT}{{\mathcal T}}
\newcommand{\CU}{{\mathcal U}}
\newcommand{\CV}{{\mathcal V}}
\newcommand{\CW}{{\mathcal W}}

\newcommand{\SC}{{\mathsf C}}

\newcommand{\SE}{{\mathsf E}}
\newcommand{\SF}{{\mathsf F}}

\newcommand{\SH}{{\mathsf H}}

\newcommand{\SJ}{{\mathsf J}}
\newcommand{\SK}{{\mathsf K}}
\newcommand{\SL}{{\mathsf L}}
\newcommand{\bL}{\bar{\SL}}

\newcommand{\SN}{{\mathsf N}}
  
\newcommand{\SP}{{\mathsf P}}  
\newcommand{\SQ}{{\mathsf Q}}  
\newcommand{\SR}{{\mathsf R}}
\renewcommand{\SS}{{\mathsf S}}
\newcommand{\ST}{{\mathsf T}}
\newcommand{\SU}{{\mathsf U}}
\newcommand{\SV}{{\mathsf V}}
\newcommand{\SW}{{\mathsf W}}

\newcommand{\ff}{v}

\newcommand{\fx}{{\mathfrak x}}

\newcommand{\sa}{{\mathsf a}}
\newcommand{\sbb}{{\mathsf b}}

\newcommand{\sg}{{\mathsf g}}
\renewcommand{\sf}{{\mathsf f}}

\newcommand{\sll}{{\mathsf l}}

\newcommand{\sq}{{\mathsf q}}
\newcommand{\spp}{{\mathsf p}}

\newcommand{\sv}{{\mathsf v}}

\newcommand{\sx}{{\mathsf x}}
\newcommand{\sz}{{\mathsf z}}

\newcommand{\FB}{{\mathfrak B}}

\newcommand{\BA}{{\mathbb A}}
\newcommand{\BF}{{\mathbb F}}
\newcommand{\BB}{{\mathbb B}}
\newcommand{\BR}{{\mathbb R}}

\newcommand{\BC}{{\mathbb C}}

\newcommand{\BS}{{\mathbb S}}
\newcommand{\BT}{{\mathbb T}}
\newcommand{\BZ}{{\mathbb Z}}

\newcommand{\raf}{\ra -\infty}

\def\ew{\hspace*{-1mm}}   \def\ppe{\hspace*{-2.5mm}}

\newcommand{\Fus}[6]{F_{{\scriptstyle #1}{\scriptstyle #2}}
  \hspace*{.3mm}\displaystyle{[} \ew \begin{array}{ll} {\scriptstyle #3 }
  \ppe & {\scriptstyle #4} \ew \\[-2mm] {\scriptstyle #5}\ppe &
  {\scriptstyle #6}\ew \end{array}\displaystyle{]}}
\newcommand{\Braid}[7]{B_{{\scriptstyle #1}{\scriptstyle #2}}^{#7}
  \hspace*{.3mm}\displaystyle{[} \ew \begin{array}{ll} {\scriptstyle #3 }
  \ppe & {\scriptstyle #4} \ppe \\[-2mm] {\scriptstyle #5}\ppe &
  {\scriptstyle #6}\ew \end{array}\displaystyle{]}}
\newcommand{\Braidg}[6]{B_{{\scriptstyle #1}{\scriptstyle #2}}^{\,{\rm G}
,\,\ep}
  \hspace*{.3mm}\displaystyle{[} \ew \begin{array}{ll} {\scriptstyle #3 }
  \ppe & {\scriptstyle #4} \ppe \\[-2mm] {\scriptstyle #5}\ppe &
  {\scriptstyle #6}\ew \end{array}\displaystyle{]}}

\newcommand{\CBls}[6]{{#6}_{{\scriptstyle #1}}^s
  \hspace*{.3mm}\displaystyle{[} \ew \begin{array}{ll} {\scriptstyle #2 }
  \ppe & {\scriptstyle #3} \ppe \\[-2mm] {\scriptstyle #4}\ppe &
  {\scriptstyle #5}\ew \end{array}\displaystyle{]}}
\newcommand{\CBlt}[6]{{#6}_{{\scriptstyle #1}}^t
  \hspace*{.3mm}\displaystyle{[} \ew \begin{array}{ll} {\scriptstyle #2 }
  \ppe & {\scriptstyle #3} \ppe \\[-2mm] {\scriptstyle #4}\ppe &
  {\scriptstyle #5}\ew \end{array}\displaystyle{]}}
\newcommand{\CfBls}[5]{\CBls{#1}{#2}{#3}{#4}{#5}{\CF}}
\newcommand{\CfBlt}[5]{\CBlt{#1}{#2}{#3}{#4}{#5}{\CF}}

\newcommand{\CpBls}[5]{\Phi_{{\scriptstyle #1}}^s
  \hspace*{.3mm}\displaystyle{[} \ew \begin{array}{ll} {\scriptstyle #2 }
  \ppe & {\scriptstyle #3} \ppe \\[-2mm] {\scriptstyle #4}\ppe &
  {\scriptstyle #5}\ew \end{array}\displaystyle{]}}

\newcommand{\CpBlt}[5]{\Phi_{{\scriptstyle #1}}^t
  \hspace*{.3mm}\displaystyle{[} \ew \begin{array}{ll} {\scriptstyle #2 }
  \ppe & {\scriptstyle #3} \ppe \\[-2mm] {\scriptstyle #4}\ppe &
  {\scriptstyle #5}\ew \end{array}\displaystyle{]}}

\newcommand{\CGC}[6]{\displaystyle{[} \,\ew \begin{array}{lll} 
  {\scriptstyle #1} \ppe
  & {\scriptstyle #3} \ppe & {\scriptstyle #5} \ew \\[-2mm] {\scriptstyle
  #2} \ppe & {\scriptstyle #4}\ppe & {\scriptstyle #6} \ew\end{array}
  \displaystyle{]}}
\newcommand{\SJS}[6]{ \displaystyle{\bigl\{ } \ew 
\begin{array}{ll} {\scriptstyle #1 }
  \ppe & {\scriptstyle #2} \ppe \\[-2mm] {\scriptstyle #3}\ppe &
  {\scriptstyle #4}\ew \end{array}\big| \ew
\begin{array}{l} {\scriptstyle #5 }
  \ppe \\[-2mm] {\scriptstyle #6}\ew  \end{array}\displaystyle{\bigr\}_b}}

\DeclareMathOperator{\sgn}{sgn}

\newcommand{\rf}[1]{(\ref{#1})}

\newcommand{\aufz}
{\begin{list}{$\bullet$}{\topsep0cm \itemsep0cm \parsep0cm}}
\newcommand{\eaufz}{\end{list}}
\newcounter{num}
\newcommand{\remlst}{\begin{list}
{(\arabic{num})}{\usecounter{num}\topsep0cm \itemsep0cm \parsep0cm}}
\begin{document}
\thispagestyle{empty}
\date{April 2001}
\bbknumero{SfB 288, No. 512} 
\title{\Huge\bf Liouville theory revisited}
\author{J. Teschner}
\address{Inst. f\"ur theoretische Physik,\\
Freie Universit\"at Berlin,\\
Arnimallee 14,\\
14195 Berlin, \\
Germany}

\begin{abstract}
In view of the importance of noncompact CFT's such as Liouville theory
and SL(2,R)-WZNW model I will reconsider some of the basic issues in such
theories as well as the important qualitative differences of these theories
as compared to RCFT in the example of Liouville theory. More specifically, 
I will give a new derivation of spectrum and fusion rules from canonical 
quantization, and try to clarify issues such as operator-state-correspondence,
normalizable vs. non-normalizable states etc. 
\end{abstract}
\maketitle

\begin{center}{\bf Abstract:}\end{center}
\begin{quote}{\small
We try to develop a coherent picture of Liouville theory as a two-dimensional 
conformal field theory that takes into account the
perspectives of path-integral approach, bootstrap, 
canonical quantization and operator approach. 
To do this, we need to develop further some of these approaches. 
This includes in particular a 
construction of general exponential field operators from a set of
covariant chiral operators. The latter are shown to satisfy 
braid relations that allow one to prove the locality of the former.} 
\end{quote}

\section{Introduction}\vspace{.1cm}

\subsection{Motivation}

Liouville theory seems to be a kind of universal building block for
a variety of models for two-dimensional gravity and non-trivial 
backgrounds in string theory. Some aspects of it were important for
understanding what the matrix models of 2D gravity actually describe
(see e.g. \cite{GM}), and it keeps popping up in sometimes unexpected
circumstances such as the the physics of membranes in string theory 
(e.g. \cite{SW}). In the context of string theory, Liouville theory
and close relatives such as the SL(2) or SL(2)/U(1) WZNW models 
seem to be the simplest examples where the new qualitative features
of nontrivial (possibly curved) non-compact backgrounds can be studied.

For all this it is crucial that Liouville theory, as indeed supported
by many investigations of this issue, can be quantized as a 
{\it conformal field theory} (CFT), implying in particular that the 
space of states forms a representation of the Virasoro algebra. 
What makes the analysis of the quantized theory much more difficult
as compared to other conformal field theories is the fact that the set
of Virasoro representations that make up the space of states is 
{\it continuous}. This can be viewed as a reflection of the 
noncompactness of the space in which the Liouville zero mode
$q\equiv \int_{0}^{2\pi}d\si \Phi(\si)$ takes values. 

Liouville theory may furthermore be seen as probably the simplest 
prototypical example for a class of conformal field theories 
called non-compact CFT which have continuous spectrum of representations
of the Virasoro algebra. It may well be expected to play a role in 
the development of a general theory of such CFT's that is analogous
to the role of the minimal models as prototype for rational 
CFT. This is in fact one of our main motivations: We believe
that other non-compact CFT will share many features
with Liouville theory that distinguish non-compact from 
rational CFT. Moreover, once the technical 
tools for the proper investigation of Liouville theory are established,   
it should not be too difficult to generalize them to other 
non-compact CFT. For example, many results from Liouville theory 
can be carried over fairly directly to the $H_3^+$-WZNW model \cite{Te3}.

\subsection{Aims and scope}

This paper focuses on the understanding of Liouville theory
on a (space-time) cylinder with circumference $2\pi$,
time-coordinate $t$ and (periodic) space-coordinate $\si$
as a two dimensional quantum field theory in its own right. 
(Semi-)classically
the theory is defined by the action
\begin{equation}\label{classact}
S_c\;=\;\int\limits_{-\infty}^{\infty}dt\int\limits_{0}^{2\pi}d\si\biggl(
\frac{1}{16\pi}\bigl((\pa_{t}\vf)^2-(\pa_{\si}\vf)^2\bigr)
-\mu_c e^{\vf}\biggr).
\end{equation}
We will attempt to describe the corresponding quantum theory that will depend
on the parameter $\hbar$ which we will write as $\hbar=b^2$. It turns out that
there are two interesting regimes for the parameter $b$ that one 
may consider: $b\in(0,1]\subset\BR$ (``weak coupling'') and $b=e^{i\vf}$,
$\vf\in[0,\frac{\pi}{2})$ (``strong coupling''). One 
aim is to construct a quantum Liouville field $\phi$ that in the
semi-classical limit $b\ra 0$ corresponds to $\vf$ via the rescaling 
$\phi\underset{b\ra 0}{\sim}\frac{1}{2b}\vf$. 

This paper grew out of an attempt to write a review of 
Liouville theory. We tried to develop a coherent picture that takes into
account the different perspectives of path-integral approach, bootstrap,
canonical quantization and operator approach.
In doing so we felt forced to
go beyond the existing literature as the listed approaches 
had not been sufficiently developed to allow for their mutual
comparison. 
The character of this paper will therefore be 
somewhat intermediate between review, preview and original contribution:
It has the character of a review in the sense that we try to exhibit
the main ideas and results rather than all technical details.
However, since many of the results and arguments
presented here have not appeared
in the literature yet, we have included short presentations of at least
some key technical points. 

We will not try to
cover Liouville theory in all of its aspects and ramifications. We will 
{\it not} discuss important and interesting topics such as 
\begin{itemize}
\item {\sc Other boundary conditions and supersymmetric versions 
of Liouville theory} $\frac{\quad}{}$ Instead of periodic 
boundary conditions in the spatial direction one may also
consider Liouville theory on the
real line \cite{dHJ}, or the strip with boundary conditions that
preserve conformal invariance at its end-points, 
see e.g. \cite{GN2}\cite{CG} and 
references therein for early work, and  
\cite{FZZ}\cite{TB}\cite{ZZ2} for more recent progress. Some exact results
on $N=1$ Liouville theory have been obtained in \cite{RS}\cite{Po}.
It would be quite important to have similar results for $N=2$ 
Liouville (for some discussion see e.g. \cite{KS}). 
\item {\sc Liouville theory on the lattice} $\frac{\quad}{}$ 
Early works include \cite{FT}\cite{B1}\cite{B2}. 
Substantial progress in this direction was made quite recently,
see \cite{FKV} and references therein,
but we felt unable to explain the precise connections to the
material present here at the present stage of development.
Nevertheless, we would like to at least mention the very recent 
result 
of L. Faddeev and R. Kashaev \cite{FK}, 
where the spectrum of the Hamiltonian in
the lattice Liouville theory \cite{FKV} is determined 
on the basis of \cite{Ka2}. This result nicely confirms
our claim that the spectrum is purely continuous 
both in the weak and strong coupling regimes.
\item {\sc Connections with Teichm\"uller theory and quantization of 
Teichm\"uller space} $\frac{\quad}{}$ (For the former see 
e.g. the review \cite{Ta}, on the latter topic see e.g. 
\cite{V}\cite{Ka1}\cite{CF}\cite{Ka2} and references therein). 
Quantization of Teichm\"uller space is expected \cite{V} to encode
topological information on
the space of conformal blocks of Liouville theory, which should be equivalent
(see \cite{Ka2} for strong  
evidence) to the description of the duality transformations on 
spaces of conformal blocks that was proposed in \cite{PT1} and which will
be discussed in Part \ref{boots}. 
\item {\sc Non-unitary spectra, ``elliptic sector''}
There exist proposals for versions of quantum Liouville theory 
that have spectra involving non-unitary representations of the 
Virasoro algebra, see e.g. \cite{GS3} and references therein. 
Related to the lack of unitarity one finds unusual hermiticity
properties of the exponential field operators. In
the present paper we will be exclusively concerned with the 
possibility of having a quantization of Liouville theory
that preserves unitarity.
\item {\sc Closed discrete sub-algebras of the algebra
of vertex operators that create unitary representations} $\frac{\quad}{}$ 
It was proposed in \cite{G2} (see also
\cite{GR1}\cite{GR2}) that for certain special values of the parameter $b$
spectrum and vertex operator algebra of Liouville theory
admit a ``unitary truncation'' in the following sense:
There exists a Hilbert space $\CH_d$ spanned by a discrete set of unitary
representations of the Virasoro algebra, equipped with a realization of a
discrete family of local vertex operators with real positive 
conformal dimensions. This would allow to construct non-critical
bosonic string models. 
Let us remark, however, that our discussion {\it will} apply to 
the values of $b$, for which the proposal of \cite{G2} was made. 
Nevertheless, it could be that the operator algebra that we discuss
here shows some kind of ``reducibility'' for certain values of $b$,
which would lead to the proposal of \cite{G2} $\frac{\quad}{}$
see also our remark in Part \ref{boots}, Subsection \ref{strc_rem}.
\item {\sc Applications to models of 2D gravity} $\frac{\quad}{}$ 
(see e.g. \cite{GM})
\end{itemize}
The author will be grateful to anybody pointing out further omissions
or missing references concerning the material that is covered
in the present paper.

\subsection{Overview}

If a conformal field theory fits into the framework introduced in
\cite{BPZ}, it is essentially fully characterized by its spectrum of 
primary fields and the full set of three point functions. 
What we will call the ``DOZZ-proposal'' (where DOZZ stands for
Dorn, Otto, Al. and A. Zamolodchikov) amounts to the 
proposal that Liouville theory fits into the framework of
\cite{BPZ}, together with an explicit formula for the 
three point functions of the primary fields \cite{DO}\cite{ZZ}.

And indeed, as we will discuss in more detail, one can show 
that essentially all of Liouville theory is encoded in these
pieces of information: Thanks to conformal symmetry it is 
possible to reconstruct all correlation functions of primary fields
from the three-point function by summing over intermediate states. 
Primary fields and the energy momentum tensor generate the operator
algebra of the theory. Mixed correlation functions involving
the  energy momentum tensor together with primary fields are 
reduced to correlation functions of only primary fields
by the conformal Ward identities. 

From that point of view the main problem for establishing 
the validity of the DOZZ-proposal 
is to show crossing symmetry of the amplitudes that can be 
reconstructed by means of conformal symmetry in terms of the particular
formula for the 
three point functions that was proposed in \cite{DO}\cite{ZZ}. 
Once this is established, one has ample reason to
view the theory that is characterized by the DOZZ-proposal 
as a quantization of classical Liouville theory: Let us only mention
that the semi-classical limit of 
the formula proposed in \cite{DO}\cite{ZZ} indeed matches the result
of direct semi-classical calculations \cite{ZZ}. 

In the present paper we will try to explain why the DOZZ-proposal
works and what the resulting picture of the physics of quantum Liouville
theory looks like. 

In Part \ref{pathint} we begin by formulating more precisely what 
we refer to as the ``DOZZ-proposal''. Afterwards some of the 
original motivation for that proposal will be 
discussed which came from the path-integral approach
to Liouville theory as initiated in \cite{GL}. 
Our main objective in that Part \ref{pathint} is to explain how 
Liouville theory can be reconstructed on the basis of the DOZZ-proposal:
How to reconstruct the Hilbert space, operators corresponding to the
fields, their correlation functions etc.. Some features
arise that are unfamiliar from rational conformal field theories: 
For example, one does not find the $SL(2,\BC)$-invariant state $|0\ket$
in the spectrum. However, it will be shown that a distributional
interpretation of the ``state'' $|0\ket$ is not only natural in a theory
with continuous spectrum, but also makes operator-state correspondence
and the interpretation of correlation functions as 
``vacuum expectation values'' work in much the same way 
as in rational conformal field theories.

The following Part \ref{canq1} discusses some aspects
of the quantum Liouville physics as encoded in the DOZZ-proposal.
We begin by discussing to what extend the DOZZ-proposal can be shown 
to represent a canonical quantization of Liouville theory: 
It is possible to reconstruct a field $\phi$ that weakly
(within matrix elements) satisfies the canonical commutation 
relations and 
a natural quantum version of the Liouville equation of motion. 
One might then hope to have a representation where the Liouville zero mode
$q$ is diagonal, so that one could describe the 
physics of Liouville theory in terms of wave-functions $\psi(q)$ on 
``target-space''. And indeed, such a representation would allow one 
to get natural interpretations for many features of the DOZZ-proposal
in terms of scattering off the Liouville potential. 
However, at present we only have good control over such a representation
in the region corresponding to zero mode $q\raf$, 
the ``asymptotic boundary of target space'' where 
the Liouville interaction vanishes. This is good enough for setting up the
scattering picture that we had mentioned above.
But it is not clear so far how 
to describe Liouville physics for finite values 
of $q$, in the ``bulk of target space''. It is not even 
clear whether such a representation with diagonal zero mode $q$ exists
at all: The zero mode operator 
$q$ has a rather complicated description in a representation where
the Hamiltonian is diagonal, making it difficult to control properties
such as self-adjointness.


Our third part is devoted to the operator approach to Liouville theory.
Classically, one has a canonical transformation from free field theory 
to the interacting Liouville theory. In the operator approach one tries
to quantize this transformation, i.e. to construct Liouville 
field operators in terms of the quantized free field. We 
outline recent results that lead to a complete construction of 
general exponential Liouville fields within such a framework 
(more details will be given in a forthcoming publication). 
Their matrix elements are given by the three point function 
proposed in \cite{DO}\cite{ZZ}.
This construction is based on the construction of a class of covariant
chiral operators that form the building blocks of the exponential fields.
Locality of the exponential fields can be controlled thanks to the 
existence of exchange- (braid-) relations that are satisfied by the covariant
chiral operators. These results form the technical core for the 
construction of Liouville theory that we propose. 

Conceptually it is interesting to compare this 
approach with the discussion in Part \ref{canq1}: 
This furnishes another interpretation for
the reflection operator $\SR$ that describes the scattering of wave-packets
off the Liouville-potential ``wall'': Classically one finds that the 
map from the free field to the Liouville field is two-to-one. This ambiguity
is expressed in the quantum theory by the existence of an operator $\SS$
that commutes with the Liouville field operators. The operator 
$\SS$ coincides precisely with the reflection operator $\SR$. 
This observation may lead one to identify the free field theory on which the 
operator approach is based with the free field theory that 
describes the asymptotic in- and out-states of the scattering off
the Liouville potential.
Alternatively, one may view 
the operator approach as providing a reconstruction
of Liouville theory from the free field theory ``living on the 
asymptotic boundary of target space''.

We finally discuss in Part \ref{boots} how Liouville theory fits
into a ``chiral bootstrap'' framework such as the one introduced for
rational conformal field theories in \cite{MS}\cite{FFK}. 
This not only yields insight into the mathematics behind
the consistency of Liouville theory (fusion of unitary Virasoro
representations, relation to quantum group representation theory), 
but also provides a useful framework to complete the verification
of locality and crossing symmetry for the exponential Liouville fields
constructed in Part \ref{gervnev}.
  
\subsection{Notational conventions}

Throughout we will use the convention that for a local field 
$F$, the notation $F(z,\bz)$ denotes the euclidean field on the 
Riemann sphere, $F(\tau,\si)$ or $F(w,\bw)$, $w=\tau+i\si$
its counterpart on the euclidean
cylinder, $F(t,\si)$ the minkowskian version, and 
$F(\si)\equiv F(0,\si)$. 

\newpage

\part{The DOZZ proposal} \label{pathint}\vspace{0.5cm}

We will start by discussing what we  call the 
DOZZ-proposal.
In a nutshell it consists of two ingredients: (i) Liouville theory
fits into a rather mild generalization of the BPZ-formalism for 
two-dimensional conformal field theories, and (ii) a proposal for
an explicit representation of the three point function (``DOZZ-formula'').
The DOZZ-proposal will be formulated more more explicitly in the next
section. 
Afterwards we will present some motivation for the 
DOZZ-formula from the path-integral point of view. 
In the remainder of this part we will try to explain how really 
all of Liouville theory is encoded in these two pieces of 
information.

\section{Conformal symmetry and three point function}

Within a formalism for conformal field theories such as that 
introduced in \cite{BPZ} one assumes Liouville theory to be fully 
characterized
by the set of all vacuum expectation values of the form
\begin{equation}\label{VEV}
\Om\Biggl(\prod_{r=1}^R T(w_r)\prod_{s=1}^S \bar{T}(\bw_r)
\prod_{i=1}^NV_{\al_i}(z_i,\bz_i)\Biggl).
\end{equation}
$T(z)$ and $\bar{T}(\bz)$ are the holomorphic and antiholomorphic
components of the energy-momentum tensor respectively, and 
the $V_{\al}(z)$, $\al\in\BC$ are the primary fields. 
The vacuum expectation values \rf{VEV} are assumed to satisfy the
following conditions:
\begin{enumerate}
\item {\sc Conformal Ward identities} $\frac{\quad}{}$ 
\[\begin{aligned}
\Om\Biggl( T(w)& \prod_{r=1}^R T(w_r)\prod_{s=1}^S \bar{T}(\bw_s)
\prod_{i=1}^NV_{\al_i}(z_i,\bz_i)\Biggl)=\\
= \; & \sum_{r=1}^R\Om\Biggl(T(w_{R})\dots 
\bigl\{T(w)T(w_r)\bigr\}\dots T(w_{1}) \prod_{s=1}^S 
\bar{T}(\bw_s)
\prod_{i=1}^N V_{\al_i}(z_i,\bz_i)\Biggl)\\
 + &  \sum_{i=1}^N\Om\Biggl( \prod_{r=1}^R T(w_r)\prod_{s=1}^S \bar{T}(\bw_s)
\; V_{\al_{N}}(z_{N},\bz_{N})\dots
\bigl\{T(w)V_{\al_i}(z_i,\bz_i)\bigr\}\dots
V_{\al_{1}}(z_{1},\bz_{1})\Biggl),
\end{aligned}
\]
where $\bigl\{T(w)T(w)\bigr\}$ and $\bigl\{T(w)V_{\al}(z,\bz)\bigr\}$
are defined as
\begin{equation*}\label{TT_OPE}\begin{aligned}
\bigl\{T(w)T(w)\bigr\}\;=& \; \frac{c}{2(z-w)^4}+\frac{2}{(z-w)^2}T(w)+
\frac{1}{z-w}\pa_wT(w), \\
\bigl\{T(w)V_{\al}(z,\bz)\bigr\}\;= & 
\;\frac{\De_{\al}}{(z-w)^2}V_{\al}(w)+
\frac{1}{z-w}\pa_w V_{\al}(w),
\end{aligned}\end{equation*}
together with a similar equation for $\Om(\bar{T}(\bw)\dots)$, and furthermore:
\item {\sc Global $SL(2,\BC)$-invariance} $\frac{\quad}{}$   
\begin{equation*}
\Om\Biggl(   \;\prod_{i=1}^N V_{\al_i}(z_i)\;\Biggl)\;=\; 
 \Om\Biggl(   \;\prod_{i=1}^N 
\bigl|\be z_i+\de \bigr|^{-4\De_{\al_i}}
V_{\al}\Bigl(\frac{\al z_i+\ga}{\be z_i+\de}\Bigr)\Biggl).
\end{equation*}
\end{enumerate}

The verification of the conformal Ward identities is 
possible, but nontrivial in low orders of the semiclassical expansion 
for the path-integral \cite{Ta}, where 
it is possible to identify $T(z)$, $\bar{T}(\bz)$  with the 
following functions of the Liouville field $\phi(z,\bz)$:
\begin{equation*}\begin{aligned} 
T(z) & \underset{b\ra 0}{\simeq} -(\pa_z\phi)^2+b^{-1}\pa_z^2\phi,\\
\bar{T}(\bz) & \underset{b\ra 0}{\simeq} 
-(\pa_{\bz}\phi)^2+b^{-1}\pa_{\bz}^2\phi,
\end{aligned}\quad\text{and}
\quad
V_{\al}(z,\bz)\underset{b\ra 0}{\simeq} e^{2\al\phi}.
\end{equation*}
From that point of view we assume here that $T(z)$, $\bar{T}(\bz)$ and 
the $V_{\al}$, $\al\in\BC$ have quantum counterparts for which the 
above formulation of conformal invariance still holds up to 
quantum corrections of the parameters $c$ and $\De_{\al}$. 
Let us anticipate the following relations between the parameters 
$b$, $\al$ and $c$, $\De_{\al}$:
\begin{equation}
c\;=\;1+6Q^2\qquad \De_{\al}\;=\;\al(Q-\al), \qquad Q=b+b^{-1}
\end{equation}

\begin{rem} Let us remember two simple consequences of these
assumptions: First, the conformal Ward identities can be read as 
a rule that allows to recursively express general vacuum expectation values
\rf{VEV} in terms of those which contain only the field $V_{\al}$.  
Second, 
the property of global $SL(2,\BC)$-invariance
allows one to determine part of the dependence of the 
vacuum expectation values on the variables $z_i$, in particular 
for $n=1,2,3,4$:
\begin{equation}\label{globalsl2}\begin{aligned}
\Om\bigl(V_{\al_1}(z_1)\bigr)\;\,=\;\,& 0\\
\Om\bigl(V_{\al_2}(z_2)V_{\al_1}(z_1)\bigr)
\;\,=\;\,&|z_{21}|^{-4\De_1}\bigl(N(\al_1)\de_{\al_2,Q-\al_1}+
\de_{\al_2,\al_1}B(\al_1)\bigr) \\
\Om\bigl(V_{\al_3}(z_3)\ldots V_{\al_1}(z_1)\bigr)\;\,=\;\,& 
|z_{32}|^{2\De_{32}}|z_{31}|^{2\De_{31}}|z_{21}|^{2\De_{21}}
C(\al_3,\al_2,\al_1)\\
\Om\bigl(V_{\al_4}(z_4)\ldots V_{\al_1}(z_1)\bigr)
\;\,=\;\,&
|z_{43}|^{2(\De_2+\De_1-\De_4-\De_3)}|z_{42}|^{-4\De_2}
|z_{41}|^{2(\De_3+\De_2-\De_4-\De_1)}\\
&|z_{32}|^{2(\De_4-\De_1-\De_2-\De_3)}G_{\al_4,\al_3,\al_2,\al_1}(z,\bz),
\end{aligned}\end{equation}
where $z_{ij}=z_i-z_j$, $\De_i=\De_{\al_i}$, 
$\De_{ij}=\De_k-\De_i-\De_j$ 
if $i\neq j$, $j\neq k$, $k\neq i$
and
$z=\frac{z_{43}z_{21}}{z_{42}z_{32}}$.
\end{rem}

\subsection{Descendants}

It will also be useful to keep in mind that further fields
can be generated from the field $V_{\al}(z,\bz)$  
by ``acting on it with $T(z)$, $\bar{T}(\bz)$''. More precisely,
let $\CW_{\al}$ be the representation of the 
Virasoro algebra that is generated by
acting with the generators $\SL_{n}$, $\bL_{n}$ $n<0$ on the vector $v_{\al}$
that satisfies $\SL_0v_{\al}=\bL_0v_{\al}=\De_{\al}v_{\al}$, 
$\SL_nv_{\al}=0=\bL_nv_{\al}$ for $n>0$. Clearly 
$\CW_{\al}\equiv \CV_{\al}\ot\CV_{\al}$ where 
$\CV_{\al}$ is a Verma-module over the Virasoro 
algebra (see Section \ref{Verma} for a summary
of some relevant results on the representation theory of the Virasoro 
algebra). 

One may then define fields $V_{\al}(\zeta|z,\bz)$ parameterized by vectors
$\zeta\in\CW_{\al}$ 
by means of the following recursive definition:
Let $V_{\al}(v_{\al}|z,\bz)\equiv V_{\al}(z,\bz)$, and 
extend the definition to a basis for $\CW_{\al}$ 
by means of the recursion relation
\begin{equation}\label{TO-OPE}
T(w)V_{\al}(\zeta|z,\bz)\;=\;
\sum_{n=-N(\zeta_{\al})}^{\infty}(w-z)^{n-2}V_{\al}(\SL_{-n}\zeta|z,\bz),
\end{equation}
and its obvious counterpart with $\bT(\bz)$. The 
number $N(\zeta)$ appearing in \rf{TO-OPE}
is the smallest positive integer
such that $\SL_n\zeta=0$ for any 
$n>N(\zeta)$. Equation \rf{TO-OPE} is to be understood
as definition of $V_{\al}(\SL_{-n}\zeta|z,\bz)$ in terms 
of $V_{\al}(\SL_{-m}\zeta|z,\bz)$, $m<n$ by shuffling the terms with 
$m<n$ to the left hand side of the equation, taking $n-2$
derivatives w.r.t. $w$ and finally the limit $w\ra z$,
explicitly:
\begin{equation}\begin{aligned}
(n-2)!\;V_{\al} & (\SL_{-n}\zeta|z,\bz)\;=\; \\
= & \;\lim_{w\ra z}
\pa_w^{n-2}
\left(
T(w)V_{\al}(\zeta|z,\bz)-
\sum_{m=-N(\zeta_{\al})}^{n-1}(w-z)^{m-2}V_{\al}(\SL_{-m}\zeta|z,\bz)\right).
\end{aligned}\end{equation}
It is clear from the definition \rf{TO-OPE} of the descendants
that the conformal Ward identities allow one to express the vacuum 
expectation values of fields $V_{\al}(\zeta|z,\bz)$ in terms of those
for $V_{\al}(z,\bz)$. It will also be useful to note that this fact 
together with global $SL(2,\BC)$-invariance imply the existence of 
\begin{equation}\label{vinfty}
\Om\bigl(V_{\al}(\zeta|\infty)\dots\bigr)\;\equiv\;
\lim_{z\ra\infty}z^{2\De(\zeta)}\bz^{2\bar{\De}(\zeta)}
\Om\bigl(V_{\al}(\zeta|z,\bz)\dots\bigr)
\end{equation}
whenever 
$\zeta\in\CW_{\al}$ is such that both
$\SL_0$ and $\bL_0$ act diagonally 
according to $\SL_0 \zeta=\De(\zeta)\zeta$ and 
$\bL_0 \zeta=\bar{\De}(\zeta)\zeta$.

\subsection{The DOZZ-formula}

The second main ingredient of the DOZZ-proposal is an explicit formula
for the function $C(\al_3,\al_2,\al_1)$ that represents 
the part of the three point function which is not determined by 
$SL(2,\BC)$-invariance:

Let us define the special function $\up(x)$ by the integral representation
\begin{equation}\label{upint}
\log\Upsilon(x)=
\int\limits_0^\infty\frac{dt}{t}\;\biggl[\biggl(\frac{Q}{2}-x\biggr)^2e^{-t}
-\frac{\sinh^2\bigl(\frac{Q}{2}-x\bigr)\frac{t}{2}}{\sinh\frac{bt}{2}
\sinh\frac{t}{2b}}\biggr]
\end{equation}
The formula as proposed in \cite{ZZ} is the following: 
\begin{equation}\label{ZZform}\begin{aligned}
C(\alpha_1, & \alpha_2,\alpha_3)=\left[\pi\mu\gamma(b^2)b^{2-2b^2}
\right]^{(Q-\sum_{i=1}^3\alpha_i)/b}\times\\
& \times \frac{\Upsilon_0\Upsilon(2\alpha_1)\Upsilon(2\alpha_2)
\Upsilon(2\alpha_3)}{
\Upsilon(\alpha_1+\alpha_2+\alpha_3-Q)
\Upsilon(\alpha_1+\alpha_2-\alpha_3)\Upsilon(\alpha_2+\alpha_3-\alpha_1)
\Upsilon(\alpha_3+\alpha_1-\alpha_2)}.
\end{aligned}\end{equation}
For later use let us summarize a couple of relevant properties of
the $\up$-function. First, it is useful to note that
it can also be constructed out of the Barnes Double
Gamma function $\Ga_2(x|\om_1,\om_2)$ \cite{Ba} as
\begin{equation}\label{updef}\begin{aligned}
{} & \up_b(x)=
\frac{1}{\Ga_b(x)\Ga_b(Q-x)},\qquad \Ga_b(x)=\Ga_2(x|b,b^{-1}), \quad 
\text{where}\\
& \log\Ga_2(s|\om_1,\om_2)=  \Biggl(\frac{\pa}{\pa t}\sum_{n_1,n_2=0}^{\infty}
(s+n_1\om_1+n_2\om_2)^{-t}\Biggr)_{t=0}.
\end{aligned}\end{equation}
One may thereby benefit from the existence of some literature on the 
Barnes Double
Gamma function, cf. e.g. \cite{Ba,Sh}. 

The following basic properties follow then directly from the integral 
representation \rf{upint} or results on $\Ga_b$ found in \cite{Sh}:

\noindent{\sc Functional equations} $\frac{\quad}{}$
\begin{equation}\label{upfunrel}
\Upsilon(x+b)=\gamma(bx)b^{1-2bx}\Upsilon(x).
\end{equation}
\noindent{\sc Self-duality} $\frac{\quad}{}$
\begin{equation}
\up_b(x)=\up_{b^{-1}}(x).
\end{equation}
\noindent{\sc Reflection property} $\frac{\quad}{}$
\begin{equation}\label{reflprop}
\up_b(x)=\up_b(Q-x).
\end{equation}
\noindent{\sc Analyticity} $\frac{\quad}{}$
$\up_b(x)$ is entire analytic with zeros at
$x=-nb-mb^{-1}$ and $x=Q+nb+mb^{-1}$, $n,m\in\BZ^{\geq 0}$.

\noindent{\sc Asymptotic behavior} $\frac{\quad}{}$
\begin{equation}\label{SGas}
\up_b(x) \sim x^2\log x+\fr{3}{2}x^2\mp \pi i x^2 +Qx\log x +\CO(x)\quad
\text{for $\Im(x)\ra\pm \infty$}. 
\end{equation}

\section{Path integral approach} \label{pathsect}

We will now discuss some of the motivation that has led Dorn, Otto
and Al.B., A.B. Zamolodchikov to propose the formula \rf{ZZform}:
Let us consider fields $V_{\al}(w,\bw)$ on the euclidean cylinder, which
correspond to the classical functions $e^{2\al\phi}$ of the 
Liouville field. One looks for a representation for the correlation 
functions of the fields $V_{\al}(w,\bw)$ 
as an integral over all possible field
configurations, 
\begin{equation} \label{corr_pi} \begin{aligned}
\Om\bigl(V_{\al_N}(w_N,\bw_N) \ldots V_{\al_1}
(w_1,\bw_1)\bigr)
\;
= \; \lim_{T\ra\infty}\;
\int[\CD\phi]\;\,e^{-S_T[\phi]}\;\prod_{i=1}^N
e^{2\al_i\phi(w_i,\bw_i)},
\end{aligned}\end{equation} 
where the configuration $\phi(\si,\tau)$ is 
weighted with a measure 
$[\CD\phi]e^{-S_T[\phi]}$ which is written in terms 
of the euclidean action
\begin{equation}\label{euclact}
S_T[\phi]\;=\;\int\limits_{-T}^{T}d\tau\int\limits_{0}^{2\pi}d\si\; 
\biggl(
\frac{1}{\pi}|\pa_w\phi|^2+\mu e^{2b\phi}\biggr).
\end{equation}
We refer to \cite{Ta} for a discussion of conformal symmetry in the
perturbative path integral framework, with interesting links to 
Teichm\"uller theory. Here we restrict ourselves to a discussion
of the information that can be obtained when 
the measure $[\CD\phi]$ factorizes as $[\CD\phi]=d\phi_0
\bigl[\CD\bar{\phi}\bigr]$,
where the Liouville field has been split
into zero mode and oscillator parts, 
$\phi(w,\bw)=\phi_0+\bar{\phi}(w,\bw)$. This approach goes back to \cite{GL}
and was further developed by \cite{D2}\cite{DK}.

\subsection{Path integral on the Riemann sphere}

For many purposes it is convenient not to perform the path integral
over fields $\phi$ defined on the cylinder, but instead to integrate over
configurations that are defined on the Riemann-sphere, which is related
to the cylinder by the conformal mapping $z=e^w$. A bit of care is
needed when transforming the action \rf{euclact}. 

Corresponding to a conformal transformation $z=z(w)$ one may consider the
following transformation law of the Liouville field $\phi$:
\begin{equation}\label{trsf}
\phi(z)=\phi'(w(z))-\frac{Q}{2}\log\biggl|\frac{dz}{dw}\biggr|^2.
\end{equation}
If the parameter $Q$
is chosen equal to $Q_{\rm c}=b^{-1}$, one has {\it invariance} of the
classical action \rf{euclact} {\it up to boundary terms}.  
In the case of the transformation $z=e^w$ one finds that 
\rf{euclact} is transformed into the following expression:
\begin{equation}\label{sphereact}
S_T[\phi]\;\;=\;\;\frac{1}{4\pi}
\int\limits_{A_R}d^2z\;\left[(\partial_a\phi)^2+
4\pi\mu e^{2b\phi}\right]\;
+\;\frac{Q}{\pi R}\int\limits_{\partial A_R}dl\;\phi \;+\; 2Q^2\log R,
\end{equation}
where $A_R$ is an annulus around $z=0$ with outer and inner radii
given by $R=e^T$ and $1/R$ respectively. 
The boundary term is often interpreted as describing the
effect of a ``background charge $-Q$ at infinity''. In the case of the
quantum theory one will have to consider values  
of $Q$ that differ from the classical value $Q_{\rm c}=b^{-1}$. 
 
\subsection{Scaling behavior}

We will now assume that the measure $[\CD\phi]e^{-S[\phi]}$ factorizes as
\begin{equation}\label{measure_fact}
[\CD\phi]e^{-S[\phi]}=d\phi_0[\CD\bar{\phi}]_{\mu,\phi_0},
\end{equation}
and that the measure for the integration over 
the zero mode $\phi_0$ is translationally invariant. The latter
assumption does not only require translational invariance of
$d\phi_0$, but also that 
\begin{equation}
[\CD\bar{\phi}]_{\mu,\phi_0}^{}\;=\;[\CD\bar{\phi}]_{e^{2ba}\mu,\phi_0-a},
\end{equation}
as is satisfied by the weight $e^{\mu \int d^2z e^{2b\phi}}$. 
By introducing a new integration variable 
$\phi'_0=\phi_0+\frac{1}{2b}\ln\mu$
it is then possible to extract the $\mu$-dependence of 
vacuum expectation values
\rf{corr_pi}:
\begin{equation} \label{scaling}\begin{aligned}
\Om\bigl(V_{\al_N}  (z_N,\bz_N) & \ldots V_{\al_1}
(z_1,\bz_1)\bigr)_{\mu}^{}
\;=\; \\
=\;\mu^{s}\; & \Om\bigl(V_{\al_N}(z_N,\bz_N) 
\ldots 
V_{\al_1}(z_1,\bz_1)\bigr)_{\mu=1}^{},\end{aligned}\qquad
s\;\equiv\;\frac{1}{b}\Bigl(Q-\sum_{i=1}^N\al_i\Bigr).
\end{equation}
The fact that the dependence of
vacuum expectation values on $\mu$ is generically not analytic in $\mu$ 
indicates that perturbation theory w.r.t. the variable $\mu$ 
can not lead to convergent series expansions for the expectation values.

\subsection{Relation to free field?} \label{freef:3}

Let us consider the question of convergence of the integration
over $\phi_0$. We will assume that the upper limit in the
$\phi_0$ does yield any problems due to the factor 
$e^{-\mu \int d^2z e^{2b\phi_0}}$ coming from the Liouville-interaction.
The leading asymptotic behavior for $\phi_0\raf$ is proportional
to $e^{-2sb\phi_0}$, so we expect convergence of the integration
over $\phi_0$ as long as $\Re(s)<0$.

The fact that the dependence of the fields $e^{2\al\phi}$ on the 
variable $\al$ is analytic suggests that the dependence 
of the vacuum expectation values on the variable $s$ might be 
analytic as long as the $\phi_0$-integration converges. In the same
spirit one might hope to get a {\it meromorphic} continuation
to $\Re(s)>0$ by standard regularization of the $\phi_0$-integration
\cite{GS}: This would be possible if the asymptotic 
behavior of $[\CD\bar{\phi}]_{\mu,\phi_0}$ for
$\phi_0\raf$ was known. The vanishing of the Liouville interaction
for $\phi_0\raf$ leads one to suspect that $[\CD\bar{\phi}]_{\mu,\phi_0}$
is asymptotic to the Gaussian measure of the path integral for
a free field theory, with corrections given by the interaction term:
\begin{equation}\label{measure_as}
{[}\CD\bar{\phi}{]}_{\mu,\phi_0}^{}\;\underset{q\ra-\infty}{\sim}\;
{[}\CD\bar{\phi}{]}_{{\rm F},Q}^{}\;
\sum_{n=0}^{\infty}\frac{(-\mu)^n e^{2bn\phi_0}}{n!} 
\biggl(\int
d^2z \,e^{2b\bar{\phi}}\biggr)^n.
\end{equation} 
We use the notation ${[}\CD\bar{\phi}{]}_{{\rm F},Q}^{}$ to denote the
standard Gaussian measure in free field theory with background charge
$-Q$, see \cite{DF1} or equations \rf{DFint}, \rf{freecorrs} below for 
its definition. 
We have included in \rf{measure_as} 
corrections which are subleading for $\phi_0\raf$. These corrections 
come from the expansion of the weight $e^{\mu \int d^2z e^{2b\phi}}$
as power series in $\mu$. One should note that
the previous argument indicating
the failure of perturbation theory w.r.t. $\mu$ does not apply here: It 
amounts to the statement that it does not make sense to consider $\mu$ as
small as long as it may be changed by shifting $\phi_0$. Here one
may take $\mu e^{2b\phi_0}$ as small variable that is 
invariant under $(\mu,\phi_0)\ra(e^{2ba}\mu,\phi_0-a)$.

The integration over $\phi_0$ can then be  
regularized in a standard way by subtracting the leading divergencies: 
\begin{equation}\begin{aligned}
\Om\bigl(V_{\al_N} &(z_N,\bz_N)\ldots 
V_{\al_1}
(z_1,\bz_1)\bigr)
\;=\;\\
= & \; \lim_{q_0\ra -\infty} \;\Biggl(\;\,
\int\limits_{q_0}^{\infty}d\phi_0\;e^{-2sb\phi_0}\;
\int
{[}\CD\bar{\phi}{]}_{\mu,\phi_0}^{}\;\prod_{i=1}^N
e^{2\al_i\bar{\phi}(z_i,\bz_i)}\\
& \qquad -\sum_{n=0}^{\infty} 
\frac{(-\mu)^n}{n!} \frac{e^{-2(s-n)bq_0}}{2b(s-n)}\int
{[}\CD\bar{\phi}{]}_{{\rm F},Q}^{}\prod_{i=1}^N
e^{2\al_i\bar{\phi}(z_i,\bz_i)}
\biggl(\int
d^2z \,e^{2b\bar{\phi}}\biggr)^n\Biggr).
\end{aligned}\end{equation}
Poles w.r.t. the variable $s$ are explicitly exhibited. Their residues
are given by path integrals in free field theory with a 
background charge $-Q$.

\subsection{Evaluation of the residues}\label{Residues}

Let us study the path-integrals that describe the residues
of Liouville expectation values at $s=n$:
\begin{equation}\label{}
\CG^n_{\al_N,\ldots,\al_1} (z_N,\ldots,z_1)\;=\;
\frac{(-\mu)^n}{2b n!}\int
{[}\CD\bar{\phi}{]}_{{\rm F},Q}^{}\prod_{i=1}^N
e^{2\al_i\bar{\phi}(z_i,\bz_i)}
\biggl(\int
d^2z \,e^{2b\bar{\phi}}\biggr)^n,
\end{equation}
where $\al_1,\dots,\al_N$ are subject to the relation $s=n$. 
It is well-known how to define the path-integrals on the right hand side 
\cite{DF1}: The result can be represented in the form
\begin{equation} \label{DFint} \begin{aligned}
{}\CG^n_{\al_N,\ldots,\al_1} (z_N,\ldots,z_1)\;\,= 
\frac{(-\mu)^n}{n!}\int d^2t_n\ldots d^2t_1\;
\bra  0|\!& :\!e^{2\al_N\phi(z_N)}\!:   \ldots 
:\!e^{2\al_1\phi(z_1)}\!:\\
& :\!e^{2b\phi(t_n)}\!:\ldots 
:\!e^{2b\phi(t_1)}\!:\!|0\ket^{}_{{\rm F},Q},\end{aligned}
\end{equation}
where correlators of the form $\bra  0|\!:\!e^{2\al_N\phi(z_N)}\!:   \ldots 
e^{2\al_1\phi(z_1)}\!:\!|0\ket^{}_{{\rm F},Q}$ are nonvanishing only 
if $Q-\sum_{i=1}^N\al_i=0$, and in that case given by
\begin{equation}\label{freecorrs}
\bra  0|\!:\!e^{2\al_N\phi(z_N)}\!:   \ldots 
:e^{2\al_1\phi(z_1)}\!:\!|0\ket^{}_{{\rm F},Q}\;=\;
\prod_{i>j}|z_i-z_j|^{-4\al_i\al_j}.
\end{equation}
If the integrals \rf{DFint} are to represent the residues of correlation
functions that satisfy the conformal Ward identities irrespective
of the choice of the $\al_i$, $i=1,\dots,N$, one evidently 
needs that the operator $:\!e^{2b\phi(t)}\!:$ that appears in 
\rf{DFint} transforms as a tensor of weight $(1,1)$ under conformal
transformations, so that $\int d^2t:\!e^{2b\phi(t)}\!:$ is conformally
invariant. This is the case iff the parameters $b$ and $Q$ are related
by $Q=b+b^{-1}$ \cite{DF1}, as will be assumed from now on.

The integrals \rf{DFint} are too complicated to carry out in general,
but in a case of fundamental importance,
namely $N=3$, Dotsenko and Fateev have been able to compute
the integrals \rf{DFint} explicitly and found the following result
\cite{DF2} (see \cite{D1} 
for an explanation of the techniques to calculate
such integrals)
\begin{equation}\label{threeptallg}
\CG^n_{\al_3,\al_2,\al_1} (z_3,z_2,z_1)\;\,=\;\,
|z_{32}|^{2\De_{32}}|z_{31}|^{2\De_{31}}|z_{21}|^{2\De_{21}}I_n(\al_3,\al_2,\al_1)_{\sum_{i=1}^3\al_i=Q-nb},
\end{equation}
where $z_{ij}=z_i-z_j$, $\De_{ij}=\De(\al_k)-\De(\al_i)-\De(\al_j)$ 
if $i\neq j$, $j\neq k$, $k\neq i$, and furthermore 
\begin{equation}\label{DFint3}
I_n(\alpha_1,\alpha_2,\alpha_3)=\left(\frac{-\pi\mu}{\gamma(-b^2)}\right)^n
\frac{\prod_{j=1}^n\gamma(-jb^2)}
{\prod_{k=0}^{n-1}\left[\gamma(2\alpha_1b+kb^2)
\gamma(2\alpha_2b+kb^2)\gamma(2\alpha_3b+kb^2)\right]}.
\end{equation} 
In \rf{DFint3} we have used the notation
\begin{equation}
\gamma(x)=\Gamma(x)/\Gamma(1-x).
\end{equation}

\subsection{Continuation in $s$} 

It is then natural to look for a ``continuation'' of the results
for $s\in\BZ^{\geq 0}$ to all of $\BC$ as a candidate for the three point 
function at generic values of $s$. 
More precisely, 
the task is to find an expression that depends meromorphically
on $s$ and has poles for $s\in\BZ^{\geq 0}$ with residues given 
by \rf{threeptallg}\rf{DFint3}. It is natural and important to
demand that this continuation preserves the $z_i$-dependence of the
three point function as given in \rf{threeptallg}, i.e. that
it takes the general $SL(2,\BC)$-invariant form
\begin{equation}
\label{threeptallg+}
\Om\bigl(V_{\al_3}(z_3,\bz_3) 
V_{\al_2}(z_2,\bz_2)
V_{\al_1}(z_1,\bz_1)\bigr)\;\,=\;\,
|z_{32}|^{2\De_{32}}|z_{31}|^{2\De_{31}}|z_{21}|^{2\De_{21}}
C(\al_3,\al_2,\al_1).
\end{equation}

The task is then to find an expression for $C(\al_3,\al_2,\al_1)$ that
has poles for $s\in\BZ^{\geq 0}$ with residues given by \rf{DFint3}.
A very promising candidate 
for such a continuation has been proposed by Dorn and Otto \cite{DO}
and independently by AL.B. and A.B. Zamolodchikov \cite{ZZ}.
A possible starting point for finding their proposal may be 
the observation that the known result for $s\in\BZ^{\geq 0}$ 
has an interesting recursive structure w.r.t. shifts of one
of the arguments by an amount of $b$:
\begin{equation} \begin{aligned}\label{Crecrel}
 {} & \frac{C(\al_3,\al_2,\al_1+b)}{C(\al_3,\al_2,\al_1)}
=\\
&= -\frac{\ga(-b^2)}{\pi\mu}
\frac{\ga(b(2\al_1+b))\ga(2b\al_1)\ga(b(\al_2+\al_3-\al_1-b))}
{\ga(b(\al_1+\al_2+\al_3-Q))\ga(b(\al_1+\al_2-\al_3))
\ga(b(\al_1+\al_3-\al_2))}.
\end{aligned}\end{equation}
The functional relation \rf{Crecrel} is a priori only known for 
$\al_3,\al_2,\al_1$ subject to the constraint $\sum_{i=1}^3\al_i=Q-sb$
for some $s\in\BZ^{\geq 0}$. However, it seems to be a natural guess that
this recursive relation may even be valid for general values 
of $\al_3,\al_2,\al_1$.  
One may then easily convince oneself that
a solution to \rf{Crecrel} can be assembled if one was 
given as building block 
a function called $\up(x)$ that satisfies the functional 
equation \rf{upfunrel} of the $\up$-function introduced in \rf{updef}.

These facts alone can hardly be considered as strong motivation
for considering the DOZZ-proposal as a promising candidate for 
the three point function of exponential fields in Liouville theory.
One should therefore emphasize that this formula has passed a couple 
of rather nontrivial checks: 
For example, it was shown to imply a quantum version of the 
Liouville equation of motion in \cite{DO} (see also our discussion
in Part \ref{canq1}). The checks performed in \cite{ZZ}
include comparison with semiclassical calculations in two 
different limiting regimes, comparison with results from the 
thermodynamic Bethe ansatz for the sinh-Gordon model (which is
related to Liouville theory in the ultraviolet limit), as 
well as a numerical check of crossing symmetry for the 
four-point functions constructed from the three point functions via 
factorization. 
 
\subsection{Duality $b\ra b^{-1}$}

The expression \rf{ZZform} has (at least) one amazing feature: It is left
invariant if one replaces $b\ra \tilde{b}\equiv b^{-1}$ and furthermore 
\begin{equation}\label{dualiz}
\mu\ra \tilde{\mu}\quad \text{where $\tilde{\mu}$ is defined by}\quad
\pi \tilde{\mu}\ga(\tilde{b}^2)=\bigl(\pi \mu \ga(b^2)\bigr)^{b^{-2}}. 
\end{equation}
Remembering that $b^2$ was playing the role of a coupling 
constant or alternatively the role of $\hbar$, this 
indicates a rather remarkable and profound
self-duality of Liouville theory.

But it also raises a puzzle: The expression \rf{ZZform} has more
poles than expected on the basis of our previous discussion in Subsection
\ref{freef:3}. Poles occur if 
\begin{equation}\label{DOZZpoles}
Q-\al_1-\al_2-\al_3=nb+mb^{-1}, \qquad n,m\in\BZ^{\geq 0}
\end{equation} and
all cases obtained by the reflections $\al_i\ra Q-\al_i$, $i=1,2,3$ from
\rf{DOZZpoles}. Is there a way to explain these poles from the path-integral
point of view? 
According to our arguments in Subsection
\ref{freef:3} one has a relation between the asymptotic 
behavior of the path integral measure $[\CD\bar{\phi}]_{\mu,\phi_0}$
for $\phi_0\raf$ and the poles of the three point function. The 
additional poles in \rf{DOZZpoles} should therefore be attributed to a 
modification of the asymptotic expansion \rf{measure_as}. In fact,
if one replaces the right hand side of \rf{measure_as} by 
\begin{equation}\label{measure_as'}
{[}\CD\bar{\phi}{]}_{{\rm F},Q}^{}\;
\sum_{m,n=0}^{\infty}\frac{(-\mu)^n(-\tilde{\mu})^m 
e^{2(bn+b^{-1}m)\phi_0}}{m!n!} 
\biggl(\int
d^2z \,e^{2b\bar{\phi}}\biggr)^n\biggl(\int
d^2z' \,e^{2\tilde{b}\bar{\phi}}\biggr)^m, 
\end{equation}
and continues as in subsections \ref{freef:3} and \ref{Residues}, 
one would find additional poles at \rf{DOZZpoles}
with residues represented
by the Dotsenko-Fateev integrals \cite{DF2} 
\begin{equation}\label{DFint+}\begin{aligned}
{} \CG^{n,m}_{\al_n,\ldots,\al_1} (z_n,\ldots,z_1)\;
=\; & \frac{(-\mu)^n(-\tilde{\mu})^m}{m!n!} 
\int\limits_{\BC} d^2t_n\ldots d^2t_1\;
\int\limits_{\BC} d^2s_m\ldots d^2s_1\;\\
& \qquad\bra  0|\prod_{i=1}^3:e^{2\al_i\phi(z_i)}:
\prod_{j=1}^n:e^{2b\phi(t_j)}:
\prod_{k=1}^m:e^{2\tilde{b}\phi(s_k)}:
|0\ket^{}_{{\rm F},Q}.\end{aligned}
\end{equation}
And indeed, the residues of the 
proposal \rf{DOZZpoles} coincide precisely with the result of 
the explicit evaluation of \rf{DFint+} performed in \cite{DF2}.
This was observed in \cite{DO}\cite{ZZ}, and elaborated upon in 
\cite{OPS1}.

It seems natural to interpret the modification \rf{measure_as'}
as describing a quantum correction to the path integral measure
that could be described by adding a second interaction term
$\tilde{\mu}\int d^2z e^{2b^{-1}\phi}$ to the action.
Such a modification is compatible with 
conformal invariance due to the fact \cite{DF1} that the normal ordered 
exponential $:e^{2b^{-1}\phi}:$ transforms under 
conformal transformations the same way as
$:e^{2b\phi}:$, namely as $(1,1)$-tensor field.
The modified action is clearly self-dual under $b\ra b^{-1}$,
$\mu\ra\tilde{\mu}$.

\subsection{Measure in the bulk of $\phi_0$-space?}

It has become clear that the appearance of poles in the
dependence of vacuum expectation values in their dependence w.r.t.
the parameters $\al_N,\dots,\al_1$ and the explicit form
of the corresponding residues are entirely explained in terms of 
the asymptotic behavior of the path-integral measure for zero mode 
$\phi_0\raf$. The free-field vacuum expectation values that
represent the residues may be called ``resonant amplitudes''
following \cite{DK}.
Are there any hints concerning the 
definition of the path-integral for the ``non-resonant'' amplitudes? 

The following observation from \cite{OPS2} is intriguing:
Let us tentatively assume that the measure 
$d\phi_0[\CD\bar{\phi}]_{\mu,\phi_0}$ is of the form 
\begin{equation}\label{OPShyp}
d\phi_0[\CD\bar{\phi}]_{\mu,\phi_0}=d\phi_0[\CD\bar{\phi}]'
e^{-\int d^2z (\mu 
e^{2b\phi(z)}+\tilde{\mu}
e^{2\tilde{b}\phi(z)})},
\end{equation}
where $[\CD\bar{\phi}]'$ is independent of $\phi_0$. In order to reproduce
the scaling behavior \rf{scaling}, we will require it to be 
$\mu$-independent as well \footnote{Note that the inclusion of 
the ``dual interaction''
$\tilde{\mu}\int d^2z e^{2b^{-1}\phi}$ preserves the scaling
\rf{scaling} if $\tilde{\mu}$ and $\mu$ are related by 
\rf{dualiz}}. 
$[\CD\bar{\phi}]'$ must then coincide with 
${[}\CD\bar{\phi}{]}_{{\rm F},Q}$ for $d\phi_0[\CD\bar{\phi}]_{\mu,\phi_0}$
to have asymptotics given by \rf{measure_as'}. 
One may then rewrite the condition of translation invariance of the
$\phi_0$-measure, $0=\int_{-\infty}^{\infty}d\phi_0\pa_{\phi_0}$,  
as a relation between vacuum expectation values with 
different numbers of operator insertions:
\begin{equation}\label{sumrule}\begin{aligned}
2b\mu\int\limits_{\BC}d^2z & \;\Om\Biggl(V_b(z,\bz)\prod_{i=1}^N
V_{\al_i}(z_i,\bz_i)\Biggl)+
2b^{-1}
\tilde{\mu}\int\limits_{\BC}d^2z\; \Om\Biggl(V_{b^{-1}}(z,\bz)\prod_{i=1}^N
V_{\al_i}(z_i,\bz_i)\Biggl)\;=\\
& =\;-2\biggl(\sum_{i=1}^N\al_i-Q\biggr)
\Om\Biggl(\prod_{i=1}^N V_{\al_i}(z_i,\bz_i)\Biggl).
\end{aligned}\end{equation}
This condition can be evaluated more explicitly in the case N=2 \cite{OPS2}.
One gets a relation between $C(b,\al,\al)$, $C(b^{-1},\al,\al)$ and 
the two-point function
\begin{equation}\label{sp_sumrule}
\Om\bigl(V_{\al_2}(z_2,\bz_2)V_{\al_1}(z_1,\bz_1)\bigr)\;=\;
\bigl(2\pi\de(\al_2+\al_1-Q)+S(\al)\de(\al_2-\al_1)\bigr)
|z_2-z_1|^{-4\De_{\al_1}}, 
\end{equation} 
which was observed to be fulfilled by the DOZZ-proposal 
in \cite{OPS2}.\footnote{In fact, it is argued in \cite{OPS2} 
that the DOZZ-proposal
is the unique expression that satisfies \rf{sumrule} and has 
residues given by the Dotsenko-Fateev integrals \rf{DFint+}. We did
not understand the assumptions underlying the uniqueness
argument of \cite{OPS2} well enough to include a discussion here.}
The objects that appear in \rf{sp_sumrule} are 
not expected to be determined by the ($\phi_0\raf$)-asymptotics of the 
measure alone. The fact that the DOZZ-proposal satisfies the condition
in the case $N=2$ 
may therefore be taken as a hint that \rf{OPShyp} is correct,
i.e. that there are no further corrections to the measure besides adding
the dual interaction $\tilde{\mu}\int d^2z e^{2b^{-1}\phi}$.

Nevertheless it is not clear to us what conclusion to draw from 
these observations. First, it is not clear whether one should 
expect the ``sum-rule'' \rf{sumrule} to be valid in general. The case
$N=2$ for which consistency with the DOZZ-proposal was verified 
is still somewhat special. Moreover, one should observe that 
the compatibility of the DOZZ-proposal with a literal interpretation
of \rf{OPShyp} is not obvious: The former seems to
imply that $\tilde{\mu}$ 
as given in \rf{dualiz} can become negative for certain values of $b$.
But this would imply trouble with the {\it upper} limit of the 
integration over $\phi_0$!

We will come back to this problem from a different point of view
in Part \ref{canq1}.

\section{Reconstruction} \label{reconst}\vspace{.5cm}

For rational conformal field theories it is well-known that the
two- and three point functions of the set $\{V_{\imath};\imath\in\CI\}$
of all primary fields
suffices to reconstruct the Hilbert-space $\CH$ of the theory
and to fully characterize
the operators $\SV_{\imath}$ that correspond to the fields $V_{\imath}$.
One may therefore suspect that really all that one might want to 
know about Liouville theory can (at least in principle) be extracted
from the DOZZ formula and conformal invariance. In the present
section we will discuss an
adaption of the reconstruction procedure from 
rational conformal field theories to Liouville theory. 
The DOZZ-proposal will be found to encode 
the spectrum of Liouville theory in a natural way. Moreover, the 
identification between three point functions of fields $V_{\al}$ 
and matrix elements of the corresponding operators $\SV_{\al}$ 
will be found to work straightforwardly for $\al$ with $Q>\Re(\al)>0$. 
However, subtleties that are unfamiliar from rational conformal field theories
arise due to the fact that the $SL(2,\BC)$-invariant state $|0\ket$
is not found in the spectrum. Furthermore, 
the description of the operators $\SV_{\al}$ in terms 
of matrix elements turns out to be more subtle in the cases where 
$\Re(\al)$ is not in $(0,Q)$.

We believe the following considerations to be important for understanding
how particular features of the DOZZ-formula (like analyticity, pole structure,
values of residues) are crucial for the success of such a reconstruction
procedure.

\subsection{Reconstruction of rational conformal field theories}

Consider a rational conformal field theory $C$ 
with set $\CP_{C}\equiv\{V_{\imath};\imath\in\CI\}$
of all primary fields. 
Let us assume for simplicity that left- and right conformal dimensions of
the primary field $V_{\imath}$ coincide: $\De_{\imath}=\bar{\De}_{\imath}$.
It is often convenient to assume that 
$\CP_{C}$ includes the identity corresponding to the label $\imath=0$. 
Assume furthermore to be given the set of all 
three-point functions of the form
\begin{equation}\label{globalsl2_rat}
\Om\bigl(V_{\imath_3}(z_3)\ldots V_{\imath_1}(z_1)\bigr)\;\,=\;\,
|z_{32}|^{2\De_{32}}|z_{31}|^{2\De_{31}}|z_{21}|^{2\De_{21}}
C(\imath_3,\imath_2,\imath_1),
\end{equation}
where $C(\imath_3,\imath_2,\imath_1)$ are real numbers.
The two-point functions are obtained by setting any of 
$\imath_3,\imath_2,\imath_1$ to $0$.  
The conformal Ward identities allow one to 
recover two- and three point functions of all descendants 
$V_{\imath}(\zeta|z)$, $\zeta\in\CR_{\De_{\imath}}\ot\CR_{\De_{\imath}}$
from the data \rf{globalsl2_rat}, where $\CR_{\De}$ denotes the
irreducible representation of the Virasoro algebra with highest 
weight $\De$. One wants to identify 
\begin{equation}\label{VEVident}
\Om\bigl(V_{\imath_n}(\zeta_n|z_n)\dots V_{\imath_1}(\zeta_1|z_1)\bigr)
\;\equiv\;
\bra 0|\SV_{\imath_n}(\zeta_n|z_n)\dots \SV_{\imath_1}(\zeta_1|z_1)|0\ket.
\end{equation}
The vacuum state $|0\ket$ must have the property $\SL_n|0\ket=\bL_n|0\ket=0$
$n=-1,0,1$ for the right hand side of \rf{VEVident} to share the 
property of $SL(2,\BC)$-invariance. This property of $|0\ket $
assures existence of the following limits:
\begin{equation}\label{obst1}
|{\imath},\zeta\ket_{\rm in}^{}\;\equiv\;
\lim_{z\ra 0}\SV_{\imath}(\zeta|z)|0\ket  ,
\qquad
{}_{\rm out}^{}\bra{\imath},\zeta|\;\equiv\;
\lim_{z\ra \infty} z^{2\De(\zeta)}\bz^{2\bar{\De}(\zeta)}
\bra 0|\SV_{\imath}(\zeta|z),
\end{equation} 
where it was assumed that $\zeta\in\CR_{\De_{\imath}}\ot\CR_{\De_{\imath}}$
is an eigenvector of both 
$\SL_0$ and $\bL_0$ with eigenvalues $\De(\zeta)$ and 
$\bar{\De}(\zeta)$ respectively.
The identification \rf{VEVident} together with operator-state correspondence
therefore allow one to recover matrix elements of $\SV_{\imath}$ as
\begin{equation}\label{mat_el1}\begin{aligned}
{}_{\rm out}^{}\bra\imath_3,\zeta_3|\SV_{\imath_2}(\zeta_2|z_2)& 
|\imath_1,\zeta_1\ket_{\rm in}^{}
\;\equiv\;\\ 
\;\equiv\; & \lim_{z_1\ra 0}\lim_{z_3\ra \infty} 
z_3^{2\De(\zeta_3)}\bz_3^{2\bar{\De}(\zeta_3)}
\Om\bigl(V_{\imath_3}(\zeta_3|z_3)V_{\imath_2}(\zeta_2|z_2)
V_{\imath_1}(\zeta_1|z_1)\bigr).
\end{aligned}\end{equation}
Let us next observe that compatibility of 
\[ 
\bigl( {}_{\rm out}^{}\bra\imath_3,\zeta_3|\SV_{\imath_2}(\zeta_2|z_2)
|\imath_1,\zeta_1\ket_{\rm in}^{}\bigr)^*={}_{\rm in}^{}\bra\imath_1,\zeta_1|
\bigl(\SV_{\imath_2}(\zeta_2|z_2)\bigr)^{\dagger}
|\imath_3,\zeta_3\ket_{\rm out}^{}\] 
with \rf{globalsl2_rat} requires 
that 
\begin{equation}
\label{herm1}
\bigl(V_{\imath}(\zeta|z)\bigr)^\dagger\;=\; 
\bz^{-2\De(\zeta)}z^{-2\bar{\De}(\zeta)}
V_{\imath}\bigl(\bar{\zeta}|\bz^{-1}\bigr),
\end{equation}
where $\bar{\zeta}$ is the complex conjugate of the vector $\zeta$.
This leads to the following relation between in- and out states:
\begin{equation}\label{indaggerout}\begin{aligned}
{}_{{\rm in}}\bra \imath,\zeta|\;\equiv\;
\bigl(|\imath,\zeta\ket_{{\rm in}}\bigr)^\dagger\;=\; &
\bigl(\lim_{z\ra 0}\; V_{\imath}(\zeta|z)|0\ket \bigr)^\dagger\;=\;
\lim_{z\ra 0}\; \bra 0|\bigl(V_{\imath}(\zeta|z)\bigr)^{\dagger}\\
\;=\; & \lim_{z\ra 0}\;\bra 0|\,\bz^{-2\De(\zeta)}z^{-2\bar{\De}(\zeta)}
V_{\imath}\bigl(\bar{\zeta}|\fr{1}{\bz}\bigr)\\
\;=\; & {}_{{\rm out}}\bra\imath,\bar{\zeta}|. 
\end{aligned}\end{equation}
Taken together this means that the Hilbert space $\CH_C$ is given as
\begin{equation}
\CH_C\;=\;\bigoplus_{\imath\in\CI}
\CV_{\De_{\imath}}^{}\ot\CV_{\De_{\imath}}^{},
\end{equation}
with scalar product given by
\begin{equation}
{}_{\rm in}\bra \imath_2,\zeta_2|\imath_1,\zeta_1\ket_{\rm in}^{}\;=\;
\lim_{z_1\ra 0}\lim_{z_2\ra \infty} 
z_2^{2\De(\zeta_2)}\bz_2^{2\bar{\De}(\zeta_2)}
\Om\bigl(V_{\imath_2}(\bar{\zeta}_2|z_2)V_{\imath_1}(\zeta_1|z_1)\bigr).
\end{equation}
The matrix elements of the operators $V_{\imath}(\zeta|z)$ are finally 
recovered as in \rf{mat_el1}, taking into account \rf{indaggerout}.

\subsection{Preliminaries}\label{prelim}

To prepare for our discussion of the reconstruction procedure in the 
case of Liouville theory let us make two observations concerning the 
DOZZ formula:

First, it satisfies reflection relations such as 
\begin{equation}\label{refl}
C(\al_3,\al_2,\al_1)\; = \; S(\al_3)C(Q-\al_3,\al_2,\al_1), 
\end{equation}
where the the {\it reflection amplitude} $S(\al)$ is given as
\begin{equation}
S(\al)\;\,=\;\,\frac{\bigl(\pi \mu\ga(b^2)\bigr)^{b^{-1}(Q-2\al)}}{b^2}
\frac{\ga(2b\al-b^2)}{\ga(2-2b^{-1}\al+b^{-2})}.
\end{equation}
By symmetry of $C(\al_3,\al_2,\al_1)$ one finds the corresponding relations 
for the other arguments. This allows one to restrict attention to values
of the variables $\al_i$, $i=1,2,3$ that satisfy the so-called 
Seiberg-bound \cite{Se}:
\begin{equation}\label{Seibd}
\Re(\al_i)\leq \fr{Q}{2},\quad i=1,2,3.
\end{equation} 

Second, let us observe that the DOZZ-formula is indeed {\it analytic} as long
as the condition for convergence of the zero mode integration in the 
path-integral is satisfied, namely 
\begin{equation}\label{convcond}
-b\Re(s)\;=\;\Re(\al_1+\al_2+\al_3)-Q\;>\;0, 
\end{equation}
but has singularities
otherwise. This suggests that an interpretation of the 
tree point functions as a matrix element,
\begin{equation} \label{C-matel}
\Om\bigl(V_{\al_3}(z_3)V_{\al_2}(z_2)V_{\al_1}(z_1)\bigr)
\;\equiv\;
\bra 0|\SV_{\al_3}(z_3)\SV_{\al_2}(z_2)\SV_{\al_1}(z_1)|0\ket
\end{equation}
will be most straightforward if one starts with the range given 
by \rf{convcond}. Taking into account \rf{Seibd} one finds that 
\rf{convcond} can only be satisfied if 
$0<\Re(\al_i)\leq \fr{Q}{2}$ $i=1,2,3$. 

\subsection{Two-point function?}

So can we define states $|\al\ket_{\rm in}^{}$, ${}_{\rm out}^{}\bra\al|$ 
that are created via 
\begin{equation}\label{obst2}
|\al\ket_{\rm in}^{}\;\equiv\;
\lim_{z\ra 0}\SV_{\al}(z)|0\ket  ,
\qquad
{}_{\rm out}^{}\bra \al|\;\equiv\;
\lim_{z\ra \infty} |z|^{4\De_{\al}}
\bra 0|\SV_{\al}(z)\;?
\end{equation}
When trying to adapt the reconstruction procedure from rational conformal
field theories to the present case, the first question to address is:
What is the unit field, or equivalently: How to recover the 
two-point functions from $C(\al_3,\al_2,\al_1)$?
The unit field should clearly have vanishing conformal 
dimension $\De_{\al}=\al(Q-\al)$. Under the restriction \rf{Seibd} 
this is only found for $\al=0$ which lies on the boundary of our 
``allowed'' region $0<\Re(\al_i)\leq \fr{Q}{2}$. So let us consider 
the behavior of $C(\al_2,\ep,\al_1)$ for small $|\ep|$ with $\Re(\ep)>0$:
It is given as 
\begin{equation}\label{singtwopt}
C(\al_2,\ep,\al_1)\;\simeq\;\frac{2\ep S(\al_1)}
{(\al_2-\al_1+\ep)(\al_1-\al_2+\ep)}+
\frac{2\ep}
{(Q-\al_2+\al_1+\ep)(\al_1+\al_2-Q+\ep)},
\end{equation}
where $S(\al)$ is the reflection amplitude introduced above.
$C(\al_2,\ep,\al_1)$ vanishes for $\ep\ra 0$ unless 
$\al_2=\al_1$ or $\al_2=Q-\al_1$ and becomes infinite otherwise.
The two-point function can therefore only be defined in the distributional 
sense. To actually define it as a distribution from \rf{singtwopt} one needs to
specify a contour over which $\al_2$, $\al_1$ are supposed to be integrated.
One clearly can only get distributions proportional to
delta-distributions, but the precise pre-factors depend on the direction
from which $\ep$ approaches zero.

The distributional character of the two-point function
is of course just what was to be expected when having 
continuous sets of primary fields, since the vanishing of the 
two-point function for 
$\al_2\neq\al_1$ and  $\al_2\neq Q-\al_1$ is required by conformal 
invariance. This also implies that we can not hope to find any state that 
is normalizable in the strict sense, but only normalizability in the
distributional sense.

\subsection{Scalar product}

Let us therefore ask for which values of $\al$ it is possible 
to define a reasonable scalar product from the two-point function.
To this aim we need to know the hermiticity properties of $\SV_{\al}$:
It follows from 
$(C(\al_3,\al_2,\al_1))^*=C(\al_3^*,\al_2^*,\al_1^*)$ that $\SV_{\al}$
behaves as follows under hermitian conjugation: 
\begin{equation}\label{herm2}
\bigl(\SV_{\al}(z)\bigr)^\dagger\;=\; |z|^{-4\De(\al)}
V_{\al^*}\bigl(\bz^{-1}).
\end{equation}
We should therefore obtain the 
scalar product from the three-point function as
\begin{equation}\label{scprod2}
{}_{{\rm in}}\bra \al_2|\al_1\ket_{{\rm in}}\;=\;
\lim_{\al\ra 0}\; C(\al_2^*,\al,\al_1).
\end{equation}
As the right hand side vanishes unless $\al_2^*=\al_1$ or $\al_2^*=Q-\al_1$,
it is easy to see that no reasonable scalar product can be obtained
unless either $\al_i\in\BR$ or $\al_i\in\frac{Q}{2}+i\BR$.
This would also follow when considering the extension of 
the scalar product \rf{scprod2} 
to states generated from
descendants: The representations $\CV_{\al_i}$ are unitary
only if $\al_i\in\BR$ or $\al_i\in\frac{Q}{2}+i\BR$.

In the first case one would need to choose in \rf{singtwopt} an $\ep$ with 
$\Im(\ep)\neq 0$, which would lead to 
\begin{equation}\label{scprod3}
{}_{\rm in}\bra \al_2|\al_1\ket_{{\rm in}}\;=\;
\pm 2\pi iS(\al_1)\de(\al_2-\al_1).
\end{equation}
Due to the unavoidable factor of $i$ in \rf{scprod3} one does not get a 
reasonable scalar product for $|\al\ket$ with $\al\in\BR$. 

The other case $\al_i=\frac{Q}{2}+iP_i$, $P_i\in\BR$
is better: The reflection property \rf{refl} allows one to restrict 
to $P_i>0$.
One then needs to choose $\ep$ with $\Re(\ep)>0$,
in which case \rf{singtwopt} gives 
\begin{equation}\label{scprod4}
{}_{{\rm in}}
\bra \fr{Q}{2}+iP_2|\fr{Q}{2}+iP_1\ket_{{\rm in}}\;=\;
2\pi \de(P_2-P_1).
\end{equation}
Let us identify $|P\ket\equiv |\fr{Q}{2}+iP_1\ket_{\rm in}$. 
It is straightforward to generalize the discussion to states $|P,\zeta\ket$,
$\zeta\in\CW_P\equiv\CW_{\frac{Q}{2}+iP}$
that are created by the vertex operators
$\SV_{\frac{Q}{2}+iP}$: The scalar product is then given as
\begin{equation}\label{scprod5}
\bra P',\zeta_2|P,\zeta_1\ket\;=\;
2\pi \de(P'-P)(\zeta_2,\zeta_1)_{\CW_P}^{},
\end{equation}
where $(\zeta_2,\zeta_1)_{\CW_P}^{}$ denotes the scalar product in
$\CW_P$ that is normalized such that $(v_P,v_P)_{\CW_P}^{}=1$
if $v_P$ is the highest weight vector in $\CW_P$. 

We conclude that conformal symmetry and DOZZ-formula imply that the
Liouville Hilbert space takes the form
\begin{equation}\label{DOZZspec}
\CH\;\simeq\;\int\limits_{\BR^+}^{\oplus}\frac{dP}{2\pi}\;\CW_P.
\end{equation}

\begin{rem}
Let us note that a calculation like \rf{indaggerout} now implies that 
\begin{equation}
|P,\bar{\zeta}\ket_{\rm out} \;=\; |-P,\zeta\ket_{\rm in} \;=\;R(-P)
|P,\zeta\ket_{\rm in},
\end{equation} 
where $R(-P)\equiv S(\frac{Q}{2}-iP)$ is the 
reflection amplitude encountered
earlier. But this means that the {\it scattering operator} $\SS$ that relates
in- and out states is diagonal in the basis 
$\{|P,\zeta\ket;P\in\BR^+,\zeta\in\CW_P\}$ 
and given by multiplication with $R(-P)$. 
The unitarity of $\SS$ 
follows from $|R(p)|^2=1$. We will see later that $\SS$ indeed 
has an interpretation as a scattering operator that describes the
scattering of wave-packets off the Liouville potential.
\end{rem}

\subsection{Matrix elements} \label{matelssec}

Having identified the set of normalizable states, we may then 
recover the matrix elements of operators $\SV_{\al}$, with $\al$ 
in the range $0<\Re(\al)\leq \fr{Q}{2}$ identified in subsection
\rf{prelim}, as follows:
\begin{equation}\label{mat_el}\begin{aligned}
\bra P_3,\zeta_3|\SV_{\al_2}(z_2)& 
|P_1,\zeta_1\ket
\;\equiv\;\\ 
\;\equiv\; & \lim_{z_1\ra 0}\lim_{z_3\ra \infty} 
z_3^{2\De(\zeta_3)}\bz_3^{2\bar{\De}(\zeta_3)}
\Om\bigl(V_{\bal_3}(\zeta_3|z_3)V_{\al_2}(z_2)
V_{\al_1}(\zeta_1|z_1)\bigr).
\end{aligned}\end{equation}
We use the notation $\al_i=\frac{Q}{2}+iP_i$,
$\bal_i\equiv Q-\al_i=\frac{Q}{2}-iP_i$, $i=1,2,\dots$. 
Knowing the matrix elements of the operators $\SV_{\al}$ should of course allow
one to represent the matrix elements of products of these operators
by summing over intermediate states: 
Let $\BB_P$ be a basis for $\CW_P$, which may be chosen to consist of
vectors $\zeta\in\CW_P$ that diagonalize $\SL_0$, $\bL_0$. 
Each element $\zeta\in\BB_P$ has a
unique ``dual'' $\zeta^t$ which is the vector defined by the property 
$(\zeta^t,\zeta')_{\CW_P}=\de_{\zeta,\zeta'}$ for all $\zeta'\in\BB_P$.
An example for the representation of matrix elements by summing over 
intermediate states can then be written as follows:
\begin{equation}\label{mat_el_prod}\begin{aligned}
\bra P_4|\SV_{\al_3}(z_3)\SV_{\al_2}(z_2)& 
|P_1\ket
\;\equiv\;\\ 
\;\equiv\; &
\int_{\BR^+}\frac{dP}{2\pi}\;\sum_{\zeta\in\BB_P}\;
\bra P_4|\SV_{\al_3}(z_3) 
|P,\zeta\ket\bra P,\zeta^t|\SV_{\al_2}(z_2) 
|P_1\ket,
\end{aligned}
\end{equation}
We may furthermore assume that the elements of $\BB_P$ 
factorize as $\zeta=\xi_2\ot\xi_1$. 
This leads to a factorized representation of the matrix element 
\rf{mat_el_prod} of the following form:
 \begin{equation}\label{mat_el_prod2}\begin{aligned}
\bra  P_4|\SV_{\al_3}(z_3)\SV_{\al_2}(z_2) 
|P_1\ket
\;\equiv\; \int_{\BS}\frac{d\al}{2\pi}\;  C\bigl(\bal_4,\al_3,\al) & 
C\bigl(\bal,\al_2,\al_1\bigr)\ti\\
\ti & \CF_{\al}^s\bigl[{}_{\al_4}^{\al_3}{}_{\al_1}^{\al_2}\bigr](z_3,z_2)
\CF_{\al}^s\bigl[{}_{\al_4}^{\al_3}{}_{\al_1}^{\al_2}\bigr](\bz_3,\bz_2)
,
\end{aligned}
\end{equation}
where the variable $\al$ was introduced by $\al=\frac{Q}{2}+iP$, such that 
the integral over $P\in\BR^+$ becomes an integral over $\al\in\BS\equiv
\frac{Q}{2}+i\BR^+$. The conformal blocks 
$\CF_{\al}^s\bigl[{}_{\al_4}^{\al_3}{}_{\al_1}^{\al_2}\bigr](z_3,z_2)$ 
are given by the following power
series:
\begin{equation}\label{CGpwser}\begin{aligned}
\CF_{\al}^s\bigl[{}_{\al_4}^{\al_3}{}_{\al_1}^{\al_2}\bigr](z_3,z_2)
\;=\;  
z_3^{\De_{\al_4}-\De_{\al}-\De_{\al_3}}
z_2^{\De_{\al}-\De_{\al_2}-\De_{\al_1}}
 \sum_{n=0}^{\infty}\;
\Bigl(\frac{z_2}{z_3}\Bigr)^n 
\CF_{\al}^s\bigl[{}_{\al_4}^{\al_3}{}_{\al_1}^{\al_2}\bigr](n).
\end{aligned}\end{equation}
The coefficients 
$\CF_{\al}^s\bigl[{}_{\al_4}^{\al_3}{}_{\al_1}^{\al_2}\bigr](n)$  are given 
by sums over vectors $\xi\in\CV_{\al}$
with fixed eigenvalues $\De_{\al}+n$.
Some more information concerning the definition of the 
conformal blocks can be found in Subsection \ref{analcfbl}.
Here let us only note that
the series that represent $\CF_{\al}$ actually converge for
$|z_2|<|z_3|$. 

\subsection{Meromorphic continuation of matrix elements}\label{mer_cont}

It will be important in the following to 
note that the matrix elements such as \rf{mat_el_prod} admit a meromorphic
continuation to arbitrary complex values of $P_4$, $P_1$, $\al_3,\al_2$. 
For notational convenience let us identify $|\al\ket\equiv |P\ket$
if $\al$ and $P$ are related by $\al=\frac{Q}{2}+iP$. We will consider 
the example of $\bra \al_4|\SV_{\al_3}(w)\SV_{\al_2}(z) 
|\al_1\ket$. It is shown in Section \ref{techapp} below 
that this matrix element has a 
meromorphic continuation w.r.t. all four variables $\al_4,\dots,\al_1$ with 
poles if and only if 
\begin{equation}
Q+\sum_{i=1}^4 s_i\bigl(\al_i-\fr{Q}{2}\bigr)\;=\;-nb-mb^{-1},\quad
s_i\in\{1,-1\}. 
\end{equation} 
The poles with $s_i=1$ $i=1,\dots,4$ are in 
precise correspondence with the singularities that one would expect on the 
basis of the path-integral arguments discussed in section \ref{pathsect}.
All others are generated by reflection relations like \rf{refl}. 

This meromorphic continuation can be represented as in \rf{mat_el_prod} 
as long as the conditions
\begin{equation}\label{funran}\begin{aligned}
{} &|\Re(\alpha_1 - \alpha_2)| < Q/2;\qquad 
    |\Re(Q - \alpha_1 - \alpha_2)| < Q/2;\\ 
   &|\Re(\alpha_3 - \alpha_4)| < Q/2;\qquad 
    |\Re(Q - \alpha_3 - \alpha_4)| < Q/2,
\end{aligned}\end{equation}
are satisfied. Otherwise one has a representation of the form
\begin{equation}\label{mat_el_prod_allg}\begin{aligned}
\bra P_4|\SV_{\al_3}(z_3)\SV_{\al_2}(z_2)& 
|P_1\ket
\;\equiv\;\\ 
\;\equiv\; &
\frac{1}{2}\int_{\CC}\frac{dP}{2\pi}\;\sum_{\zeta\in\BB_P}\;
\bra P_4|\SV_{\al_3}(z_3) 
|P,\zeta\ket\bra P,\zeta^t|\SV_{\al_2}(z_2) 
|P_1\ket,
\end{aligned}
\end{equation}
where the contour $\CC$ can 
generically be taken as $\BR$ plus a finite sum of small circles around 
certain poles of the
three-point functions that appear in \rf{mat_el_prod}.

\subsection{Normalizable vs. non-normalizable states} \label{micmac}

Our identification of the set of normalizable states seems to create puzzles:
For example, the states $|P\ket$ have conformal dimensions larger than $Q^2/4$,
so one does not find the state $|0\ket$ among them. But if $|0\ket$ is not
in the spectrum, what meaning does the state-operator correspondence 
have?

To start with, let us recall that $|P\ket\notin\CH$. This means in particular
that scalar products such as $\bra P|\psi\ket$ will not be defined for
all $|\psi\ket\in\CH$ (square-integrability only requires $\bra P|\psi\ket$
to be defined up to $P$ from a set of measure zero). One needs to 
consider subspaces $\CT\subset\CH$ which are such that $\bra P|\psi\ket$
is defined for any $|\psi\ket\in\CT$. 
$|P\ket$ is then interpreted as 
an element of the hermitian dual $\CT^{\dagger}$ of $\CT$, the 
space of all anti-linear forms on $\CT$. The triple of
spaces $\CT\subset\CH\subset\CT^{\dagger}$ is often called a
``Gelfand triple''. Let us remark that
for elements of the spectrum, 
$|P\ket$ with $P\in\BR^+$, it does not matter which subspace $\CT\subset\CH$
one chooses. 

But one may also consider particularly ``nice'' subspaces $\CT$ which are such 
that the wave-function $\psi(P)\equiv\bra P|\psi\ket$ admits an 
analytic continuation into some region $R\subset\BC$ around $\BR^+$. On such 
a subspace one may of course consider the forms $|P\ket$ 
to be defined by $\bra\psi|P\ket\equiv (\psi(P))^*$ for any 
complex $P\in R$. 
So let us try to see whether such a distributional 
interpretation of the states $|P\ket$ for $P\in\BC$ 
is useful in the present context.

The matrix element $\bra P_2|\SV_{\al}(z)|P_1\ket$, $\Re(\al)>0$, $|z|<1$, 
can be interpreted as the wave-function  
of the state $\SP_0\SV_{\al}(z)|P_1\ket$, where $\SP_0$ denotes the
projection onto the subspace $\CH_0\subset\CH$ of vectors that satisfy 
$\SL_n|\psi\ket=0=\bL_n|\psi\ket$.
The very existence
of that wave-function means that the operator $\SV_{\al}(z)$ must 
act {\it smoothing} on states $|\psi\ket$: It creates a state with 
smooth wave-function when acting on the
distribution $|P_1\ket$, which would have a 
delta-distribution as ``wave-function''. 
For $\Re(\al)>0$, the wave-function
$\bra P_2|\SV_{\al}(z)|P_1\ket$ is not only smooth w.r.t. $P_2$, 
but even analytic within the strip $\{P_2\in\BC; |\Im(P_2)|<\Re(\al)\}$,
as follows from the DOZZ formula.
The distributional interpretation of the states $|P\ket$ with 
$P\notin\BR$ is therefore very natural in this context.

However,  if the $|P\ket$ are to be interpreted in a distributional 
sense both for $P\in\BR$ and $P\notin\BR$, why can't the 
$|P\ket$ with $P\notin\BR$ appear in the spectrum? 
It can be shown on rather general functional analytic grounds
that a state such as $|P\ket$ can appear in a spectral decomposition 
only if the distribution $\bra P|$ is defined on {\it any} space
$\CT\subset\CH$ of test functions that allows the
wave-functions $\bra P|\psi\ket$ to be {\it pointwise} defined
for any $\psi\in\CT$, not only up to a set of measure zero 
\footnote{For a more precise mathematical discussion and proofs
see e.g. the introduction and first section of \cite{Be}.}. 
This is clearly not the case for the $|P\ket$ with $P\notin\BR$: 
The condition of analyticity in some strip that determines
the domain of $\bra P|$ is much too strong. A related 
fact is that ``scalar products'' like $\bra P'|P\ket$ can be defined
in the distributional sense as long as $P,P'\in\BR$, but do not have 
a canonical definition as soon as $P$ or $P'$ have non-vanishing imaginary
part. 

\subsection{The vacuum $|0\ket$}

Let us now check that the distributional interpretation of $|P\ket$ for
complex $P$ yields a self-consistent understanding of the 
$SL(2,\BC)$ -invariant ``state'' $|0\ket$. To this aim we need 
to check if  (i) the state $\SV_{\al_3}(z_3)\SV_{\al_2}(z_2)|P_1\ket$
is in the domain of $\bra 0|$, and
(ii) The value of $\bra 0|\SV_{\al_3}(z_3)\SV_{\al_2}(z_2)|P_1\ket$
is given by 
\begin{equation}
\bra 0|\SV_{\al_3}(z_3)\SV_{\al_2}(z_2)|P_1\ket\;=\;
\lim_{z_1\ra 0}\Om\bigl(\SV_{\al_3}(z_3)\SV_{\al_2}(z_2)\SV_{\al_1}(z_1)\bigr),
\end{equation}
where $\al_1=\frac{Q}{2}+iP_1$. 
 
So let us consider $\bra P_4|\SV_{\al_3}(z_3)\SV_{\al_2}(z_2)|P_1\ket$, 
the wave-function of the state 
$\SP_0\SV_{\al_3}(z_3)\SV_{\al_2}(z_2)|P_1\ket$. It follows from our 
observation in Subsection \ref{mer_cont} that 
$\bra P_4|\SV_{\al_3}(z_3)\SV_{\al_2}(z_2)|P_1\ket$ is analytic w.r.t.
$P_4$ in a strip of width $\Re(\al_1+\al_2)$. So if 
$\Re(\al_1+\al_2)>\frac{Q}{2}$ one indeed finds that the 
state $\SP_0\SV_{\al_3}(z_3)\SV_{\al_2}(z_2)|P_1\ket$ is in the domain 
of $\bra 0|\equiv \bra P|_{P=-i\frac{Q}{2}}$.

In order to determine its value, one needs to return to the representation
\rf{mat_el_prod} for $\bra P_4|\SV_{\al_3}(z_3)\SV_{\al_2}(z_2)|P_1\ket$.
When continuing $P_4$ to $-i\frac{Q}{2}$ one necessarily leaves 
the range \rf{funran}, so that the more general representation of the 
form \rf{mat_el_prod_allg} has to be used. The residual terms include
the contribution from the pole 
of $\bra P_4|\SV_{\al_3}(z_3) 
|P,\zeta\ket$
at $\frac{Q}{2}-iP_4=\al_3+\frac{Q}{2}+iP$, which yields 
a contribution that can be identified with 
$\lim_{z_1\ra 0}
\Om\bigl(\SV_{\al_3}(z_3)\SV_{\al_2}(z_2)\SV_{\al_1}(z_1)\bigr)$.
All other contributions are found to vanish due to the zero of 
$C(\al_3,\al_2,\al_1)$ at $\al_3=0$.

So for $\Re(\al_1+\al_2)>\frac{Q}{2}$, the smoothing effect 
of $\SV_{\al_3}(z_3)\SV_{\al_2}(z_2)$ is strong enough to map 
$|P_1\ket$ into the domain of $\bra 0|$. This discussion is easily 
generalized to the case where $\Re(\al_1)\neq \frac{Q}{2}$, 
$\al_1=\frac{Q}{2}+iP_1$: $\SV_{\al_3}(z_3)\SV_{\al_2}(z_2)|\al_1\ket$
will be in the domain of $\bra 0|$ if $\Re(\al_1+\al_2+\al_3)>Q$.
Moreover, by 
a very similar reasoning one would find that
the smoothing effect of $\SV_{\al_3}(z_3)\SV_{\al_2}(z_2)\SV_{\al_1}(z_1)$
is strong enough to map $|0\ket$ into the domain of $\bra 0|$.
We have thereby obtained the precise meaning of the relation
\begin{equation} \label{C-matel2}
\Om\bigl(\SV_{\al_3}(\zeta_3|z_3)\SV_{\al_2}(\zeta_2|z_2)
\SV_{\al_1}(\zeta_2|z_1)\bigr)
\;\equiv\;
\bra 0|\SV_{\al_3}(\zeta_3|z_3)\SV_{\al_2}(\zeta_2|z_2)
\SV_{\al_1}(\zeta_1|z_1)|0\ket.
\end{equation}
It is straightforward to interpret vacuum expectation
values of more than three operators along these lines. Let us note
that the condition for $\bra 0|$ to be in the domain of the state
created by acting with $\prod_{i=1}^n\SV_{\al_i}(z_i)$ on $|0\ket$
is precisely the condition for convergence of zero mode integration 
found in Subsection \ref{freef:3}.

\subsection{Operators $\SV_{\al}$ with $\Re(\al)\leq 0$}

So far we had only discussed the connection between fields
$V_{\al}$ and operators $\SV_{\al}$ for $0\leq \Re(\al)\leq \frac{Q}{2}$.
Let us now generalize to $\Re(\al)\leq 0$.
 
The main point is best seen when considering 
the wave-function $\bra P_2|\SV_{\al}(z)|\psi\ket$ of the state
created by acting with $\SP_0\SV_{\al}(z)$ on a state $|\psi\ket$
of the form $|\psi\ket=\frac{1}{2\pi}\int_{0}^{\infty}dP_1\psi(P_1)|P_1\ket$. 
It is clearly well-defined as long as $0< \Re(\al)\leq \frac{Q}{2}$ and 
given by 
\begin{equation}\label{wfVpsi}
\bra P_2|\SV_{\al}(z)|\psi\ket\;=\;
\int_{\BR^+}\frac{dP_1}{2\pi}\; \psi(P_1)\;\bra P_2|\SV_{\al}(z)|P_1\ket.
\end{equation}
When $\Re(\al)\ra 0$,  it will generically cease to be defined due to
the poles of 
$\bra P_2|\SV_{\al}(z)|P_1\ket$ at 
$\frac{Q}{2}\pm iP_2=\al+\frac{Q}{2}\pm iP_1$. One may, however, 
consider states $|\psi\ket$ whose wave-functions
$\psi(P)$ admit an analytic continuation
into a strip of width $w$. For such $|\psi\ket$ it is possible 
to analytically continue the expression \rf{wfVpsi} to all 
$\al$ with $-w<\Re(\al)\leq 0$:
This is done by first using the
reflection property $|P\ket=R(P)|-P\ket$ to extend the 
integration over $\BR^+$ to an integral over $\BR$,
and then deforming the contour of integration over $P_1$
suitably around the poles of $\bra P_2|\SV_{\al}(z)|P_1\ket$ that 
cross the axis $\BR$ when $\Re(\al)$ becomes negative.
We conclude that $\SV_{\al}(z)$ for
$\Re(\al)\leq 0$ makes sense 
as an unbounded operator with domain restricted to states 
with wave-functions analytic in strips of width larger or equal 
to $|\Re(\al)|$.

The following alternative point of view is useful:
Consider $\SP_0\SV_{\al_2}(z_1)|P_1\ket$,
which for $0< \Re(\al)\leq \frac{Q}{2}$ can be represented as
\begin{equation}\label{V-Pcont}
\SP_0\SV_{\al_2}(z_1)|P_1\ket\;=\;\frac{1}{2}\int_{\BR}\frac{dP}{2\pi} \;|P\ket\bra P|
\SV_{\al_2}(z_1)|P_1\ket. 
\end{equation}
The matrix element $\bra P|
\SV_{\al_2}(z_1)|P_1\ket$ has poles at 
$\frac{Q}{2}\pm iP=\al_2+\frac{Q}{2}\pm iP_1$,
which approach the contour $\BR$ of integration in \rf{V-Pcont}
when $\Re(\al)\ra 0$.

On a formal level it is straightforward to perform the continuation 
to $\Re(\al)\leq 0$: We had previously discussed the continuation
of the $|P\ket $ to complex values of $P$ in the
sense of distributions in $\CT^{\dagger}$. In this spirit one 
would perform the continuation of \rf{V-Pcont} simply by again
deforming the contour of integration 
over $P$. This generically yields a representation of the 
form \rf{V-Pcont} but with contour $\BR$ replaced by $\BR$ plus 
a finite sum of circles around the poles of $\bra P|
\SV_{\al_2}(z_1)|P_1\ket$ that have crossed the contour $\BR$. 
For example, if $-b<\Re(\al)\leq 0$ one would get instead of
\rf{V-Pcont} the expression
\begin{equation}\label{V-Pcont'}
\SP_0\SV_{\al_2}(z_2)|P_1\ket\;=|P_1-i\al_2\ket
\;+\;\frac{1}{2}\int_{\BR}\frac{dP}{2\pi} \;|P\ket\bra P|
\SV_{\al_2}(z_2)|P_1\ket,
\end{equation}
where the value of the residues has been worked out from
the DOZZ-formula taking into account the reflection property \rf{refl}.
Formula \rf{V-Pcont'} clearly makes sense as an equation 
in $\CT^{\dagger}$ if $\CT$ is such that
the wave-functions $\bra P|\psi\ket$ 
of $|\psi\ket \in\CT$
are analytic in a strip of width larger than $|\Re(\al_2)|$. The unboundedness
of $\SV_{\al_2}(z_2)$ for $\Re(\al_2)\leq 0$ is now reflected in the 
appearance of non-normalizable states such as e.g. the state 
$|P_1-i\al_2\ket$ in \rf{V-Pcont'}. 

\subsection{Null vector decoupling}

Something interesting happens if one considers $\SV_{\al}(z)$
for $2\al=2\al_{m,n}\equiv -mb-nb^{-1}$, $n,m\in\BZ^{\geq 0}$, where the 
Verma module $\CV_{\al}$ contains a singular vector $s_{n,m}$. 
The matrix element $\bra P_3|
\SV_{\al_2}(z)|P_1\ket$ is proportional to $\up(2\al_2)$, which vanishes
for $\al_2=\al_{m,n}$. This means that the expansion \rf{V-Pcont'}
of $\SP_0\SV_{\al_2}(z)|P_1\ket$ over states $|P\ket$ does not 
contain the part represented as an integral over $\BR$. 
What is non-vanishing, however, are the residue terms that 
were picked up in the process of defining the analytic continuation. 
These are found to be given by an expression of the form 
\begin{equation}\label{V-Pcont''}
\SP_0\SV_{\al}(z)|P\ket\;=\;\sum_{r=0}^{m}\sum_{s=0}^n\;
C_{r,s}^{m,n}(P)\;|P-i(\al+rb+sb^{-1})\ket.
\end{equation}
The values of $P-i(\al+rb+sb^{-1})$ that appear in \rf{V-Pcont''}
are in precise correspondence to the so-called 
fusion rules for decoupling of the 
null vector in the representation $\CV_{\al_2}$: It follows from a
theorem of Feigin and Fuchs \cite{FF}, which is quoted below in 
Section \ref{Verma} as Theorem 1, that 
\begin{equation}
\SV_{\al}(s_{n,m}\ot v_{n,m}|z)\;=\;0,\qquad
\SV_{\al}(v_{n,m}\ot s_{n,m}|z)\;=\;0,
\end{equation}
if $v_{n,m}$ and $s_{n,m}$ are the highest weight and the singular vectors
of the Verma module $\CV_{\al_{m,n}}$ respectively. Vacuum expectation
values that contain these fields will satisfy differential 
equations of the type discussed in \cite{BPZ}.

We find it remarkable to see how deep information on the 
representation theory of the Virasoro algebra is encoded in the 
analytic structure of the DOZZ three point function.

\begin{rem}
A special case of this result reproduces the identification of the
identity operator as the primary field with vanishing conformal dimension: 
\begin{equation}\label{ident}
\lim_{\al\ra 0}\; \SV_{\al}(z)\;\,=\;\,\id,
\end{equation}
which may be seen as an example of null vector decoupling in the case $n=0=m$.
\end{rem}

\section{Locality and crossing symmetry?}

A major issue of consistency arises: If follows from the 
preceding discussions that vacuum expectation values such as 
$\bra 0|\prod_i^n \SV_{\al_i}(z_i)|0\ket$ are uniquely given
by conformal symmetry and the DOZZ formula: Summing over intermediate states
as in \rf{mat_el_prod} produces power series with coefficients that can
all be expressed in terms of the 
matrix elements $\bra P_3,\zeta_3|\SV_{\al_2}(z_2) 
|P_1,\zeta_1\ket$, which characterize the operators $\SV_{\al}$ uniquely.
But are the operators that are characterized in such a way also local,
i.e. do they satisfy 
$\SV_{\al_2}(z_2)\SV_{\al_1}(z_1)=\SV_{\al_1}(z_1)\SV_{\al_2}(z_2)$?

It clearly suffices to consider the case of $n=4$ which corresponds to 
the matrix element $\bra P_4|\SV_{\al_3}(z_3)\SV_{\al_2}(z_2) 
|P_1\ket$. In view of the power series expansions for this matrix element
discussed in Subsection \ref{matelssec} one sees that this question
involves the following issues: 
First, it was stated there that the power series that 
represent $\bra P_4|\SV_{\al_3}(z_3)\SV_{\al_2}(z_2) 
|P_1\ket$ and 
$\bra P_4|\SV_{\al_2}(z_2)\SV_{\al_3}(z_3) 
|P_1\ket$ are convergent for $|z_2|<|z_3|$ and $|z_3|<|z_2|$ respectively.
In order for the question of locality to have any sense, one evidently
needs that the conformal blocks can be analytically continued into 
$|z_2|>|z_3|$ and $|z_3|>|z_2|$ respectively.
Given that such an analytic continuation exists, locality 
amounts to a highly nontrivial identity between the conformal 
blocks, which are fully given by conformal symmetry, and the coefficients
$C\bigl(\al_3,\al_2,\al_1\bigr)$ that represent the measure with which 
the conformal blocks are weighted in \rf{mat_el_prod2}.  

By now we believe to have a proof for this crucial property, which will
be sketched in Parts \ref{gervnev} and \ref{boots}, 
with details to be presented elsewhere.
Until then we will simply assume that the DOZZ-proposal indeed
describes local operators $\SV_{\al}$.

\subsection{Crossing symmetry}

Locality is closely related (almost equivalent) to another property
of vacuum expectation values $\bra 0|\prod_i^n \SV_{\al_i}(z_i)|0\ket$
that is usually called crossing symmetry.
Let us again restrict to the case $n=4$, which is good enough.
Inserting a complete set of intermediate states yields an expansion
\begin{equation}\label{mat_el_prod+}\begin{aligned}
\bra 0|\SV_{\al_4}(z_4)\SV_{\al_3}(z_3) & \SV_{\al_2}(z_2) 
\SV_{\al_1}(z_1)|0\ket
\;\equiv\;\\ 
\;\equiv\; &
\int_{\BS}\frac{d\al}{2\pi}\;\sum_{\zeta\in\BB_{\al}}\;
\bra 0|\SV_{\al_4}(z_4)\SV_{\al_3}(z_3) 
|\al,\zeta\ket\bra \al,\zeta^t|\SV_{\al_2}(z_2)\SV_{\al_1}(z_1) 
|0\ket.
\end{aligned}
\end{equation}
Alternatively one may use locality to move $\SV_{\al_1}$ to the 
right of $\SV_{\al_4}$, and then insert a complete set of states 
between $\SV_{\al_4}\SV_{\al_1}$ and $\SV_{\al_3}\SV_{\al_2}$.
Noting that $\SV_{\al}(z) 
|0\ket=e^{zL_{-1}+\bz\bL_{-1}}|\al\ket$ one may rewrite the 
resulting expansion as follows:
\begin{equation}\label{mat_el_prodt}\begin{aligned}
\bra 0|\SV_{\al_4}(z_4)\dots
\SV_{\al_1}(z_1)|0\ket
\;\equiv\;
\int_{\BS}\frac{d\al}{2\pi}\;\sum_{\zeta\in\BB_{\al}}\;
& \Om\bigl(V_{\al_4}(z_4)V_{\al_1}(z_1) 
V_{\al}(\zeta|z_2)\bigr)\ti\\
 & \quad\ti \Om\bigl(
V_{\bal}(\zeta^t|\infty)V_{\al_3}(z_3-z_2)V_{\al_2}(0) \bigr),
\end{aligned}
\end{equation}
where we used the identification \rf{C-matel2} and the notation \rf{vinfty}.
If one finally uses locality of $\SV_{\al_1}(z_1)$ and 
$\SV_{\al}(\zeta|z_2)$ one gets an expansion 
that can be read as the result of expanding 
the product of operators $\SV_{\al_3}(z_3)\SV_{\al_2}(z_2)$ according to
the OPE:
\begin{equation}\label{OPE}  
\SV_{\al_3}(z_3)\SV_{\al_2}(z_2)
\;=\;\int_{\BS}\frac{d\al}{2\pi}\;\sum_{\zeta\in\BB_{\al}}\;
\SV_{\al}(\zeta|z_2)\;\Om\bigl(
V_{\bal}(\zeta^t|\infty)V_{\al_3}(z_3-z_2)V_{\al_2}(0) \bigr).
\end{equation}
The fact that the power series expansion obtained by 
inserting \rf{OPE} into $\Om\bigl(\SV_{\al_4}(z_4)\dots
\SV_{\al_1}(z_1)\bigr)$ also yields a valid representation for that
vacuum expectation value is usually referred to as the property of
{\it crossing symmetry}. It can be written as 
\begin{equation}\label{t-factor}\begin{aligned}
\Om\bigl(V_{\al_4}(\infty) & V_{\al_3}(1) 
V_{\al_2}(z,\bz)V_{\al_1}(0)\bigr)\;=\;\\
=\; & \int\limits_{\BS}\frac{d\al}{2\pi}\; 
C(\al_4,\al,\al_1)C(\bal,\al_3,\al_2)\;
\CfBlt{\al}{\al_3}{\al_2}{\al_4}{\al_1}(z)
\CfBlt{\al}{\al_3}{\al_2}{\al_4}{\al_1}(\bz),
\end{aligned}\end{equation}
where the  {\it t-channel conformal blocks} 
$\CF_{\al}^t$ (as opposed to the {\it s-channel 
conformal blocks}  $\CF_{\al}^s$ that appear in \rf{mat_el_prod2})
are given as power series expansions
of the form
\begin{equation}\label{t-confbl}
\CfBls{\al}{\al_3}{\al_2}{\al_4}{\al_1}(z)\;=\;
(1-z)^{\De_{\al}-\De_{\al_3}-\De_{\al_2}}
\sum_{n=0}^{\infty}\;(1-z)^n\;  
\CfBlt{\al}{\al_3}{\al_2}{\al_4}{\al_1}(n).
\end{equation}
This relationship can be generalized to the expressions obtained by using
the operator product expansion iteratively. One gets expansions of the 
form
\[\begin{aligned}
\Om\bigl(\dots  V_{\al_k}(\zeta_k|z_k)\dots  V_{\al_l}(\zeta_l|z_l)\dots\bigr)
=
\int\limits_{\BS}\frac{d\al}{2\pi}\,
\sum_{\zeta\in\BB_{\al}} 
\,
& \Om\bigl(\dots  V_{\al}(\zeta|z_l)\dots\bigr)\\
\ti\;\, & \Om\bigl(V_{Q-\al}(\zeta^t|\infty)
V_{\al_k}(\zeta_k|z_{kl})\dots  V_{\al_l}(\zeta_l|0)\bigr),
\end{aligned}\]
where $z_{kl}\equiv z_k-z_l$.

\section{Appendix A: Verma modules of the Virasoro algebra}\label{Verma}

\subsection{Verma modules}\label{subVerma}

Let $\CV$ be the infinite dimensional vector space with basis
$\CB=\{v_\nu ;\nu\in\CT\}$, where $\CT$ is the set of all tuples
$\nu=(r_1,\dots,r_i,\dots)$
with all but finitely many $r_i$ being zero. The element of $\CB$
that corresponds to the 
tuple with $r_i\equiv 0$
for all $i$ will be denoted $v$. 
$\CV$ is the direct sum of finite 
dimensional vector spaces $\CV[m]$ with fixed ``level''
$m$, which are spanned by the vectors $v_\nu$ with $n(\nu)=m$,
where $n(\nu)=\sum_{i=0}^{\infty}ir_i$.

There is a standard family $\CV_{\al}$, $\al\in\BC$
of representations of the Virasoro
algebra that can be defined on $\CV$: It is uniquely defined by the 
requirements that
\begin{itemize} 
\item[(i)] $L_nv=0$ for $n>0$ and $L_0v_{0}^{}=\De_{\al}v$, where 
$\De_{\al}=\al(Q-\al)$, and 
\item[(ii)] $v_{\nu}\;=\;\prod_{i=1}^{\infty} (L_{-i})^{r_i}v$ if 
$\nu=(r_1,\dots,r_i,\dots)$.
\end{itemize}
The representation $\CV_{\al}$
depends of course on the {\it conformal dimension} 
$\De_{\al}$ only\footnote{We mostly consider the central
charge $c$ 
as fixed parameter in what follows}, so $\CV_{\al}\equiv\CV_{Q-\al}$.
There is a standard bilinear form $\bra.,.\ket_{\al}^{}$ on 
$\CV$ which is defined
by $\bra v,v\ket_{\al}^{}=1$ and 
$\bra L_{-n}\xi,\zeta\ket_{\al}^{}=\bra\xi,L_n\zeta\ket_{\al}^{}$.
The representation $\CV_{\al}$ is {\it irreducible} if and only if 
$\bra.,.\ket_{\al}^{}$
is non-degenerate. A criterion for the latter
is vanishing of the determinant of the matrix with elements
$B_{\nu\mu}(\al)=\bra v_{\nu}^{},
v_{\mu}^{}\ket_{\al}^{}$. This matrix is block-diagonal with 
blocks $B_{\nu\mu}(\al,n)$ for each subspace $\CV_{\al}[n]$
($\equiv \CV[n]$ as a vector space).
The formula for the determinant $D_n(\al)$
of the matrix $B_{\nu\mu}(\al,n)$
was conjectured in \cite{K1} and proven in \cite{FF}. 
It may be written as
\begin{equation}\label{kacdet}\begin{aligned}
{} & D_n(\al)\;= \;C\prod_{r,s=0}^{\infty}(\De_{\al}-\De_{r,s})^{p(n-rs)},
\;\text{where}\\
 & \De_{r,s}= \al_{r,s}(Q-\al_{r,s}), \qquad
\al_{r,s}=-\fr{b}{2}r-\fr{1}{2b}s,
\end{aligned}\end{equation}
where $C$ is a constant independent of $\al$, $c$, and 
$p(n)$ denotes the dimension of $\CV[n]$.

With the help of the determinant formula \rf{kacdet} it is possible 
to determine the cases where one has a scalar product $(.,.)_{\al}$
on $\CV_{\al}$ such that 
$( L_{-n}\xi,\zeta)_{\al}=(\xi,L_n\zeta)_{\al}$
(unitarity).
We are interested in the case $c>1$, in which the necessary and sufficient
condition for unitarity of the representation $\CV_{\al}$ 
was found to be $\De_{\al}>0$
\cite{K2}.

If $\al\neq\al_{r,s}$ one has a unique 
basis $\CB_{\al}^t$ that is dual to $\CB$ w.r.t. $\bra.,.\ket_{\al}$: 
It has elements $v_{\al,\nu}^t$ 
that are 
defined by $\bra v_{\al,\mu}^t,v_{\nu}^{}\ket_{\al}^{}=
\de_{\mu,\nu}^{}$. 
The expansion of the vectors $v_{\al,\nu}^t$ 
w.r.t. the canonical basis for $\CV$ can be written in terms of the 
inverse $B^{\nu\mu}(\al)$ of the matrix 
$B_{\nu\mu}(\al)$:
\begin{equation}\label{projexp}
v_{\al,\nu}^{t}\;=\;\sum_{\mu\in\CT(\nu)}
\;B^{\nu\mu}(\al)\,v_{\mu}^{},
\end{equation}
where $\CT(\nu)$ is the set of all tuples $\mu=(r_1,\dots,r_i,\dots)$ with
$n(\mu)\equiv n(\nu)$. It is clear that the dependence of the
$v_{\al,\nu}^t$ on $\al$ is rational with poles at $\al=\al_{r,s}$. We
will need to know the singular behavior at $\al=\al_{r,s}$ more precisely.

To this aim let us first consider $\CV_{r,s}\equiv \CV_{\al_{r,s}}$.
Vanishing of the Kac-determinant $D_n(\al)$ for $\al=\al_{r,s}$.
is equivalent 
to the existence of a subspace 
$\CS_{r,s}^{}\subset\CV_{r,s}{}$ that consists of vectors
$\xi$ with the property $\bra\xi,\zeta\ket_{\al_{r,s}}^{}=0$ for all $\zeta\in
\CV_{r,s}$. The singular subspace $\CS_{r,s}\subset\CV_{r,s}$ is generated
from a so-called null-vector $s_{r,s}^{}\in\CV_{r,s}^{} $ that 
satisfies the highest weight property $L_ns_{r,s}^{}=0$ for $n>0$
and $L_0s_{r,s}^{}=(\De_{r,s}+rs)s_{r,s}^{}$.

We would now like to show that
$ v^t_{\al,\nu}$ has a pole of first order at $\al=\al_{r,s}$ and  
$\lim_{\al\ra\al_{r,s}}(\al-\al_{r,s})\;v^t_{\al,\nu}\in
\CS_{r,s}$. To verify the claim, one may argue
as follows: Since $v^t_{\al,\nu}$ is rational there exists an integer
$p_{r,s}(\nu)>0$ such that the limit
$v_{r,s;\nu}^t\equiv 
\lim_{\al\ra\al_{r,s}}(\al-\al_{r,s})^{p_{r,s}(\nu)}\;v^t_{\al,\nu}$
exists. One finds $\bra v^t_{r,s;\nu},v_{\mu}^{}\ket_{\al_{r,s}}^{}=0$ for all 
$\mu\in\CT$ by combining the definition of $v_{r,s;\nu}^t$
with $\bra v^t_{\al,\nu},v_{\mu}^{}\ket_{\al}^{}
=\de_{\mu,\nu}$. Therefore $v_{r,s;\nu}^t\in \CS_{r,s}$. 
It remains to show that $p_{r,s}(\nu)=1$ for all 
$\nu\in\CT$. But this follows from the fact that 
the order of the zero of $D_n(\al)$ at $\al=\al_{r,s}$ coincides 
with the dimension of $\CS_{r,s}^n\equiv \CS_{r,s}\cap \CV[n]$.

\subsection{The trilinear form $\rho$}

Let us define a family of trilinear forms 
$\rho^{\al_3,\al_2,\al_1}_{z_3,z_2,z_1}:
\CV_{\al_3}\ot\CV_{\al_2}\ot\CV_{\al_1}\ra \BC$ such that
\newcommand{\bxi}{\bar{\xi}^{}}
\begin{equation*}\begin{aligned}
\Om\bigl(
V_{\al_3}(\xi^{}_3, \bxi_3|
z_3) & V_{\al_2}(\xi^{}_2,  \bxi_2|z_2)
V_{\al_1}(\xi^{}_1, \bxi_1|z_1) \bigr)
\;=\; \\
& \;=\; C(\al_3,\al_2,\al_1)
\rho^{\al_3,\al_2,\al_1}_{z_3,z_2,z_1}(\xi^{}_3,\xi^{}_2,\xi^{}_1)
\rho^{\al_3,\al_2,\al_1}_{z_3,z_2,z_1}(\bxi_3,\bxi_2,\bxi_1).
\end{aligned}\end{equation*}
Recall that the left hand side is defined by the conformal 
Ward identities and \rf{globalsl2}. Let us 
spell out the corresponding 
definition of $\rho(\xi^{}_3,\xi^{}_2,\xi^{}_1)$ a bit more explicitly:
One first of all needs to have  
\begin{equation}\begin{aligned}
\rho^{\al_3,\al_2,\al_1}_{z_3,z_2,z_1} & 
(v, v, v)
\;=\; \\
& \;=\;
(z_3-z_2)^{\De_{\al_1}-\De_{\al_2}-\De_{\al_3}}
(z_3-z_1)^{\De_{\al_2}-\De_{\al_1}-\De_{\al_3}}
(z_2-z_1)^{\De_{\al_3}-\De_{\al_2}-\De_{\al_1}}.
\end{aligned}\end{equation}
The conformal Ward identities imply rules of the form
\begin{equation*}\begin{aligned}
(n-2)!
\rho^{\al_3,\al_2,\al_1}_{z_3,z_2,z_1}& (L_{-n}^{}\xi^{}_3,\xi^{}_2,\xi^{}_1)
\;=\;\\
\;=\;  \rho^{\al_3,\al_2,\al_1}_{z_3,z_2,z_1} &
\bigl(\xi^{}_3,\pa_{z_3}^{n-2}T_{>}(z_3-z_2)\xi^{}_2,\xi^{}_1\bigr)
+\rho^{\al_3,\al_2,\al_1}_{z_3,z_2,z_1}
\bigl(\xi^{}_3,\xi^{}_2,\pa_{z_3}^{n-2}T_{>}(z_3-z_1)\xi^{}_1\bigr),
\end{aligned}\end{equation*}
where $n>1$, $T_>(z)=\sum_{n=-1}^{\infty}L_n z^{-n-2}$, 
together with analogous equations for 
$\rho^{\al_3,\al_2,\al_1}_{z_3,z_2,z_1}(\xi^{}_3, L_{-n}^{}\xi^{}_2,\xi^{}_1)$
and $\rho^{\al_3,\al_2,\al_1}_{z_3,z_2,z_1}(\xi^{}_3,\xi^{}_2, 
L_{-n}^{}\xi^{}_1)$.
These rules allow one to express $
\rho^{\al_3,\al_2,\al_1}_{z_3,z_2,z_1}(\xi^{}_3,\xi^{}_2,\xi^{}_1)$ 
in terms of $\rho^{\al_3,\al_2,\al_1}_{z_3,z_2,z_1}  
(L_{-1}^{n_3}v, L_{-1}^{n_2}v, L_{-1}^{n_1}v)$.
The evaluation of $\rho^{\al_3,\al_2,\al_1}_{z_3,z_2,z_1}$ is 
therefore completed by noting that
\begin{equation}\begin{aligned}
\rho^{\al_3,\al_2,\al_1}_{z_3,z_2,z_1}(L_{-1}\xi^{}_3,\xi^{}_2,\xi^{}_1)
\;=&\;\pa_{z_3}^{}
\rho^{\al_3,\al_2,\al_1}_{z_3,z_2,z_1}(\xi^{}_3,\xi^{}_2,\xi^{}_1),
\end{aligned}\end{equation}
and analogously for $\pa_{z_2}\rho$ and $\pa_{z_1}\rho$.
It is a consequence of these definitions that 
\begin{equation}\begin{aligned}
\rho^{\al_3,\al_2,\al_1}_{\infty,z_2,0} 
(v_{\nu_3}, v_{\nu_2}, v_{\nu_1})
\;\equiv \; \lim_{z_3\ra\infty}\lim_{z_1\ra 0}z_3^{2\De_{\al_3}^{\nu_3}}
\rho^{\al_3,\al_2,\al_1}_{z_3,z_2,z_1}(v_{\nu_3}, v_{\nu_2}, v_{\nu_1})\\
\; = \; z_2^{\De_{\al_3}^{\nu_3}-\De_{\al_2}^{\nu_2}-\De_{\al_1}^{\nu_1}}
\rho^{\al_3,\al_2,\al_1} 
(v_{\nu_3}, v_{\nu_2}, v_{\nu_1}),
\end{aligned}\end{equation}
 where $\De_{\al}^\nu\equiv\De_{\al}+n(\nu)$ and 
$\rho^{\al_3,\al_2,\al_1} 
(v_{\nu_3}, v_{\nu_2}, v_{\nu_1})$
is a polynomial in $\al_3,\al_2,\al_1$ and $c$.
The following important result is proven in \cite{FF}:
\begin{thm} {\sc Null vector decoupling:}\\ 
Let $i,j,k\in\{1,2,3\}$ be chosen such that $j\neq i$, $k\neq i$, 
$j\neq k$. Assume that 
(i) $\al_i=\al_{r,s}$, and 
(ii) $\xi_i$ lies in the singular subspace $\CS_{r,s}$.
One then finds that
$\rho^{\al_3,\al_2,\al_1}_{z_3,z_2,z_1} 
(\xi^{}_3,\xi^{}_2,\xi^{}_1)=0$ if and only if 
$\al_j$ and $\al_k$ satisfy the fusion-rules
$\De_{\al_k}=\De_{\al_j+mb+nb^{-1}}$, where 
$m\in\{-\frac{r}{2},-\frac{r}{2}+1,\dots,\frac{r}{2}\}$,
$n\in\{-\frac{s}{2},-\frac{s}{2}+1,\dots,\frac{s}{2}\}$.
\end{thm}

\section{Appendix B: Meromorphic continuation of matrix elements} 
\label{techapp}

\subsection{Analytic properties of conformal blocks} \label{analcfbl}

Our definition of $\rho$ allows us to write the 
coefficients of the power series \rf{CGpwser} that define 
the conformal blocks as follows:
\begin{equation}
\CF_{\al}^s\bigl[{}_{\al_4}^{\al_3}{}_{\al_1}^{\al_2}\bigr](n)\;=\;
\sum_{\mu,\nu\in\CT_n}\;\rho^{\al_4,\al_3,\al}(v,v,v_{\mu})\;
B^{\mu\nu}(\al)\; 
\rho^{\al,\al_2,\al_1}(v_{\nu},v,v).
\end{equation}

The convergence of the power series \rf{CGpwser} for 
$|z_2|<|z_3|$ can be shown \cite{TO} by means of
the free field representation for the chiral vertex 
operators (cf. Part \ref{gervnev}, Subsection \ref{chirvert} and 
Part \ref{boots}, Subsection \ref{freeCVO}).
This fact yields important information on the dependence of the
conformal blocks $\CF_{\al}^s$ w.r.t. the variables $\al_1,\dots,\al_4$ 
and $\al$:

{\it The conformal blocks $\CF_{\al}^s$ are entire analytic as functions
of $\al_1,\dots,\al_4$ and meromorphic as function of $\al$, with
poles for $\De_{\al}=\De_{r,s}$. The residues of these poles vanish
iff either $(\al_4,\al_3,\al)$ or $(\al,\al_2,\al_1)$ satisfy the
fusion rules from Theorem 1.}

By convergence of the power series \rf{CGpwser} it suffices 
to verify the corresponding claims for the coefficients $
\CF_{\al}^s\bigl[{}_{\al_4}^{\al_3}{}_{\al_1}^{\al_2}\bigr](n)$.
The product $\rho^{\al_4,\al_3,\al}(v,v,v_{\mu})
\rho^{\al,\al_2,\al_1}(v_{\nu},v,v)$ is a polynomial in $\al$
and $\al_1,\dots,\al_4$. $B^{\mu\nu}(\al)$ depends rationally on 
$\al$ with poles iff $\De_{\al}=\De_{r,s}$. By our observation at 
the end of Subsection \ref{subVerma} one may express the 
resulting residues of $
\CF_{\al}^s\bigl[{}_{\al_4}^{\al_3}{}_{\al_1}^{\al_2}\bigr](n)$
as sum of terms containing either $\rho^{\al_4,\al_3,\al}_{\infty,1,0} 
(v, v, s)$ or $
\rho^{\al,\al_2,\al_1}_{\infty,1,0} 
(s, v, v)$, where $s$ is an element of the singular subspace $\CS_{r,s}$.
The claim concerning vanishing of the residues is therefore
a direct consequence of Theorem 1.

\subsection{Proof of meromorphic continuation of correlation
functions} \label{mercont+} 

To establish the meromorphic continuation of the correlation
function $\bra  P_4|\SV_{\al_3}(z_3)\SV_{\al_2}(z_2) 
|P_1\ket$ let us first
consider the integrand in 
\rf{mat_el_prod}: 
By the remark at the end of the previous subsection
and the analytic properties of the $\up$-function  
it takes the form of a product of functions that are meromorphic w.r.t.
$\al_4,\dots,\al_1$ and $\al$. 
The double poles from 
the conformal blocks $\CF_{\al}$ are cancelled 
against zeros of the factors $\up(Q+2iP)$ and $\up(Q-2iP)$. 
One is therefore
left with poles coming from the DOZZ three point functions. 
The resulting pattern of poles is as follows:
Let us use the variable $\al\equiv \frac{Q}{2}+iP$ instead of $P$. One 
has strings of poles at
\begin{equation}\label{fourptpoles}\begin{aligned}
\al\;=& \; \si_1(\al_1-Q/2)+\si_2(\al_2-Q/2)-nb-mb^{-1},\\
\al\;=& \; \si_3(\al_3-Q/2)+\si_4(\al_4-Q/2)-nb-mb^{-1},\\
\al\;=& \; Q-\si_1(\al_1-Q/2)-\si_2(\al_2-Q/2)+nb+mb^{-1},\\
\al\;=& \; Q-\si_3(\al_3-Q/2)-\si_4(\al_4-Q/2)+nb+mb^{-1},
\end{aligned}
\end{equation} where $n,m\in\BZ^{\geq 0}$, 
$\si_i\in\{+,-\}$, $i=1,\dots,4$. As long as
\begin{equation}\label{funran2}
\begin{aligned}
{} &|\Re(\alpha_1 - \alpha_2)| < Q/2;\qquad 
    |\Re(Q - \alpha_1 - \alpha_2)| < Q/2;\\ 
   &|\Re(\alpha_3 - \alpha_4)| < Q/2;\qquad 
    |\Re(Q - \alpha_3 - \alpha_4)| < Q/2,
\end{aligned}\end{equation}
one finds that all the poles in \rf{fourptpoles} are strictly to the
left or right of the contour $\frac{Q}{2}+i\BR^+$ of integration over $\al$.
The integral over $\al$ is analytic w.r.t. $\al_4,\dots,\al_1$ 
in this case. When continuing outside \rf{funran2} one finds that
poles from \rf{fourptpoles} would cross the axis $\frac{Q}{2}+i\BR$. To 
define the meromorphic continuation one may use the 
reflection property \rf{refl} together with the fact that 
$ \CF_{\al}=\CF_{Q-\al}$, to 
``unfold'' the $\al$-integration to an integral over the full axis 
$\frac{Q}{2}+i\BR$. Let us furthermore introduce 
\begin{equation}
\al_{21}^{\pm}=\fr{Q}{2}-\al_1 \pm\bigl(\fr{Q}{2}-\al_2\bigr),\qquad
\al_{43}^{\pm}=\fr{Q}{2}-\al_3 \pm\bigl(\fr{Q}{2}-\al_4\bigr).
\end{equation}
Thanks to the reflection symmetry $\al_i\ra Q-\al_i$
it suffices to consider the case that
$\arg(\al_{21}^{\pm})\in[0,\frac{\pi}{2}]$ and
$\arg(\al_{43}^{\pm})\in[0,\frac{\pi}{2}]$.  
The
definition of the meromorphic continuation is straightforward in the
case that the imaginary parts of $\al_{21}^{\pm}$ and $\al_{43}^{\pm}$ 
are all different from zero and from each other. In this case one simply
has to deform the original contour $\frac{Q}{2}+i\BR$ to a contour
that is indented around the strings of poles that have crossed 
$\frac{Q}{2}+i\BR$. Equivalently one may use a contour that is the
sum of $\frac{Q}{2}+i\BR$ and a finite sum of small circles around the 
poles just mentioned. For concreteness let us consider the
case that $\Re(\al_{21}^\pm)>0$. One then has a contour 
consisting of $\frac{Q}{2}+i\BR$ and small circles around the poles
at
\begin{equation}\begin{aligned}
\al& =\al_{21}^{\pm}-nb-mb^{-1};\quad n,m\in\BZ^{\geq 0}, 
\quad\Re(\al)>\fr{Q}{2},\\
\al& =\al_{43}^{\pm}-nb-mb^{-1};\quad n,m\in\BZ^{\geq 0}, 
\quad\Re(\al)>\fr{Q}{2},
\end{aligned}\end{equation}
together with their ``reflected partners'' obtained by $\al\ra Q-\al$.

It remains to consider the following 
cases:
\begin{enumerate}
\item {\it The imaginary part of 
one of $\al_{21}^{\pm}$, $\al_{43}^{\pm}$ 
becomes zero.} It is enough to consider the case $\Im(\al_{21}^-)\ra 0$.
One finds 
a collision of the poles $\al =\al_{21}^{-}-nb-mb^{-1}$ that have 
crossed $\frac{Q}{2}+i\BR$ from the left with the poles $\al=Q-\al_{21}^{-}
+n'b+m'b^{-1}$ only if $2\al_{21}^-=Q+(n+n')b+(m+m')b^{-1}$,
so that $2\al=(1+n'-n)b+(1+m'-m)b^{-1}$. 
The Verma module $\CV_{\al}$ will therefore generically
contain a singular subspace. However, in these cases it is easy to
check that $\al_2$ and $\al_1$ satisfy the fusion rules of Theorem 1. 
By our discussion in the previous subsection one observes  
that the conformal blocks $\CF_{\al}$ remain nonsingular at these
values of $(\al,\al_2,\al_1)$. But this means that all of the poles 
that can potentially arise will be cancelled by the zeros of the factors
$\up(2\al)$ and $\up(2Q-2\al)$ contained in 
$C(\al_4,\al_3,\al)C(\bal,\al_2,\al_1)$.
\item  $\Im(\al_{21}^+)=\Im(\al_{21}^-)$ or 
$\Im(\al_{43}^+)=\Im(\al_{43}^-)$: The first case requires
$\De_{\al_1}=\De_{r,s}$ for some $r,s\in\BZ^{\geq 0}$. 
The zeros of the factor $\up(2\al_1)$ in $C(\bal,\al_2,\al_1)$
prevent the occurrence of a singularity. The second case is treated
analogously.
\item $\Im(\al_{43}^{\pm})=\Im(\al_{21}^{\pm})$: Here it suffices
to consider $\Im(\al_{43}^{+})=\Im(\al_{21}^{-})$, all other cases
being related to this one by reflections $\al_i\ra Q-\al_i$. 
Collision of poles $\al =\al_{21}^{+}-nb-mb^{-1}$ with 
$\al=Q-\al_{43}^{+}+n'b+m'b^{-1}$ occurs if 
$\sum_{i=1}^4 \al_i=Q-rb-sb^{-1}$ for some $r,s\in\BZ^{\geq 0}$.
These cases indeed produce poles of the matrix element.
\end{enumerate}
\newpage
\part{Canonical quantization}\label{canq1}\vspace{.5cm}

One may view the DOZZ-proposal as providing a complete 
description of Liouville theory in the energy representation: 
It is trivial to rewrite $\CH\simeq \int_{\BR^+}\CW_P$ 
as decomposition of $\CH$ into eigenspaces of the Hamiltonian 
$\SH=\SL_0+\bL_0$. However, it is difficult to get insight into
the physics of Liouville theory on the basis of this representation
only: One would like to have something like a Schr\"odinger- or coordinate
representation, where states in Liouville theory are represented
by wave-functions on ``target-space''. 
We are trying to address the following two questions in 
the present Part \ref{canq1} of our paper:
\begin{itemize}
\item[(i)] Does such a representation exist?
\item[(ii)] What can be learned from it?
\end{itemize}
The basis for our discussion will be the description of Liouville theory
as provided by our previous discussion of the DOZZ-proposal. 
We will propose a certain picture, many consequences of which are found to
be consistent with the DOZZ-proposal. Moreover, it provides  
a more intuitive interpretation for some of the otherwise 
mysterious consequences of the DOZZ-proposal, like the reflection property
and the associated Seiberg-bound. 

However, our discussion will remain in 
some respects inconclusive. In fact, we regard 
some of the
questions that we are going to discuss as extremely interesting
open problems for the future study of quantum Liouville theory.

\section{The problem of canonical quantization}

Let us start by formulating what we will mean when speaking of
``canonical quantization'' in the context of Liouville theory.

\subsection{Classical theory} \label{can_form}

Classically one may introduce the 
canonical formalism by starting from 
the action \rf{classact} and defining the momentum conjugate
to $\vf$ as $\Pi_{\vf}=\frac{1}{8\pi}\pa_{t}\vf$. By introducing
the Poisson bracket
\begin{equation}\label{canPB}
\{ \Pi_{\vf}(\si),\vf(\si')\}\;=\;\de(\si-\si')
\end{equation}
and the canonical Hamiltonian $H$  
\begin{equation}
H\;=\;\int\limits_{0}^{2\pi}d\si\biggl(4\pi\Pi_{\vf}^2+
\frac{1}{16\pi}(\pa_{\si}\vf)^2
+\mu_c e^{\vf}\biggr)
\end{equation}
one may recast the Liouville equation of motion in the Hamiltonian
form
\begin{equation}
\pa_{t}\vf(\si,t) \;=\; \{H,\vf(\si,t)\},\qquad
\pa_{t}\Pi_{\vf}(\si,t) \;=\; \{H,\Pi_{\vf}(\si,t)\}.
\end{equation}

\subsection{Notion of canonical quantization}

Naively one would
want to define the algebra of observables to be generated by
operators
$\vf(\si)$ and $\Pi_{\vf}(\si)$ that satisfy commutation 
relations obtained from \rf{canPB} by replacing $\{ \, ,\,\}\ra
\frac{i}{\hbar}[\, ,\,]$. Recall that we write 
$\hbar$ as $\hbar=b^2$ and use the rescaled fields 
$\phi=\frac{1}{2b}\vf$ and $\Pi_{\phi}=\frac{2}{b}\Pi_{\vf}$ 
when discussing the quantum theory. The canonical commutation 
relations would take the form $\bigl[\phi(\si),\phi(\si')\bigr]=0$,
$\bigl[\Pi_{\phi}(\si),
\Pi_{\phi}(\si')\bigr]=0$ and
\begin{equation}
\bigl[\phi(\si)\;,\;\Pi_{\phi}(\si')\bigr]= i \de(\si-\si'). 
\end{equation}
The fields for nonzero time will then 
be given as solutions of the quantum equation of motion, which one would 
expect to be of the form 
\begin{equation}
(\pa_t^2-\pa_{\si}^2)\phi=-4\pi\mu b \;[e^{2b\phi}]_b^{}.
\end{equation}
The notation $[O]_b^{}$ is supposed to indicate 
the quantum corrections (e.g. normal ordering, other
renormalizations) that are necessary to properly define an operator
$[O]_b^{}$ that corresponds to the classical observable 
$O$ upon taking the semi-classical limit $b\ra 0$.
We have introduced $\mu=b^{-2}\mu_c$.

There are of course many well-known
subtleties associated with such a formulation:
Due to short-distance singularities one can not expect 
$\phi(x)$, $\Pi_{\phi}(x)$, $x=(t,\si)$ to represent 
well-defined operators. 

\renewcommand{\ba}{\sbb}
One might, however, hope that the situation is better for the Fourier-modes 
of $\phi(\si)$, $\Pi_{\phi}(\si)$, which are good operators at least in the
case of the free bosonic field field theory in two dimensions. 
The Fourier modes  
$\sq$, $\spp$, $\sa_n$, $\ba_n$ of $\phi(\si)$,
$\Pi_{\phi}(\si)$ will be introduced such that
\begin{equation} 
\phi(\si)\;=\; \sq + i\sum_{n\neq 0}\frac{1}{n}
\bigl( \sa_{n} e^{-in\si} +\ba_{n} e^{in\si} \bigr)\qquad
\Pi_{\phi}(\si) \;=\; 2\spp+\sum_{n\neq 0}
\bigl( \sa_n e^{-in\si} +\ba_n e^{in\si} \bigr).
\end{equation}
Instead of the 
canonical commutation relations for $\phi(\si)$,
$\Pi_{\phi}(\si)$ one would then consider the following 
commutation relations: 
\begin{equation}\label{osc}
{[}\spp,\sq{]}=-\fr{i}{2}\qquad {[}\sa_n,\sa_m{]}=\fr{n}{2}\de_{n,-m}
\qquad {[}\ba_n,\ba_m{]}=\fr{n}{2}\de_{n,-m}, 
\end{equation}
together with the hermiticity relations
\begin{equation}\label{herm_osc}
\sq^{\dagger}=\sq\qquad \spp^{\dagger}=\spp, \qquad
\sa_n^{\dagger}=\sa_{-n}^{},\qquad \ba_n^{\dagger}=\ba_{-n}^{}.
\end{equation}
The Liouville Hilbert space $\CH$ would be required to form 
a representation of the commutation relations \rf{osc}, with
dynamics being generated by a Hamiltonian of the form
\begin{equation}\label{Ham}
 \SH=2\spp^2+
2\sum_{k>0}[\sa_{-k}\sa_k+\ba_{-k}\ba_k]_b^{}\;+\;\mu  
\int\limits_{0}^{2\pi} d\si\;{[}e^{2b\phi(\si)}{]}_b^{}.
\end{equation}

\subsection{Representation?} \label{repr?}

But how to choose $\CH$? 
The most natural choice might seem to be $\CH^{\rm F}=
L^2(\BR)\ot\CF\ot\CF$, 
where the zero modes
$\sq$ and $\spp$ are realized on $L^2(\BR)$ 
as multiplication operator and $-\frac{i}{2}\pa_q$ 
respectively, and $\CF\ot\CF$ is the 
Fock-space generated 
by acting with the oscillators $\sa_{-n}$, $\ba_{-n}$,  $n>0$ 
on the Fock-vacuum $\Om$.
But it is well-know that there exist many inequivalent
unitary representations of the canonical commutation relations. 
A large class of such representations may e.g. be obtained by improper
Bogoliubov transformations of the Fock-space representation \cite{Bz}.
Moreover, 
for massive quantum field theories it is known to be impossible
to define an interacting quantum dynamics when using  
a Fock-space representation for the canonical commutation
relations (see e.g. \cite{Ha}, p.55).

This issue is of course closely related to the problem of
defining the interaction term $\mu  
\int_{0}^{2\pi} d\si\;{[}e^{2b\phi(\si)}{]}_b^{}$. How 
to choose the ordering of the $\sa_{-n}$, $\ba_{-n}$? Are there other 
quantum corrections (renormalizations of parameters, counterterms)
necessary to define $\SH$? It could also be that there are 
many ways to define $(\CH,\SH)$ which would all represent canonical 
quantizations of Liouville theory in the sense of the previous subsection.
Here, however, 
we are looking for a very particular one: 
$\CH$ should form a representation of two commuting 
Virasoro algebras and $\SH=\SL_0+\bL_0$. If conformal invariance 
is taken as the primary requirement, it is not clear whether 
it is possible to insist on the strict interpretation of 
``the rules of canonical quantization''. It may become necessary to
consider weak forms of the above requirement like only demanding that
the canonical commutation relations hold between a dense set of states
in $\CH$.

To complete the confusion, let us 
note that it is not clear what 
the precise sense of the hermiticity 
relations \rf{herm_osc} should be. For example, $\spp$ might 
be symmetric, but not self-adjoint: Just think of quantum mechanics on 
the half-line, where $\spp^2$ can be made self-adjoint, but $\spp$ can't
(otherwise one could ``leave'' the half-line by means of some
translation $e^{it\spp}$). It could even be much worse:
The operators $\sq$, $\spp$, $\sa_n$, $\ba_n$ 
might be too singular to have a dense domain of definition. 

\begin{rem}
For readers considering such discussions as pedantic let us make the
following comment: A string background can abstractly be defined
as BRST-cohomology
of a collection of conformal field theories with total central charge
26 or 10, tensored with ghosts. Scattering amplitudes are constructed from 
correlation functions of the conformal field theories. One would
of course like to have an interpretation of the amplitudes as coming
from the perturbative expansion of a string field theory around 
some target space that is supposed to be described by the collection
of conformal field theories. What is the target space to a 
given collection of abstractly defined conformal field theories? 
One might try to get coordinates for the target space from the operators 
that describe the motion of the center of mass of the string, the 
zero modes. But what if these operators cease to be well-defined in the
case of interacting conformal field theories? This could indicate 
some obstruction to point-like localization in target space, some
``fuzziness'', ``stringy uncertainty''  or 
``non-commutativity of target space''.
\end{rem}

\section{Canonical quantization from DOZZ-proposal?}

Let us now examine to what extend one may view the 
DOZZ-proposal as providing a canonical quantization of 
Liouville theory.
The first task is of course to reconstruct the Liouville field itself.
The identification $V_{\al}=e^{2\al \vf}$ that one has semi-classically
suggests that $\SV_{\al}^{}(x)=[e^{2\al\phi}]_b(x)$, $x=(t,\si)$, 
and furthermore
\begin{equation}\label{lioureco1}
\phi(\si)\;=\; \fr{1}{2}\pa_{\al}^{}\SV_{\al}^{}(\si)|_{\al=0}^{},\qquad
\Pi_\phi(\si)\;=\; \fr{1}{4\pi}\pa_{t}^{}\bigl(\pa_\al^{}
\SV_{\al}^{}(t,\si)|_{\al=0}^{}\bigr)_{t=0}^{}.
\end{equation}
We will see that this definition 
produces fields $\phi(\si)$, $\Pi_\phi(\si)$ that can be shown to represent
the canonical commutation relations in a weak sense
i.e. between states from a dense subset 
of $\CH$. Moreover, the {\it euclidean} field $\phi(z,\bz)$ 
weakly solves a natural quantum version of the Liouville 
equation of motion.

\subsection{Canonical fields} \label{minkphi}

First of all, we need to recover 
fields $\SV_{\al}^{}(\tau,\si)$ 
on the euclidean cylinder from the fields $\SV_{\al}^{}(z,\bz)$
on the Riemann sphere
that are furnished by the DOZZ-proposal. Fields $\SV_{\al}^{}(\tau,\si)$, 
on the euclidean cylinder are recovered from 
 $\SV_{\al}^{}(z,\bz)$ by means of the 
conformal mapping $z=e^{\tau+i\si}$: $\SV_{\al}^{}(\tau,\si)=
|z|^{2\De_{\al}}\SV_{\al}^{}(z,\bz)$. The fields 
$\SV_{\al}^{}(\tau,\si)$ are well-defined as operators for 
negative euclidean time $\tau<0$. We may then define euclidean fields 
$\phi(\tau,\si)$, $\Pi_\phi(\tau,\si)$ as follows
\begin{equation}\label{lioureco2}
\phi(\tau,\si)\;=\; \fr{1}{2}\pa_{\al}^{}\SV_{\al}^{}(\tau,\si)|_{\al=0}^{},
\qquad
\Pi_\phi(\tau,\si)\;=\; \fr{i}{4\pi}\pa_{\tau}^{}\bigl(\pa_\al^{}
\SV_{\al}^{}(\tau,\si)|_{\al=0}^{}\bigr).
\end{equation}
Matrix elements such as 
$\bra\psi^{}_2|\phi(\tau_2,\si_2)\phi(\tau_1,\si_1)|\psi^{}_1\ket$,
$\bra\psi^{}_2|\Pi_\phi(\tau_2,\si_2)\phi(\tau_1,\si_1)|\psi^{}_1\ket$
etc. will be well-defined for $\tau_2>\tau_1$ and can be 
recovered from the matrix elements 
$\bra\psi^{}_2|\SV_{\al_2}(z_2,\bz_2)\SV_{\al_1}(z_1,\bz_1)|\psi^{}_1\ket$.

Moreover, it is possible to recover the matrix elements of the Liouville field
from the DOZZ three point function:
Let us start by considering 
$\bra P| \phi(\tau,\si)| \psi\ket$, where $|\psi\ket$ is of the form 
$|\psi\ket=\frac{1}{2\pi}\int_{\BR^+}dP\psi(P)|P\ket$. 
This matrix element is by our definition \rf{lioureco2}
represented as
\[
\bra P| \phi(\tau,\si)| \psi \ket\;=\;
\frac{1}{4}\lim_{\al\ra 0}\;\pa_{\al}^{}\!
\int\limits_{-\infty}^{\infty}\frac{dP'}{2\pi}\; 
C\bigl(\fr{Q}{2}-iP,
\al,\fr{Q}{2}+iP'\bigr)e^{2\tau(\De(P)-\De(P'))}\;\psi(P'),
\]
Taking the limit requires some care since the contour of integration
will be pinched between poles approaching the real axis in the limit
$\al\ra 0$ (cf. our discussion in Section \ref{techapp}). 
The result may be written as
\begin{equation}\label{matphi}\begin{aligned}
\bra P|\phi(\tau,\si)| \psi \ket
=\frac{1}{4}\int\limits_{-\infty}^{\infty}\frac{dP'}{2\pi}\;
e^{2\tau(\De(P)-\De(P'))}
\Biggl(\;
\frac{\la^{\frac{i}{b}(P-P')}\up(-2iP')
\up(+2iP)}{|\up(i(P-P'))\up(i(P+P'))|^2} &\;\psi(P')\\
 -\biggl(\frac{1}{(P-P')^2}+\frac{1}{(P+P')^2}\biggl) & \psi(P)
\Biggl),
\end{aligned}\end{equation}
where $\la\equiv \pi\mu\ga(b^2) b^{2-2b^2}$.
Vacuum expectation values of arbitrary descendants of $\bra P_2|$ and
$|\psi\ket$ can then be
obtained by using the commutation relations
\begin{equation}\label{phitrsf}\begin{aligned}
{[}\SL_n,\phi(w,\bw){]} \;=\; &  e^{nw}\bigl(\pa_w \phi(w,\bw)+Qn\bigr),\\
{[}\bL_n,\phi(w,\bw){]} \;=\; &  e^{n\bw}\bigl(\pa_{\bw} \phi(w,\bw)+Qn\bigr),
\end{aligned}\end{equation}
where $w=\tau+i\si$, 
which follow from those for $V_{\al}(z,\bz)$. 


One may then define
operator-valued distributions
$\SV_{\al}^{}(\si)$ at time
$t=0$ 
by taking $\tau\uparrow 0$, in other
words: Smeared operators $\SV_{\al}^{}(f)$ 
are obtained as 
\begin{equation} \label{eucl-mink}
\SV_{\al}^{}(f)\;=\; \lim_{\tau\uparrow 0}\;
\int_{0}^{2\pi}\!d\si \;f(\si)
\SV_{\al}^{}(\tau,\si),\qquad f\in\CC^{\infty}(S^1).
\end{equation}
In a similar way one recovers operator-valued distributions
$\phi(\si)$, $\Pi_{\phi}(\si)$ from $\phi(\tau,\si)$, 
$\Pi_{\phi}(\tau,\si)$. 

\begin{rem}
It looks likely that the zero mode operator 
$\sq\equiv \lim_{\tau\uparrow 0}
\int_{0}^{2\pi}d\si \phi(\tau,\si)$ is at least densely defined,
as the norm $\lVert\sq|\psi\ket\rVert^2$ would be given by the following 
expression
\begin{equation} 
\lVert\,\sq|\psi\ket\,\rVert^2\;=\;
\lim_{\tau_2\downarrow 0}\lim_{\tau_1\uparrow 0}\;
\int_{0}^{2\pi}d\si_2d\si_1\;
\bra\psi|\phi(\tau_2,\si_2)\phi(\tau_1,\si_1)|\psi\ket,
\end{equation}
which should be finite for 
states $|\psi\ket$ which are annihilated by some power of 
$L_n$ for any $n$. 
However, the resulting description for $\sq$ will be complicated,
making it difficult to control properties such as self-adjointness,
spectrum etc..
\end{rem}

\subsection{ Equation of motion}

Again start by considering $|\psi\ket$ 
of the form 
$|\psi\ket=\int_{\BR^+}dP\psi(P)|P\ket$. The relation 
\[
\pa_w\pa_{\bw}\bra P| \phi(w,\bw)| \psi\ket \;=\; 
\pi\mu\ga(b^2) b^{2-2b^2}\frac{\up_0}{\up(2b)} 
\bra P| \SV_b(w,\bw)| \psi\ket
\]
follows from \rf{matphi} 
and the DOZZ-formula for the matrix elements of $\SV_{b}$ by a 
straightforward calculation using the functional
equations for the $\up$- and $\Ga$-functions. 
Note  furthermore that
\[ \up(2b)=b^{1-2b^2}\ga(b^2)\up(b)=b^{1-2b^2}\ga(b^2)
\lim_{\ep\ra 0}b^{1-2b\ep}\frac{\Ga(b\ep)}{\Ga(1-b\ep)}\up(\ep)
=b^{1-2b^2}\ga(b^2)\up_0.
\]
Since $\pa_w\pa_{\bw}\phi(w,\bw)$ and $V_b(w,\bw)$ transform the same way 
under
Virasoro transformations, one obtains the relation 
\[
\pa_w\pa_{\bw}\bra P,\zeta| \phi(w,\bw)| \psi\ket \;=\; 
\pi\mu b 
\bra P,\zeta| \SV_b(w,\bw)| \psi\ket
\]
for all $|\psi\ket$ 
of the form 
$|\psi\ket=\sum_{\zeta\in\CB^{\ot 2}}
\int_{0}^{\infty}
dP\psi^{}_{\zeta}(P)|P,\zeta\ket$ with $\psi^{}_{\zeta}(P)$ nonzero
for finitely many $\zeta$ only.

\subsection{Canonical commutation relations}

In the present subsection 
 it will be proved that 
\begin{equation}
\bra P_2,\zeta_2|{[}\phi(\si),\pa_t\phi(\si'){]}|P_1,\zeta_1\ket
\;=\;i\,(2\pi)^2\,  \de(P_2-P_1)\de(\si-\si')\,
\bigl(\zeta_2,\zeta_1\bigr)_{\frac{Q}{2}+iP_1}^{}.
\end{equation}
Let us note that this result will essentially be a consequence 
of locality and crossing symmetry of the four point functions.
We will begin by considering the distribution
\[ D_{P_2,P_1}(\si,\si')\equiv
\bra P_2|{[}\phi(\si),\pa_t\phi(\si'){]}|P_1\ket.
\]
The distribution $D_{P_2,P_1}(\si,\si')$ should be given
in terms of euclidean correlation 
functions $E_{P_2P_1}(z,w)=\bra P_2|\phi(z,\bz)
\phi(e^{\tau}w,e^{\tau}\bw)|P_1\ket$ as 
\[ 
\begin{aligned}
D_{P_2,P_1}(\si,\si')=& 
\lim_{\tau\uparrow 0}i\pa_{\tau}E_{P_2P_1}(z,w_{\tau})-
\lim_{\tau\downarrow 0}i\pa_{\tau}E_{P_2P_1}(w_{\tau},z),
\end{aligned}
\]
where $z=e^{i\si}$, $w_{\tau}=e^{\tau+i\si'}$. The correlation functions 
$E_{P_2P_1}(z,w)$
may be represented as
\[
E_{P_2P_1}(z,w)
= \lim_{\al_1\ra 0}\lim_{\al_2\ra 0}
\frac{1}{4}
\pa_{\al_2}^{}\pa_{\al_1}^{}F_{P_2P_1}^{\al_2\al_1}(z,w),
\]
where $F_{P_2P_1}^{\al_2\al_1}(z,w)\equiv
\bra P_2|V_{\al_1}(z,\bz)V_{\al_2}(w,\bw)|P_1\ket$.
It turns out to be useful to employ the expression for the correlator
$F$ in terms of $t$-channel conformal blocks. 
For real $\al_1$, $\al_2$ near zero one may write the expansion of $F$
into conformal blocks as (cf. Section \ref{techapp}) 
\begin{equation}\label{F-expansion}
\begin{aligned}
F_{P_2P_1}^{\al_2\al_1}(z,w)
=\int\limits_{\BS}
\frac{d\be_t}{2\pi} \;
    C(\bbe_t,\al_2,\al_1)& C(\be_2,\be_t,\be_1)\;
   \bigl|\CfBlt{\be_t}{\al_2}{\al_1}{\be_2}{\be_1}
(z,w)\bigr|^2 \\[-1ex]
+ &   C(\be_2,\al_{21},\be_1)
    \bigl|\CfBlt{\al_{21}}{\al_2}{\al_1}{\be_2}{\be_1}(z,w)\bigr|^2,
\end{aligned}
\end{equation}
where $\be_2=\frac{Q}{2}-iP_2$, $\be_1=\frac{Q}{2}+iP_1$ and
$\al_{21}=\al_2+\al_1$.

Let us note that 
$F_{P_2P_1}^{\al_2\al_1}(z,w)=F_{P_2P_1}^{\al_1\al_2}(w,z)$ 
(locality of the $\SV_{\al}$) implies 
$E_{P_2P_1} (z,w)=E_{P_2P_1} (w,z)$. This implies that the distribution
$D_{P_2P_1}(\si,\si')$ can have support for $\si=\si'$ only. Let 
us therefore focus on the singular behavior of 
$E_{P_2P_1} (z,w)=E_{P_2P_1} (w,z)$ for $z=w$.
$E_{P_2P_1}(z,w)$ can be expanded as 
$E_{P_2P_1} (z,w)=E^{c}_{P_2P_1}(z,w)+E^{d}_{P_2P_1}(z,w)$, where
\[
\begin{aligned}
E^{c}_{P_2P_1}(z,w)\;\equiv\; &\frac{1}{4}\int\limits_{\BS}
\frac{d\be_t}{2\pi}\; C(\be_2,\be_t,\be_1\bigr)\;\pa_{\al_2}^{}\pa_{\al_1}^{}
C\bigl(\bbe_t,\al_2,\al_1\bigr)_{\sst {}^{\sst \al_1=0}_{\sst \al_2=0}}
    \;\bigl|\CfBlt{\be_t}{0}{0}{\be_2}{\be_1}(z,w)\bigr|^2,\\[-1ex]
E^{d}_{P_2P_1}(z,w)\;\equiv\; & 
\frac{1}{4} \pa_{\al_2}^{}\pa_{\al_1}^{}\bigl(
    C(\be_2,\al_2+\al_1,\be_1)
    |\CfBlt{\al_{21}}{\al_2}{\al_1}{\be_2}{\be_1}(z,w)|^2
\bigr)_{{}^{\sst \al_1=0}_{\sst \al_2=0}}.
\end{aligned}
\]
We have simplified the expression for $E^c$ slightly by noting that 
terms where not both derivatives w.r.t. $\al_2$ and $\al_1$ act on 
$C(\bbe_t,\al_2,\al_1)$ vanish when taking $\al_1\ra 0$ and $\al_2\ra 0$.

There are two types of singular behavior that one must consider:
Power-like behavior of the form $|z-w|^{2\la}$ with {\it positive} $\la$, 
and logarithmic behavior of the form $\log|z-w|^2$. It is straightforward
to verify that the former does not produce contributions to 
$D_{P_2P_1}(\si,\si')$. The logarithmic short-distance 
singularities come from $E^{d}_{P_2P_1}(z,w)$ only. They are produced 
when the derivatives $\pa_{\al_2}^{}\pa_{\al_1}^{}$ both act on the factor
$|z-w|^{-4\al_1\al_2}$ in 
$|\CF_{\al_{21}}^t|^2$.
By observing that $\lim_{\al \ra 0}C(\be_2,\al,\be_1)=2\pi \de(P_2-P_1)$ 
and furthermore 
\[ 
\lim_{\tau\ra 0} \pa_{\tau} 
\bigl(\ln|z-e^{-\tau}w|^2+\ln|z-e^{\tau}w|^2\bigr)
=\lim_{\tau\ra 0} 
\frac{4\tau}{\tau^2+(\si-\si')^2}=4\pi\de(\si-\si'),
\]
one finds that 
\begin{equation}
\bra P_2|{[}\phi(\si),\pa_t\phi(\si'){]}|P_1\ket
\;=\;i\,(2\pi)^2 \, \de(P_2-P_1)\de(\si-\si').
\end{equation}
This argument can easily be generalized to 
descendants of $\bra P_2|$, $|P_1\ket$.

\section{Zero mode Schr\"odinger representation} \label{schrep}

\hspace{5cm}
\begin{minipage}{8cm} 
{\it\small 
\def\baselinestretch{0.8}
Meine S\"atze erl\"autern dadurch,
da\ss\ sie der, welcher mich versteht, am Ende als unsinnig erkennt,
wenn er durch sie - auf ihnen - \"uber sie hinausgestiegen ist. (Er
mu\ss\ sozusagen die Leiter wegwerfen, nachdem er auf ihr heraufgestiegen 
ist.) (L. Wittgenstein)}
\end{minipage}\vspace{.4cm}

We are now going to discuss an assumption concerning 
the representation of the 
operators $\sq$, $\spp$, $\sa_n$, $\ba_n$ that would lead to 
a representation of states by wave-functions on target-space:
Assume that \begin{equation}\label{Schrrep}
\CH \;\simeq \;\CH^{\rm Schr}\;=\;L^2(\BR)\ot\CF\ot\CF\;\simeq\;
\int_{\BR}^{\oplus}dq\;\CF_q\ot\CF_q,
\end{equation}
where $\sq$ and $\spp$ are represented on functions $\psi(q)\in L^2(\BR)$ 
as the operator of multiplication with $q$ and the operator 
$-\frac{i}{2}\pa_q$ 
respectively,  
and the nonzero modes $\sa_n$, $\ba_n$ are represented in $\CF\ot\CF$
by a standard Fock-representation generated from a vector $\Om\in \CF\ot\CF$
that satisfies $\sa_n\Om=0=\ba_n \Om$. 
   
Within such a Schr\"odinger representation for the zero mode one could
represent states by wave-functions $\psi(q)$ that take values in 
$\CF\ot\CF$. The scalar product would be represented as
\begin{equation}
\bra\psi^{}_2,\psi^{}_1\ket_{\CH}\;=\;\int_{\BR}dq
\;\bigl( \psi^{}_2(q),\psi^{}_1(q)\bigr)_{\CF\ot\CF}^{},
\end{equation}
where $(.,.)_{\CF\ot\CF}^{}$ denotes the scalar product 
in $\CF\ot\CF$.

Our assumption \rf{Schrrep} has fair chances to be wrong (cf. our remarks in 
Subsections \ref{repr?} and \ref{minkphi}, as well as the
discussion in Section \ref{disc} below). Nevertheless
it will help us to develop a certain picture for the ``target-space''
physics of Liouville theory. Many features of that picture
turn to be consistent with the DOZZ proposal in a rather nontrivial
way. Let us therefore adopt \rf{Schrrep} as a working hypothesis that 
is useful to discuss certain issues, but which will ultimately 
have to be replaced by some refined description.

\subsection{Asymptotic correspondence to the free field} \label{asymfree}

In the representation \rf{Schrrep} one will find the Hamiltonian
$\SH$ to be represented as a second order differential operator of 
the form 
\begin{equation}\label{Ham_Schr}
 \SH=-\frac{1}{2}\pa_q^2+N_b+
2\sum_{k>0}(\sa_{-k}\sa_k+\ba_{-k}\ba_k)\;+\;\mu  
\int_{0}^{2\pi} d\si\;{[}e^{2b\vf(\si)}{]}_b^{},
\end{equation}
where $N_b$ is some normal-ordering constant.
At least semi-classically one has $\mu  
\int_{0}^{2\pi} d\si\;{[}e^{2b\vf(\si)}{]}_b^{}\propto e^{2bq}$,
which vanishes exponentially for $q\ra -\infty$.
It seems plausible to conjecture that quantum corrections in the 
interaction term ${[}e^{2b\vf(\si)}{]}_b^{}$ will preserve the
zero mode dependence $\propto e^{2bq}$, at least in leading order for 
$q\ra -\infty$. The role 
of the interaction will therefore
become negligible if one considers wave-packets
that have support in regions with large negative values
of $q$. It should be possible to approximate the time-evolution
of such wave-packets by the time-evolution generated by the
free Hamiltonian
\begin{equation}
\SH^{\rm F}=-
\frac{1}{2}\pa_q^2+N_b+2\sum_{k>0}(\sa_{-k}\sa_k+\ba_{-k}\ba_k).
\end{equation}
In this spirit one would also
expect to have 
\begin{equation}\label{ascorr}
{[}e^{2b\phi(\si)}{]}_b^{}\;\underset{q\ra-\infty}{\sim}\;
:e^{2b\phi(\si)}:,
\end{equation}
where $:\!\CO\hspace{-.5ex}:$ 
denotes the operator obtained by the usual free field
normal ordering.
Next-to-leading order corrections in the 
asymptotics for $q\ra -\infty$ should then be represented by
\begin{equation}\label{F-Ham}
 \SH^{\rm F}_{\mu}=-\frac{1}{2}\pa_q^2+N_b+
2\sum_{k>0}(\sa_{-k}\sa_k+\ba_{-k}\ba_k )\;+\;\mu  
\int_{0}^{2\pi} d\si:e^{2b\vf(\si)}:.
\end{equation}

\subsection{Generalized eigenfunctions of $\CH$}

Let us now examine how (generalized) eigenstates of $\SH$ 
would be described in the representation \rf{Schrrep}. 
One would want to construct generalized eigenfunctions $\psi^{}_{E,\nu}(q)$
to each eigenstate $|E,\nu\ket$ of $\SH$, where $\nu$ is just some label
for the degeneracy of the eigenvalue $E$ for the moment. 
This is of course impossible unless one knows the precise definition
for $\SH$. However,
if the asymptotic correspondence with free field theory, as discussed in the
previous subsection, really holds, one may at least discuss the 
asymptotic behavior of $\psi^{}_{E,\nu}(q)$ for $q\raf$, which is 
enough to get some important information on the spectrum.

The 
asymptotic behavior of $\psi^{}_{E,\nu}(q)$ for $q\raf$
should then of course be given by
solutions $\psi_{E}^{\rm F}(q)$ of the free eigenvalue equation 
$\SH^{\rm F}\psi_{E}^{\rm F}(q)=E \psi_{E}^{\rm F}(q)$,
which take the form
\renewcommand{\ff}{{\rm f}}
\begin{equation}\label{asymallg}
\psi_{E}^{\rm F}(q)\;\,=
\;\,e^{2iPq}\ff^+_n \;+\; e^{-2iPq}\ff^-_n,\qquad \quad
E=P^2+N_b+n,\end{equation}
where ${\rm f}_n^{\pm}\in\CF\ot\CF$ are eigenstates of the number operator
$\SN=2\sum_{k>0}(\sa_{-k}\sa_k+\ba_{-k}\ba_k)$ which have
the eigenvalue $n$. Wave-functions $\psi^{}_{E,\nu}(q)$ with asymptotic
behavior \rf{asymallg} would correspond to generalized eigenstates
in the continuous spectrum $\CH^{\rm c}$
of $\SH$. However, a priori it is not at all clear 
 for which choices for the parameters $P$, $\ff^+_n$, $\ff^-_n$ 
it is possible to ``integrate'' the eigenvalue equation 
$\SH \psi^{}_{E}=E\psi^{}_{E}$ to obtain a (plane-wave) normalizable
wave-function $\psi[P,\ff^+_n,\ff^-_n](q)$
with asymptotics \rf{asymallg}. But if $\CH$ forms
a representation of the canonical commutation relations for 
the nonzero modes
one would need that existence of a wave-function
$\psi[P,\ff^+_n,\ff^-_n](q)$ implies existence of 
$\psi[P,\sa_n\ff^+_n,\sa_n\ff^-_n](q)$
and $\psi[P,\ba_n\ff^+_n,\ba_n\ff^-_n](q)$. It follows that 
$\CH^{\rm c}$ must decompose into a collection of Fock-spaces
parameterized by $P$:
\begin{equation}\label{spec:Hc}
\CH^{\rm c}\;\simeq\;  
\int_{\BR}^{\oplus}
d\mu(P)\;
\CF_P\ot\CF_P,
\end{equation}
where the subscript $P$ in the notation $\CF_P\ot\CF_P$ indicates that
the action of $\SH$ on $\CF_P\ot\CF_P$ is represented as $P^2+N_b+\SN$.

\subsection{Comparison with DOZZ-proposal}

This seems to be as far as one can get on the basis of the asymptotic 
correspondence to the free field. Let us now compare to the structure
of the spectrum as given by the DOZZ-proposal. First of all, the latter
is purely continuous. One would therefore need to identify 
$\CH^{\rm c}\equiv\CH$.
Second, instead of \rf{spec:Hc} we had found in Part \ref{pathint}
a similar expansion with Fock-spaces $\CF_P$ replaced by 
Verma modules $\CV_P$. But this is of course perfectly consistent since
$\CF_P\simeq \CV_P$ as vector spaces (we will discuss the realization
of conformal symmetry later). 

The interesting point to observe
is that comparison with \rf{DOZZspec} implies that $\BS=\BR^+$, 
only {\it half} of the spectrum of free field theory.
This also implies that for each value of $P$ it suffices to 
specify e.g. the vector ${\rm f}_n^+$ in \rf{asymallg}. The vector
${\rm f}_n^-$ must be a function of ${\rm f}_n^+$ and $P$, 
${\rm f}_n^-=\SR(P){\rm f}_n^+$.
This is very plausible from the point of view of the 
``quasi-quantum mechanical'' picture that we are developing:
Whatever possible quantum modifications of the potential may be,
as long as they do not make it vanish in the opposite
limit $q\ra +\infty$, one will have
to impose a boundary condition concerning the behavior of 
wave-functions for $q\ra +\infty$ which selects a 
particular choice of $ \ff^-_n$ in dependence of 
$\ff^+_n$ and $P$ (or vice versa). 

The resulting picture is that
the wave-functions $\psi^{}_{P,{\rm f}}(q)$ 
corresponding to generalized 
eigenstates $|P,{\rm f}\ket$ are uniquely specified by 
the coefficient ${\rm f}\in\CF\ot\CF\simeq \CV_P\ot\CV_P$ of the 
``in-going'' plane wave $e^{2iPq}$, with coefficient of the 
``reflected out-going'' plane wave $e^{-2iPq}$ being given by the 
{\it reflection operator} $\SR(P)$:
\begin{equation}\label{asym}
\psi_{P,{\rm f}}^{}(q)\underset{q\ra-\infty}{\sim}\psi_{P,{\rm f}}^{\rm F}(q),
\qquad
\psi_{P,{\rm f}}^{\rm F}(q)\;\,=
\;\,e^{2iPq}{\rm f} \;+\; e^{-2iPq}\SR(P){\rm f}.\end{equation}
The wave-functions $\psi^{}_{P,{\rm f}}(q)$ describe the 
generalized Fourier transformation from the 
zero mode Schr\"odinger representation \rf{Schrrep} to the 
spectral representation 
\begin{equation}\label{spec:H}
\CH\;\simeq\;\CH^{\rm spec}\;=\;  
\int\limits_{\BR^+}^{\oplus}
\frac{dP}{2\pi}\;
\CF_P\ot\CF_P.
\end{equation}
Let us finally consider the wave-function $\psi^{}_P(q)\equiv
\psi^{}_{P,\Om}(q)$. For both terms in \rf{asym} to produce the
same eigenvalue of $\SH\sim \SH^{\rm F}$ one needs to have 
$\SR(P)\Om=R(P)\Om$, introducing the {\it reflection amplitude} $R(P)$.
Therefore
\begin{equation}\label{asymzm} 
\psi_{P}^{}(q)\underset{q\ra-\infty}{\sim}\psi_{P}^{\rm F}(q),
\qquad
\psi_{P}^{\rm F}(q)\;\,=
\;\,(e^{2iPq}\;+\; e^{-2iPq}R(P))\Om.\end{equation}

\subsection{Reflection in the potential}\label{reflpot}

We will now try to clarify the physical interpretation of $\SR(P)$: 
We claim that $\SR(P)$ represents
the scattering operator that describes how a wave-packet
coming in from $q\raf$ for $t\raf$ is reflected into 
another wave packet that is pushed out to $q\raf$ for $t\ra\infty$.
The Liouville interaction acts like a perfectly reflecting 
potential ``wall''. To verify this statement let us consider a wave-packet
\newcommand{\RB}{{\rm B}}
\[
\psi(q,t)=e^{-i\SH t}\psi(q)\;\,=
\;\,\sum_{{\rm f}\in\RB}\;\,\int\limits_0^{\infty} \frac{dP}{2\pi} \;\,
e^{-iE_{P,{\rm f}}t}\;\bra P,{\rm f}|\psi\ket\;\psi^{}_{P,{\rm f}}(q),
\]
where we assume that the orthonormal basis $\RB$ for $\CF\ot\CF$
was chosen such that all ${\rm f}\in\RB$ are eigenstates  of the number
operator $\SN$ with eigenvalue $N({\rm f})$,
so that $\SH |P,{\rm f}\ket=E_{P,{\rm f}}|P,{\rm f}\ket$ with
$E_{P,{\rm f}}\equiv 2P^2+\frac{Q^2}{2}+N({\rm f})$.
By the method of stationary phase it is possible to see
that $\psi(q,t)$ will vanish for any finite $q$ as $t\ra-\infty$, the
wave-packet is pushed out to large negative values of $q$: In order
to pick out the mode with energy $2P^2+Q^2/2$ as the dominant 
saddle point contribution one would have to consider $\psi(2Pt,t)$.

Since the wave packet is asymptotically supported for negative infinite
values of $q$, one may approximate the $\psi^{}_{P,\ff}(q)$ by their
asymptotic behavior $e^{2iPq}\ff$ (the term with $e^{-2iPq}$ gets 
suppressed in this limit). One is thereby led to the
conclusion that the behavior of $\psi(q,t)$ for $t\ra-\infty$ 
will be represented as time evolution 
according to the free Hamiltonian $\SH^{\rm F}$, 
$\psi(q,t)\sim
e^{-i\SH^{\rm F}t}\psi^{\rm in}(q)$ with 
\begin{equation}\label{tmasym+} 
\psi^{\rm in}(q)\;\,=\;\,\sum_{\ff\in\RB}\;\,
\int\limits_0^{\infty} \frac{dP}{2\pi} \;\,
\bra P,\ff|\psi\ket \; e^{2iPq}\,
{\rm f}\quad\in\quad\CH^{\rm F}
\equiv L^2(\BR)\ot\CF\ot\CF
\end{equation}
In the other limit $t\ra+\infty$ one would similarly find 
$\psi(q,t)\sim
e^{-i\SH^{\rm F}t}\psi^{\rm out}(q)$ with
\begin{equation}\label{tmasym-} 
\psi^{\rm out}(q)\;\,=\;\,\sum_{\ff\in\RB}\;\int\limits_0^{\infty} \frac{dP}{2\pi} \;\,
\bra P,\ff|\psi\ket \;e^{-2iPq}\,\SR(P)\ff\quad\in
\quad \CH^{\rm F}
\end{equation}
One may then define 
{\it generalized wave operators} as
\[
\SW^{\pm}(\SH_0,\SH,\SJ^{\pm})=\lim_{t\ra\pm\infty}\;\;e^{i\SH_0 t}\;\SJ^{\pm}
\;e^{-i\SH t}\;,
\]
where the identification maps $\SJ^{\pm}:\CH\ra \CH^{\rm F}$ are defined by 
$\SJ^{\pm}\psi^{}_{P,{\rm f}}(q)=e^{\mp 2iPq}{\rm f} $,
and the corresponding scattering operator
$\SS=(\Om^-)^{-1}\Om^+:\CH^{\rm F}\ra \CH^{\rm F}$ is easily read off from
\rf{tmasym+} and \rf{tmasym-} to be represented by 
our stationary reflection operator $\SR(P)$. For this identification 
to be consistent we evidently need that $\SR(P)$ is {\it unitary}.

\subsection{Conformal symmetry in the quantum theory}\label{quantvir}

Let us observe that the asymptotic correspondence between Liouville 
field and a free field that was discussed in Subsection \ref{asymfree} 
would suffice to conclude that quantum Liouville theory becomes 
a conformal field theory upon choosing the normal ordering 
constant $N_b$ appropriately:

If the spectral decomposition of the Hamiltonian can be
represented as in \rf{spec:H} one may always introduce an action of two  
commuting copies of the Virasoro algebra on $\CH$ by using usual free
field representations on the Fock-spaces $\CF_P$.
The action on the first tensor factor of $\CF_P\ot\CF_P$ would 
be defined in terms
of the generators  
\begin{equation}\label{FockVir}
\begin{aligned}\SL_n^{\rm F}(P)\;=\;& 
(2P+inQ)\sa_n^{}+\sum_{k\neq 0,n}\sa_k^{} \sa_{n-k}^{},
\qquad n\neq 0, \\
\SL_0^{\rm F}(P)\;=\;&P^2 +\frac{Q^2}{4}+
2\sum_{k>0}\sa_{-k}^{}\sa_k^{},
\end{aligned}
\end{equation}
whereas the action on the second tensor factor 
is generated by operators $\bL_n$ that are obtained by replacing 
$\sa_n\ra\ba_n$ in the expressions \rf{FockVir}. The spectral 
decomposition \rf{spec:H} then allows us to define operators 
$\SL_n$, $\bL_n$ on $\CH$ by means of the relations 
\begin{equation}\label{Virdef}
\SL_n\,|P,\ff\ket=|P,\SL_{n}^{\rm F}(P)\ff\ket\qquad\qquad
\bL_n\,|P,\ff\ket=|P,\bL_{n}^{\rm F}(P)\ff\ket.
\end{equation}
The operators $\SL_n$, $\bL_n$ satisfy
the usual commutation relations of the Virasoro algebra
with central charge $c$ given in terms of the parameter $b$ by the 
relations
\begin{equation}
c\;=\;1+6Q^2, \qquad Q=b+b^{-1}.
\end{equation}

However, the crucial property for identifying conformal 
transformations as a symmetry
of the theory is the fact that the Hamiltonian is recovered from $\SL_0$,
$\bL_0$ as 
\begin{equation}
\SH\;=\;\SL_0+\bL_0.
\end{equation}
For this to be the case one just 
needs that the normal ordering constant $N_b$
is equal to $\frac{Q^2}{2}$. 
Taking into account that the Fock space representation $\CF_P$ is irreducible
for real values of $P$ \cite{Fr},
one may identify
the spectral representation \rf{spec:H} of $\SH$
with the decomposition of $\CH$ into irreducible representations 
$\equiv$ Verma modules $\CV_P\ot \CV_P$ of the Virasoro algebra.
Under this correspondence one identifies the highest weight state 
$v\ot v$ of $\CV_P\ot \CV_P$ with 
$|P\ket\equiv
|P,\Om\ket\equiv|P,v\ot v\ket$ corresponding to the wave-function
$\psi^{}_{P}(q)$ with asymptotics \rf{asymzm}.

\subsection{Consistency of conformal symmetry with reflection}\label{refl_op}

The next thing one needs to observe, however, is the fact that
the requirement of having a consistent realization of conformal symmetry 
in the Schr\"odinger representation $\CH^{\rm Schr}$
imposes strong constraints 
on the form of the reflection operator $\SR(P)$. 
If one considers
wave-packets supported for $q\raf$, one finds that 
consistency of \rf{Virdef},\rf{FockVir} with the asymptotic form
\rf{asym} of wave-functions requires that 
\begin{equation}\label{BCTvir}
\SL_n \;\underset{q\ra-\infty}{\sim}\;\SL_n^{\rm F},\;\;\text{where}
\qquad\begin{aligned}\SL_n^{\rm F}=& 
(-i\pa_q+inQ)\sa_n+\sum_{k\neq 0,n}\sa_k \sa_{n-k}\\
\SL_0^{\rm F}=&\frac{1}{4}(-\pa_q^2+Q^2)+2\sum_{k>0}\sa_{-k}\sa_k.
\end{aligned}
\end{equation} 
However, the action of the Virasoro generators $\SL_n$ on the 
second term in \rf{asym} may now be expressed in two ways: Either 
as $e^{-2iPq}\SR(P)\SL_n^{\rm F}(P)\ff$ by using \rf{Virdef} or
alternatively as $e^{-2iPq}\SL_n^{\rm F}(-P)\SR(P)\ff$ when using 
\rf{BCTvir} directly. We conclude that we must have
\begin{equation}\label{refint}
\SR(P)\SL_n^{\rm F}(P)\;=\;\SL_n^{\rm F}(-P)\SR(P)\qquad\qquad
\SR(P)\bL_n^{\rm F}(P)\;=\;\bL_n^{\rm F}(-P)\SR(P),
\end{equation}
so the reflection operator $\SR(P)$ must be an intertwining operator between
the Fock-representations $\CF_P$ and $\CF_{-P}$.

Such an operator is uniquely determined by 
this intertwining property in terms of 
the {\it reflection amplitude} $R(P)$
which characterizes the action of $\SR(P)$ on the Fock-vacuum 
$\Om$. This is easily seen by recalling that the Fock-space 
representation $\CF_P\ot \CF_P$ is isomorphic 
to $\CV_P\ot\CV_P$ for $P\in\BR$. This means 
that any $\ff\in\CF_P\ot\CF_P$
can be uniquely written as the action of some polynomial 
$\CP_{P,\ff}^{}$ in the 
variables $\SL_n(P),\bL_n(P)$ on the Fock-vacuum $\Om$, i.e.
$\ff =\CP_{P,\ff}^{}[\SL_n^{\rm F}(P),\bL_n^{\rm F}(P)]\Om$. 
The intertwining property
\rf{refint} then implies that
\[
\SR(P)\ff\;=\;\SR(P)\;\CP_{P,\ff}^{}\bigl[\SL_n^{\rm F}(P),\bL_n^{\rm F}(P)
\bigr]\Om
\;=\;R(P)\;\CP_{P,\ff}^{}\bigl[\SL_n^{\rm F}(-P),\bL_n^{\rm F}(-P)\bigr]
\Om.
\]
Conformal symmetry therefore reduces the description of the 
scattering of wave-packets in the Liouville potential to the
knowledge of a single function, the reflection amplitude $R(P)$. 
Let us note that $\SR(P)$ will be unitary iff $|R(P)|=1$.

\subsection{Macroscopic vs. microscopic states} \label{micmac2}

We had observed in Part \ref{pathint}, Subsection \ref{micmac}
that one may naturally consider the analytic continuation of states
$|P\ket$ to complex values of $P$ in some distributional sense. 
This should of course be reflected by the existence of an analytic
continuation for the corresponding wave-function $\psi^{}_{P}(q)$. 

By means of analytic continuation one may in particular compare
$\psi^{}_{P}(q)$ and $\psi^{}_{-P}(q)$: In view of the asymptotics 
\rf{asym} one would find $\psi^{}_{P}(q)=R(P)\psi^{}_{-P}(q)$. But this
is to be compared to the reflection property $|P\ket=S(\frac{Q}{2}+iP)|-P\ket$
that follows from the DOZZ-proposal. We clearly must have 
$R(P)= S(\frac{Q}{2}+iP)$, so that 
\begin{equation}\label{Refamp}
R(P)=-\bigl(\pi\mu\ga(b^2)\bigr)
^{-\frac{2iP}{b}}\frac{\Ga(1+2ibP)\Ga(1+2ib^{-1}P)}
{\Ga(1-2ibP)\Ga(1-2ib^{-1}P)}.
\end{equation}

Our discussion in Subsection \ref{micmac} can now easily be rephrased
in terms of the asymptotic behavior of the wave-functions $\psi^{}_{P}(q)$:
In view of the asymptotic behavior \rf{asymzm} one would 
expect the $\psi^{}_{P}(q)$ to be plane-wave normalizable for 
$P\in\BR$, non-normalizable for $P\notin\BR$. However, in the 
latter case one would still expect $\psi^{}_{P}(q)$ to represent
a distribution $|P\ket$ that can be defined on the subspace of $\CH$ 
which is represented by wave-functions $\psi(q)$ with sufficiently
strong exponential decay. Let us adopt the terminology ``macroscopic
states'' for the (plane-wave) normalizable states
$|P,\ff\ket$ with $P\in\BR$, and ``microscopic
states'' for the $|P,\ff\ket$ with $P\notin\BR$, as proposed by
Seiberg in \cite{Se}. 

The characterization of the domain of $\bra P|$, $P\notin\BR$ in terms
of the exponential decay properties of the wave-functions $\psi(q)$
can easily be translated into our previous characterization of the domain of
$\bra P|$ in terms of analyticity of $\bra P|\psi\ket$:

Consider a state $|\psi\ket$ that is represented 
in the zero mode Schr\"odinger representation by a 
wave-function $\psi(q)$ which decays 
for $q\ra\infty$ faster than $e^{2\la q}$ for $q\ra-\infty$.
The wave-functions of $|\psi\ket$ in the spectral representation,
\begin{equation} \label{wvrepr}
\bra P,{\rm f}|\psi\ket \;=\; \int\limits_{-\infty}^{\infty}dq
\;\bra \,\bar{\psi}_{P,{\rm f}}(q)\, ,\,\psi(q)\,\ket^{}_{\CF\ot\CF},
\end{equation}
will then be analytic in $P$ as long as the integral
in \rf{wvrepr} converges. This will be the case for $|\Im(P)|<\la$.
Conversely, if the wave-functions
$\bra P,\ff|\psi\ket$ in the spectral representation are 
analytic in some 
strip around the real axis of width larger than 
$\la$, one may get the asymptotic behavior of
\begin{equation}\label{wfpsi}
\psi(q)\;=\;\sum_{\ff\in\RB}\;
\int\limits_{0}^{\infty} \frac{dP}{2\pi} \; \psi^{}_{P,\ff}(q)\;\bra P,{\rm f}|\psi\ket,
\end{equation}
by using unitarity of $\SR(P)$ to rewrite the integral over $\BR^+$
as an integral over the contour $\BR$, and taking advantage of the 
analyticity of the integrand in \rf{wfpsi} to  
shift that contour of integration to the axis $\BR+i\la$: 
It follows that
$\psi(q)$ decays faster as $e^{2\la q}$ for $q\ra-\infty$.

\subsection{Next to leading order corrections to Virasoro generators}
\label{ntol}

Equations \rf{BCTvir} describe the representation of the 
Virasoro algebra in \rf{Schrrep} only to leading order for $q\raf$. 
In next-to-leading
order one will need corrections to \rf{BCTvir}:
In order to have a consistent realization of conformal symmetry
one needs to have generators 
$\SL_{n,\mu}^{\rm F}$, $\bL_{n,\mu}^{\rm F}$ such that 
$\SH^{\rm F}_{\mu}=
\SL_{0,\mu}^{\rm F}+\bL_{0,\mu}^{\rm F}$, where $\SH^{\rm F}_{\mu}$ 
was defined in \rf{F-Ham}.
And indeed, there exists such a 
one-parameter ``deformation'' of the representation
\rf{BCTvir} that preserves the commutation relations:
\begin{equation}
\SL_{n,\mu}^{\rm F}\;=\;\SL_n^{\rm F}+\frac{\mu}{2}\int\limits_{0}^{2\pi}d\si
\;e^{in\si}:\!e^{2b\phi(\si)}\!:,
\end{equation}
and similarly for $\bL_{n,\mu}^{\rm F}$. It was shown in \cite{CT} 
that the modified generators $\SL_{n,\mu}^{\rm F}$ indeed satisfy the same
algebra as the $\SL_{n}^{\rm F}$. It will be important to notice that 
there is a further  
deformation 
of the generators $\SL_{n,\mu}^{\rm F}$, defined by
\begin{equation}\label{doublemod}
\SL_{n,\mu,\tilde{\mu}}^{\rm F}\;=\;\SL_{n,\mu}^{\rm F}+\frac{\tilde{\mu}}{2}
\int\limits_{0}^{2\pi}d\si\;
e^{in\si}:e^{2\tilde{b}\phi(\si)}:,
\end{equation}
where $\tilde{b}=b^{-1}$, that still
satisfies the same algebra as the $\SL_{n}^{\rm F}$.  
This possibility is due to the fact that
$:e^{2\tilde{b}\phi(\si)}:$ has conformal 
dimensions $(1,1)$ just like $:e^{2b\phi(\si)}:$. There is of course a 
very similar deformation $\bL_{n,\mu,\tilde{\mu}}^{\rm F}$
of the generators $\bL_{n,\mu}^{\rm F}$.

It seems natural to regard the term proportional to $\tilde{\mu}$  
in \rf{doublemod} as a possible quantum correction in the definition
of the Virasoro generators that preserves conformal symmetry.
It would correspond to the following modification of the 
``perturbative'' Hamiltonian $\SH^{\rm F}_{\mu}$: 
\newcommand{\tb}{\tilde{b}}
\begin{equation}
\SH^{\rm F}_{\mu\tilde{\mu}}\equiv \SH^{\rm F}+\mu \SU+\tilde{\mu}\tilde{\SU},
\end{equation} 
where $\SU$ and $\tilde{\SU}$ are defined as 
\begin{equation}
 \SU\;\equiv\;
\int\limits_{0}^{2\pi}
d\si :e^{2b\phi(\si)}:,\qquad
\tilde{\SU}\;\equiv\;
\int\limits_{0}^{2\pi}d\si
:e^{2b^{-1}\phi(\si)}:.
\end{equation}

\begin{rem} \label{as_rem} There do not seem to be any other operators 
that one could add to $\SL_{n,\mu,\tilde{\mu}}^{\rm F}$ 
and $\bL_{n,\mu,\tilde{\mu}}^{\rm F}$ 
without destroying the Virasoro-algebra commutation relations.   
It is therefore tempting to identify
$\SH\equiv\SH^{\rm F}_{\mu\tilde{\mu}}$, 
$\SL_n\equiv \SL_{n,\mu,\tilde{\mu}}^{\rm F}$ 
and $\bL_n\equiv \bL_{n,\mu,\tilde{\mu}}^{\rm F}$,
with $\tilde{\mu}$ and $\mu$ related by the formula \rf{dualiz} 
required by the DOZZ-proposal. We will discuss later (Section \ref{disc})
why such an identification appears to be problematic. 
However, these problems will {\it not} at all exclude the
possibility that $\SH^{\rm F}_{\mu\tilde{\mu}}$ 
represents the asymptotic behavior 
of $\SH$ for $q\raf$ up to terms that vanish faster than 
exponentially in that limit.
\end{rem}

\section{Exponential operators}

Let us now consider the exponential operators $[e^{2\al\phi}]_b^{}(\si)$
in more detail. In the representation \rf{Schrrep} it is of course natural
to try the ansatz
\begin{equation}\label{expans}
[e^{2\al\phi}]_b(\si)\;\overset{?}{=}\;
:e^{2\al\phi(\si)}:.
\end{equation}
This would indeed yield local operators that transform 
covariantly under the Virasoro algebras generated by the
$\SL_{n,\mu,\tilde{\mu}}^{\rm F}$, $\bL_{n,\mu,\tilde{\mu}}^{\rm F}$,
\begin{equation}\begin{aligned}\label{Ecov}
\bigl[\SL_{n,\mu,\tilde{\mu}},[e^{2\al\phi}]_b(\si)\bigr]= & e^{+in\si}
\bigl(-i\pa_{\si}+n\De_\al)\bigr)\;[e^{2\al\phi}]_b(\si),\\
\bigl[\bL_{n,\mu,\tilde{\mu}},[e^{2\al\phi}]_b(\si)\bigr]= & e^{-in\si}
\bigl(+i\pa_{\si}+n\De_\al)\bigr)\;[e^{2\al\phi}]_b(\si),
\end{aligned}\end{equation}
as desired. However, as we are already expecting some 
trouble with the representation \rf{Schrrep}, it seems important
to note that all that we will really be using in the following discussion
is the assumption that
\begin{equation}\label{asycond}
[e^{2\al\phi}]_b(\tau,\si)\;\underset{q\ra-\infty}{\sim}\;
:e^{2\al\phi(\tau,\si)}:
\end{equation}
holds weakly (between wave-packets). 

\subsection{State-operator correspondence}

We had previously (Section \ref{reconst}) discussed the correspondence 
between vertex operators $\SV_{\al}$ and microscopic states 
$|\al\ket$ (in the distributional sense,
cf. Subsection \ref{micmac}). It may be formulated as follows:

{\sc State-operator correspondence} $\frac{\quad}{\quad}$ 
{\it The vertex operators $\SV_{\al}$ are in one-to-one correspondence to 
the microscopic states $|\al\ket$, and create these states via
\begin{equation}\label{st-op} \begin{aligned}
\bra\psi |\al\ket\;=&\;\lim_{z\ra 0}\;\,\bra\psi |
[e^{2\al\phi}]_b^{}(z,\bz)|0\ket \\
 \bra Q-\al|\psi\ket\; = & \;
\lim_{z\ra \infty}\;\,\bra 0|[e^{2\al\phi}]_b^{}
(z,\bz)|\psi\ket\;|z|^{4\De_{\al}}
\end{aligned}\qquad
\text{for}\quad \psi\in\CD_{\al},
\end{equation}
where $\CD_{\al}\subset\CH$ is the domain of $\bra\al|$.}

We would now like to show that the above correspondence between operators
and states holds if and only if $\Re(\al)<\frac{Q}{2}$ as a consequence
of the fact that wave-packets get pushed out to $q\ra -\infty$ for 
$t\ra\pm\infty$:

For simplicity let us consider wave-packets of the form 
$\bra\psi|=\int_0^{\infty}dP\bra P|\psi(P)$. The generalization to 
descendants thereof is straightforward.
The limit $z \ra 0$ corresponds to 
$t\ra -e^{-i\ep}\infty $ in the
minkowskian formulation. One should therefore analyze 
\begin{equation}\label{minkstop}
\lim_{t\ra-e^{-i\ep}\infty}e^{2it\De_{\al}}
\bra \psi|[e^{2\al\phi}]_b^{}(t,\si)|0\ket.
\end{equation}
The limit \rf{minkstop} may be related 
to the asymptotic behavior of wave-packets
for $t\ra -\infty$ by changing from the Heisenberg to the 
Schr\"{o}dinger picture,
$\bra\psi|[e^{2\al\phi}]_b(t,\si)|0\ket=
\bra\psi(t)|[e^{2\al\phi}]_b(\si)|0\ket$. 
As discussed in the previous section, one will
for $t\ra -e^{-i\ep}\infty$ 
find wave-packets to be supported far off the potential,
i.e. moving off to $q\raf$.
On such wave-packets one may therefore represent the microscopic state
$[e^{2\al\phi}]_b^{}(\si)|0\ket$ by
its leading $q\ra-\infty$ behavior
$:\exp(2\al\phi(\si)):\exp(-Qq)\Om$. 
A non-vanishing result can indeed only be 
obtained for $\Re(\al)\leq Q/2$ since otherwise the wave function of 
$[e^{2\al\phi}]_b^{}(\si)|0\ket$ vanishes where the wave-packet 
$\bra\psi(t)|$ is supported for $t\raf$. 
One may then represent the above matrix element in the limit $t\raf$ 
by a matrix element in $\CH$ between the states
$|0\ket_{\rm F}=e^{-Qq}\,\Om$ and 
${}_{\rm F}\bra\psi|=\int_0^{\infty}\frac{dP}{2\pi}\;\bar{\psi}(P)e^{-2iPq}
\Om^{\dagger}$: 
\[ 
\begin{aligned}
\lim_{t\ra-e^{-i\ep}\infty}\;e^{2it\De_{\al}}\,
\bra\psi(t)|[e^{2\al\phi}]_b^{}(\si)|0\ket\;=\; &
\lim_{t\ra-e^{-i\ep}\infty}\;e^{2it\De_{\al}}
\,{}_{\rm F}^{}\bra \psi(t)|:e^{2\al\phi(\si)}:
|0\ket_{\rm F}^{}\\
=\; & \lim_{t\ra-e^{-i\ep}\infty}\;e^{2it\De_{\al}}\,
{}_{\rm F}^{}\bra\psi|:e^{2\al\phi(t,\si)}:
|0\ket_{\rm F}^{}\\
=\; & \;\,
{}_{\rm F}^{}\bra\psi|e^{2\al q}|0\ket_{\rm F}^{}\;
=
\;
{}_{\rm F}^{}\bra\psi|\al\ket_{\rm F}^{}\;=\;\bra\psi|\al\ket
\end{aligned}.
\]
We conclude that
only the operators $[e^{2\al\phi}]_b^{}(\si)$ 
with $\Re(\al)\leq Q/2$ are in correspondence with 
microscopic/macroscopic states. 

\subsection{Seiberg bound} \label{seibd}

The operators $\SV_{\al}$ as characterized by the three point functions 
$C(\al_3,\al_2,\al_1)$ satisfy the remarkable reflection
property $\SV_{\al}=R(\al)\SV_{Q-\al}$. We would like to
identify the operators $\SV_{\al}$ with the
(suitably quantum corrected) exponential operators 
${[}e^{2\al\phi}{]}_b$. As identifying property for the latter 
we had required that these operators have leading
asymptotics for $q\ra-\infty$ given by the standard free field normal 
ordered exponential operators $:\exp(2\al\phi):$, cf. 
\rf{asycond}. 
But the reflection
property $\SV_{\al}=R(\al)\SV_{Q-\al}$ then 
produces a problem for identifying
operators ${[}e^{2\al\phi}{]}_b$ for $\Re(\al)>\frac{Q}{2}$: In that case
$:\exp(2(Q-\al)\phi):$ would dominate over
$:\exp(2\al\phi):$ in the asymptotics for
$q\raf$, so that one would identify 
$\SV_{\al}=R(\al){[}e^{2(Q-\al)\phi}{]}_b$. If one insists on characterizing
exponential operators ${[}e^{2\al\phi}{]}_b$ by having leading asymptotics
\rf{asycond}, this simply means that one does not find 
any operators ${[}e^{2\al\phi}{]}_b$, $\Re(\al)>\frac{Q}{2}$ among the
operators $\SV_{\al}$, $\al\in\BC$. So how about operators 
${[}e^{2\al\phi}{]}_b$, $\Re(\al)>\frac{Q}{2}$: Do they escape our methods
or don't they exist? An argument in favor of the second possibility
was given by N. Seiberg in \cite{Se}, which is why this issue goes under
the name of ``Seiberg-bound''. 

Let us translate Seiberg's argument into the 
present framework: A slight modification of the
argument used in the previous subsection gives 
\begin{equation}\label{seib1}
\bra \psi|[e^{2\al\phi}]_b^{}(\si)|0\ket\;=\;0 .
\end{equation} 
To see this, it suffices to write
\begin{equation}\begin{aligned}
\bra \psi|[e^{2\al\phi}]_b^{}(\si)|0\ket\;=& \;\lim_{t\ra-\infty}\;
\bra \psi(t)|e^{-iHt}[e^{2\al\phi}]_b^{}(\si)e^{iHt}|0\ket\\
=&\; \lim_{t\ra-\infty}\;
\bra \psi(t)|[e^{2\al\phi}]_b^{}(\si,-t)|0\ket.
\end{aligned}\end{equation}
By again using that $[e^{2\al\phi}]_b^{}(\si,-t)|0\ket$ has zero
mode dependence $\sim \exp((2\al-Q)q)$ and that the wave-function
representing $\bra \psi(t)|$ will tend to zero for any finite value of $q$
as $t\ra\infty$,
one concludes that the expression vanishes.
But now it is easy to see that \rf{seib1} indeed implies that
\begin{equation}\label{seib2}
\bra \psi^{}_2|[e^{2\al\phi}]_b^{}(\si)|\psi^{}_1\ket\;=\;0 .
\end{equation}
To this aim one only has to invoke state-operator correspondence
as discussed in the previous subsection to represent $|\psi^{}_2\ket$
as the limit for $z\ra 0$ of a state of the form 
\begin{equation}
|\psi^{}_2(z)\ket\;=\;\int\limits_0^{\infty}\frac{dP}{2\pi} 
\;\SV_{\al_P}(\zeta_P|z)|0\ket,
\quad \zeta_P \in\CV_P\ot\CV_P.
\end{equation}
By using mutual locality of $V_{\al_P}(\zeta_P|z)$ and 
$[e^{2\al\phi}]_b^{}(\si)$ one reduces \rf{seib2} to \rf{seib1}.
We conclude that the Seiberg-bound is a simple consequence of
our potential scattering picture of Liouville dynamics.

\subsection{Asymptotics vs. analyticity}

We would finally like to understand the analytic properties of the three point 
functions (meromorphic continuation, location and residues of poles) from
the point of view of the zero mode 
Sch\"odinger representation. 

Let us begin by noting that the smoothing properties of the
operators $\SV_{\al}$, $\Re(\al)>0$, that we noted in Section \ref{reconst} 
are quite easily understood by considering the asymptotic
behavior for $q\raf$. 
Matrix elements of $[e^{2\al\phi}]_b^{}(z,\bz)$, $|z|<1$, 
would in the zero mode
Schr\"odinger representation \rf{Schrrep} be represented as  
\begin{equation}\label{Matr-el}
\bra P_2,\ff_2|[e^{2\al\phi}]_b^{}(z,\bz)|P_1,\ff_1\ket\;=\;
\int\limits_{-\infty}^{\infty}dq\;\bigl( \,\psi^{}_{P_2,\ff_2}(q)\,
,\, [e^{2\al\phi}]_b^{}(z,\bz)\psi^{}_{P_1,\ff_1}(q)\,\bigr)_{\CF\ot\CF}^{},
\end{equation}
As the zero mode dependence of $[e^{2\al\phi}]_b^{}(z,\bz)$
provides an exponential damping factor for $q\raf$, one would expect the 
integration in \rf{Matr-el} to converge and define a function 
that is analytic in 
\begin{equation}
\{ (P_2,P_1)\in\BC^2\; ;\; |\Im(P_2\pm P_1)|<\Re(\al)\},
\end{equation}
which fits our discussion of the DOZZ-proposal in Section \ref{reconst}.

\subsection{Asymptotic expansion}

In order to study the analytic properties
of the matrix elements 
$\bra P_2,\ff_2|[e^{2\al\phi}]_b^{}(z,\bz)|P_1,\ff_1\ket$ 
in the case $\Re(\al)\leq 0$, one will need to 
take into account sub-leading contributions
to the asymptotic behavior of wave-functions for $q\raf$.
On the basis of our discussion in Subsection \ref{ntol}, especially Remark
\ref{as_rem}, one expects that $\SH$ may be approximated by  
$\SH^{\rm F}_{\mu\tilde{\mu}}$ when studying the asymptotic behavior
for $q\raf$.

Standard perturbation 
theory for the Hamiltonian $\SH^{\rm F}_{\mu\tilde{\mu}}$
yields a formal series expansion for eigenfunctions of that operator:
\begin{equation}\label{pertser}
\psi^{}_{P,{\rm f}}(q)\;=  \;\sum_{m,n=0}^{\infty}\;\mu^n\tilde{\mu}^m\;
\psi_{P,{\rm f}}^{(n,m)}(q),
\end{equation}
where the initial term is given by 
$\psi_{P,{\rm f}}^{(0,0)}(q)\;=\; e^{2iPq}{\rm f}+e^{-2iPq}\SR(P){\rm f}$ 
and the higher terms can be expressed in terms of the operators
$\SU(t)\equiv e^{i\SH^{\rm F}t}\SU e^{-i\SH^{\rm F}t}$, 
$\tilde{\SU}(t)\equiv e^{i\SH^{\rm F}t}\tilde{\SU} e^{-i\SH^{\rm F}t}$ as
\begin{equation}\begin{aligned}
\psi_{P,{\rm f}}^{(n,m)}(q)\;= \; \frac{(-i)^n}{n!}\frac{(-i)^m}{m!}
\int\limits_{-\infty}^0
dt_1\dots  & dt_n  d\tilde{t}_1\dots d\tilde{t}_m \ti\\
& \ti T\Bigl(\SU(t_1)\dots \SU(t_n)\tilde{\SU}(\tilde{t}_1)\dots
\tilde{\SU}(\tilde{t}_m)\Bigr)\psi_{P,{\rm f}}^{(0,0)}(q),
\end{aligned} \end{equation}
where $T(\dots)$ denotes the usual time-ordered product of operators.

\subsection{Meromorphic continuation}

To start with, one may consider the case $-b<\Re(\al)<0$. 
The natural ansatz for defining the analytic continuation of 
the representation \rf{Matr-el} for the matrix element
$\bra P_2,\ff_2|[e^{2\al\phi}]_b^{}(z,\bz)|P_1,\ff_1\ket$ would be
\begin{equation}\label{mercont2}\begin{aligned}
\bra P_2,\ff_2|  {[}e^{2\al\phi}{]}_b(z,\bz)|P_1,\ff_1\ket
=  \lim_{q_0\ra -\infty}
\Biggl(  \int_{q_0}^{\infty} & dq\; 
\bra \,\psi^{}_{P_2,\ff_2}(q)\,|\,
[e^{2\al\phi}]_b^{}(z,\bz)\, , \,\psi^{}_{P_1,\ff_1}(q)\,\ket_{\CH(q)}^{}\\
 \qquad +\;\sum_{s_1,s_2=0, 1} & 
\frac{e^{2i((1-2s_1)P_1-(1-2s_2) P_2-i\al)q_0}}
{2i((1-2s_1)P_1-(1-2s_2)P_2-i\al)}
\ti
\\ 
\times 
\bigl(&  \SR^{s_2}(P_2)\ff_2 ,:\!e^{2\al\bar{\phi}}(z,\bz)\!:\!
\SR^{s_1}(P_1)\ff_1\bigr)_{\CF\ot\CF}^{}
\Biggr),
\end{aligned}
\end{equation}
where $\bar{\phi}\equiv \phi-q$.
Poles and corresponding residues of the meromorphic continuation 
of the matrix elements $\bra P_2,\ff_2|
{[}e^{2\al\phi}{]}_b(z,\bz)|P_1,\ff_1\ket$
to $-b<\Re(\al)<0$ are explicitly exhibited in \rf{mercont2}.

In order to continue to values of $\Re(\al)$ smaller than $-b$ 
one needs to take into account sub-leading terms in the asymptotic expansion
of the wave-functions $\psi^{}_{P,{\rm f}}(q)$ as given in \rf{pertser}.
One thereby finds poles for 
\begin{equation}
\al+i(s_1P_1-s_2P_2)=-nb-mb^{-1}, 
\quad s_1,s_2=\pm 1;\quad n,m \in\BZ^{\geq 0},
\end{equation}
in precise correspondence with the poles of the matrix element
$\bra P_2,\ff_2|{[}e^{2\al\phi}{]}_b(z,\bz)|P_1,\ff_1\ket$
as given by the DOZZ-proposal. Moreover,
by some arguments that are familiar from standard derivations of 
quantum field theoretical perturbation theory one finds that the 
corresponding residues are precisely given by the Dotsenko-Fateev
integrals that were discussed in Part \ref{pathint}.

\section{Discussion} \label{disc}

\subsection{The success}

Our attempt to understand Liouville theory in terms of the
zero mode Sch\"odinger representation \rf{Schrrep}
was in some respects amazingly successful:
\begin{enumerate}
\item It gave a natural interpretation of the reflection
property $|P\ket=R(P)|-P\ket$ in terms of reflection 
of wave-packets from a potential-''wall''.
\item The Seiberg-bound followed
from the fact that wave-packets get pushed out to $q\raf$ 
for time $t\ra \pm\infty$: If the exponential $[e^{2\al\phi}]_b$ decays 
too strongly for $q\raf$ it can not have any overlap with the 
asymptotic wave-packets.
\item Considering the asymptotic expansion of eigenstates of the 
Hamiltonian $\SH$ for $q\raf$ allowed one to get a detailed
understanding of the analytic properties of matrix elements
such as $\bra P'|\SV_{\al}(z)|P\ket$ (location of poles,
residues, reflection property), in perfect agreement
with the DOZZ-proposal.  
\end{enumerate} 

\subsection{Problems}

\begin{enumerate}
\item Let us recall that the DOZZ-proposal requires that 
\begin{equation*}
\pi \tilde{\mu}\frac{\Ga(b^{-2})}{\Ga(1-b^{-2})}\;=\;
\biggl(\pi \mu\frac{\Ga(b^{2})}{\Ga(1-b^{2})}\biggr)^{b^{-2}}.
\end{equation*}
For $0<b<1$ one may observe that there exist certain ranges 
for the values of $b$ where not both of $\mu$ and $\tilde{\mu}$
can be positive: $\Ga(1-b^{-2})$ may become negative.
However, in such a case it seems impossible to have positivity
of $\SH$: One could always find regions in $q$-space where 
the interaction term
$\mu\SU+\tilde{\mu}\tilde{\SU}$ gives negative contributions
to $\SH^{\rm F}_{\mu\tilde{\mu}}$, which would contradict the
positivity of $\SH$.
\item The asymptotic expansion \rf{pertser} can be 
rewritten in terms of $(\SH^{\rm F}-E_{P,\ff})^{-1}\SU$,
$(\SH^{\rm F}-E_{P,\ff})^{-1}\tilde{\SU}$. If \rf{pertser} would provide 
a valid representation for $\psi^{}_{P,{\rm f}}(q)$ at finite 
values of $q$, one would not understand how $\psi^{}_{P,{\rm f}}(q)$
could have the nice analytic properties in its $P$-dependence
that are suggested by the correspondence with $|P,\ff\ket$: 
The operators $(\SH^{\rm F}-E_{P,\ff})^{-1}$ would introduce 
an awkward collection of poles for complex $P$.
\item If $\SH=\SH^{\rm F}_{\mu\tilde{\mu}}$ one would of course also
need to have $\SL_n\equiv \SL_{n,\mu,\tilde{\mu}}^{\rm F}$ 
and $\bL_n\equiv \bL_{n,\mu,\tilde{\mu}}^{\rm F}$. In this case nothing
would prevent us to identify
$[e^{2\al\phi}]_b(\si)\equiv
:e^{2\al\phi(\si)}:$. But this would be inconsistent
with the Seiberg bound as discussed in Subsection \ref{seibd}.
\end{enumerate}

\subsection{What to conclude?}

The description of Liouville theory in a representation such as 
\rf{Schrrep} where the zero mode $q$ is diagonal works very well
as long as only the asymptotics $q\raf$ is considered. 
We conclude that one has $\CH\sim\CH^{\rm Schr}$ and
$\SH\sim\SH^{\rm F}_{\mu\tilde{\mu}}$ up to corrections that
vanish faster than any exponential for $q\raf$. This is good enough to
support the picture of Liouville dynamics as describing 
scattering of wave-packets off some perfectly reflecting ``potential''.

All the problems that we have mentioned have to do with the 
representation of Liouville theory ``at finite q''. 
The latter fact may not be too surprising in view of our remarks
in  Subsections \ref{repr?} and \ref{minkphi}. But what precisely
goes wrong? Let us just point out two options that one might
consider:

\begin{enumerate}
\item There exists a self-adjoint zero mode operator $\sq$, but it has a spectral
representation that is less trivial than \rf{Schrrep}:
\begin{equation*}
\CH \;\simeq \;\CH^{\rm Schr}\;=\;\int_{\BR}d\mu(q)\;\CH(q),
\end{equation*}
where $\CH(q)$ must not be a {\it Fock}-representation of the 
canonical commutation relations for $\sa_n$, $\ba_n$. This would of course 
also render the representation of $\SH$ in $\CH^{\rm Schr}$ nontrivial.
However, making contact with the DOZZ-proposal would
require that $\CH(q)$ is at least weakly asymptotic to 
$\CF\ot\CF$ for $q\raf$. Moreover, possible $q$-dependent 
quantum corrections in the definition of $\CH(q)$ and $\SH$ would
have to vanish faster than any exponential.
\item The zero mode operator $\sq$ does not exist, is not densely 
defined, or ceases to be selfadjoint for other reasons. 
In this case one would seem to loose any ground for 
describing Liouville theory in terms of wave-packets localized in target
space. One could, however, look for a self-adjoint operator $\sq'$
that approximates $\sq$ in a suitable limit. There is in fact
a natural and plausible candidate for such an operator $\sq'$:
\begin{equation*}
\sq'=\fr{1}{2b}\log\sll,\qquad\sll\equiv\int_0^{2\pi}d\si \;\SV_b(\si).
\end{equation*}
\end{enumerate}
We consider the resolution of these issues as an important  
problem for future research.
\newpage
\part{Operator-approach} \label{gervnev}\vspace{.5cm}

\section{Classical integrability}

Classically one standard way of exhibiting the complete integrability 
of the Liouville equation is based on the following observation:

\subsection{Linear system}

\noindent {\bf Claim:} {\it The following statements are equivalent:\\
(i) $\vf$ satisfies the classical Liouville equation of motion
\[ \pa_+\pa_-\vf=-2\pi \mu_c e^{\vf} \]
(ii) $e^{-\frac{1}{2}\vf}$ satisfies 
\begin{equation}\label{lin} 
\begin{aligned}\pa_+^2 e^{-\frac{1}{2}\vf(x_+,x_-)}\;=&\;
T_{++}(x_+)
e^{-\frac{1}{2}\vf(x_+,x_-)}\\
\pa_-^2e^{-\frac{1}{2}\vf(x_+,x_-)}\;=&\;
T_{--}(x_-)e^{-\frac{1}{2}\vf(x_+,x_-)},
\end{aligned}\end{equation}
where $T_{++}$ (resp. $T_{--}$) depend on $x_+$ (resp. $x_-$) only.}
\begin{proof}
\underline{(i)$\Rightarrow$(ii):}\\ Any $\vf(x_+,x_-)$ will
satisfy \rf{lin} with 
\[
T_{\pm\pm}(x_+,x_-)\;\equiv\; 
\fr{1}{4}(\pa_{\pm}\vf)^2-\fr{1}{2}\pa_{\pm}^2\vf. 
\]
However, direct calculation shows that 
$\pa_{\mp}T_{\pm\pm}=0$ if the Liouville equation of motion is satisfied.\\
\underline{(ii)$\Rightarrow$(i):}\\
Equation \rf{lin} implies that
\begin{equation}\label{chirdec}
e^{-\frac{1}{2}\vf(x_+,x_-)}=\sum_{i,j=1}^2\; f^+_i(x_+)C^{}_{ij}f^-_j(x_-),
\end{equation}
where $f_i^{\pm}(x_{\pm})$, $i=1,2$ are 
two linearly independent real solutions of 
$\pa_{\pm}^2 f_i^{\pm}=T_{\pm\pm}f_i^{\pm}$, which can and will be 
normalized such that 
\begin{equation}\label{normcond}
f_2^{\pm}\pa_{\pm}f^{\pm}_1-f_1^{\pm}\pa_{\pm}f^{\pm}_2=1.
\end{equation} 
It is no loss of generality to assume that $C_{ij}=C\de_{ij}$ in 
\rf{chirdec} since this may always be achieved by 
forming suitable linear combinations of the $f_i^{\pm}(x_{\pm})$, $i=1,2$.
By direct calculation using \rf{normcond} one may then verify that
\begin{equation}\label{liouf}
\vf(x_+,x_-)\;=\;-2\log\Biggl(\sqrt{2\pi\mu_c} 
\sum_{i=1}^2\; f^+_i(x_+)f^-_i(x_-)\Biggr)
\end{equation}
satisfies the Liouville equation of motion.
\end{proof}
This means that the solution of the nonlinear
Liouville equation of motion can be reduced to the integration
of the linear equation \rf{lin}, where 
$T_{\pm\pm}(x_{\pm})$ can be considered as data,
given in terms of either initial or asymptotic conditions.\footnote{See
\cite{PP} for the solution of the initial value problem.} From these
one may then reconstruct $e^{-\frac{1}{2}\vf(x_+,x_-)}$ by solving
{\it linear} differential equations, and finally $\vf$ itself
by taking the logarithm. Any solution of the Liouville equation
is obtained in this way, and the data $T_{\pm\pm}(x_{\pm})$ 
characterize a solution $\vf(x_+,x_-)$ uniquely up to an additive
constant. 

\subsection{Boundary conditions}
We are interested in the case where $\vf(\si,t)$ is periodic,
$\vf(\si+2\pi,t)=\vf(\si,t)$. It follows that
$T_{\pm\pm}(x_\pm)$ must also be periodic.
However, the solutions $f_i^{\pm}$ will only be {\it quasi}-periodic
\begin{equation}\label{monod}
f^{\pm}(x_{\pm}\pm 2\pi)\;=\; M^{\pm}\cdot f^{\pm}(x_{\pm}),
\end{equation}
where we used matrix notation, $f^{\pm}$ being 
the transpose of the row vector $(f^{\pm}_1,f^{\pm}_2)$. 
The {\it monodromy} matrices $M^{\pm}$ must be elements of $SL(2,\BR)$ 
in order for \rf{monod} to be consistent with \rf{normcond}.
Periodicity of $\vf(\si,t)$ therefore requires that $(M^{+})^t\cdot
M^{-}= 1$.
The Liouville field \rf{liouf} is clearly unchanged under the  
transformations
\begin{equation}\label{f-gauge}
f^{\pm} \ra A^{\pm}f^{\pm},
\qquad (A^{+})^t\cdot A^{-}=1,
\end{equation}
under which $M$ transforms as $M^{\pm}\ra A^{\pm}\cdot M^{\pm}\cdot
(A^{\pm})^{-1}$.
The Liouville field therefore only depends on the conjugacy class 
of the matrix $M$.
It can be shown \cite{PP} that the Liouville field is regular if and only if
${\rm Tr}M>2$, corresponding to the so-called {\it hyperbolic}
conjugacy classes of $SL(2,\BR)$. By means of \rf{f-gauge}
one may then always bring $M$ into the form $M={\rm diag}(e^{\frac{p}{2}},
e^{-\frac{p}{2}})$.
We will henceforth assume $M$ to be of that form.

\subsection{Map to free field}

Another useful representation of the general solution to the Liouville
equation is obtained by introducing
\begin{equation}
A_\pm(x_\pm)\equiv f_2(x_\pm)/f_1(x_\pm).
\end{equation}
Note that the normalization conditions \rf{normcond} imply that $A_\pm$ are
monotonic, i.e. $\pa_\pm A_{\pm}=(f_1^\pm)^{-2}>0$.
It is therefore possible to
recover $f_i^{\pm}$ via
\begin{equation}\label{f_from_A}
f_1^\pm =(\pa_\pm A_\pm)^{-\frac{1}{2}}\qquad
f_2^\pm =(\pa_\pm A_\pm)^{-\frac{1}{2}}A_\pm,
\end{equation}
which leads to the following classical
representation for the Liouville field:
\begin{equation}\label{class_sol}
\vf(x_+,x_-)=\log\left(\sqrt{2\pi\mu_c}
\frac{\pa_+A_+\,\pa_-A_-}{(1+A_+A_-)^2}\right).
\end{equation}

However, it turns out that
the data $A_{\pm}$ are not very convenient as starting point 
for quantization: Their Poisson brackets are complicated \cite{PP}
and it is not easy to realize the condition of monotonicity on
the quantum level. The following variables appear to be better
suited: Define $\vf_\pm^{\rm F}(x_\pm)$ by
\begin{equation}\label{vfdef}
e^{\vf_\pm^{\rm F}(x_\pm)}\;\equiv\; \pa_\pm A_\pm.
\end{equation}
The quasi-periodicity of $A_\pm$ implies that
$\vf^{\rm F}(x_+,x_-) \equiv \vf_+^{\rm F}(x_+)+\vf_-^{\rm F}(x_-)$
is a solution of the free wave equation $\pa_+\pa_-\vf^{\rm F}=0$ on the
{\it cylinder}. It is easy to see that $A_\pm$ (and therefore
the Liouville field itself) can be recovered from
$\vf_\pm^{\rm F}(x_\pm)$ by means of
\begin{equation}\label{Arep}
A_\pm[\vf^{\rm F}](x_\pm)= \frac{1}{e^{\pm p}-1}\int\limits_0^{2\pi}dy_\pm\;
e^{\vf_\pm^{\rm F}(y_\pm+x_{\pm})}.
\end{equation}
These considerations yield a map from the phase space $\CP^{\rm F}$ 
into the phase space $\CP$ of Liouville theory. 

There is a problem, though: The map $\CP^{\rm F}\ra \CP$
is not one-to-one but two-to-one. 
For any given solution $\vf_1^{\rm F}(x_+,x_-)$
of the free field wave equation there exists a second one
$\vf_2^{\rm F}(x_+,x_-)$ that maps to the same solution of the 
Liouville equation. This is most easily verified by noting that \rf{class_sol}
is unchanged if one replaces $A_{\pm}\ra 1/A_{\pm}$, which corresponds
to the exchange of $f_1^{\pm}$ and $f_2^{\pm}$. The second 
free field configuration $\vf_2^{\rm F}(x_+,x_-)$ is therefore 
given in terms of $\vf_1^{\rm F}(x_+,x_-)\equiv 
\vf_{1,+}^{\rm F}(x_+)+\vf_{1,-}^{\rm F}(x_-)$ via
\begin{equation}\label{S-classdef}
\vf_2^{\rm F}(x_+,x_-)\;=\;\log\biggl(\pa_+\frac{1}{A_+[\vf_{1}]}
\pa_-\frac{1}{A_-[\vf_{1}]}\biggr).
\end{equation}
The resulting map $\CP^{\rm F}\ra\CP^{\rm F}$ will be denoted by $S$.

In order to get a unique parametrization of the Liouville phase space
in terms of free field variables one therefore has two obvious options:
Either one may note that $S$ maps the zero mode $p$ which is  
recovered from $\vf^{\rm F}$ as
\[
p=\frac{1}{2\pi}\int_0^{2\pi}d\si\;\pa_{t}\vf^{\rm F} 
\]
into its negative. The map $\CP^{\rm F}\ra\CP$ is therefore invertible 
when restricted 
e.g. to the subspaces $\CP^{\rm F}_\pm$ defined by the condition $\pm p>0$.
The corresponding maps $\CP\ra\CP^{\rm F}_\pm$ will be denoted 
$W_\pm$.
But it may sometimes be more convenient to think of $\CP^{\rm F}_+$
as $\CP^{\rm F}\!/\!\!\sim $, where two configurations $\vf_i^{\rm F}(x_+,x_-)$
$i=1,2$ are considered as equivalent (notation: $\vf_1^{\rm F}\sim
\vf_2^{\rm F}$) iff $S[\vf_1]=\vf_2$.

The importance of this ``gauge symmetry'' represented by the map $S$ was 
emphasized by Gervais and Neveu. We will identify its proper quantum 
counterpart below.

\subsection{Canonical formalism}

We had introduced the canonical formalism for Liouville theory in section
\ref{can_form}. Its counterpart for the free field $\vf^{\rm F}$ 
is obviously obtained by replacing $\vf \ra\vf^{\rm F}$  and setting
$\mu_c=0$. 
A basic result that represents important motivation for the
program of constructing quantum Liouville theory
in terms of the quantized free field is the following:

{\it The maps $W_\pm:\CP\ra\CP^{\rm F}_\pm$ are canonical. More precisely: The
canonical Poisson bracket relations 
\begin{equation}\label{PBR}
\{ \Pi_{\vf}(\si),\vf(\si')\}\;=\;\de(\si-\si')
\end{equation}
imply the same commutation relations for 
the images $\vf^{\rm F}$, $\Pi^{\rm F}_{\vf}$ of $\vf$, $\Pi_{\vf}$ under 
$W_\pm$. Conversely:
Canonical Poisson bracket relations 
\begin{equation}\label{PBRF}
\{ \Pi^{\rm F}_{\vf}(\si),\vf^{\rm F}(\si')\}\;=\;\de(\si-\si')
\end{equation}
imply \rf{PBR}. }

The inverse direction \rf{PBRF} $\Rightarrow$ \rf{PBR} can be 
shown by direct, 
but tedious calculation (see e.g. \cite{KN} for details). To go 
from \rf{PBR} $\Rightarrow$ to \rf{PBRF} is more
difficult, see \cite{PP}. It also follows that the map 
$S:\CP^{\rm F}\ra\CP^{\rm F}$ will be canonical.

\subsection{Conformal symmetry}

It is quite important to note that the integrable 
structure of Liouville theory (as represented by the maps 
$W_\pm:\CP\ra\CP^{\rm F}_\pm$) is compatible with conformal symmetry:
Let us first remark that
canonical generators of conformal transformations in the free field theory 
are easily identified as 
\begin{equation}
T_{\pm\pm}^{\rm F}=
\fr{1}{4}(\pa_\pm \vf^{\rm F})^2-\fr{1}{2}\pa_{\pm}^2\vf^{\rm F}.
\end{equation}
The modes $l_n^{\pm}=\frac{1}{2\pi}\int_{0}^{2\pi}d\si e^{in\si}
T_{\pm\pm}^{\rm F}(\si)$ will then satisfy a Poisson-counterpart of the
Virasoro algebra:
\begin{equation}
\{l_n^{\pm},l_m^{\pm}\}\;=\;i(n-m)l_{n+m}^{\pm}+\fr{i}{2}n^3\de_{n+m}. 
\end{equation}

Compatibility of conformal symmetry with
the integrable 
structure of Liouville theory is a
consequence of the fact that the maps $W_\pm:\CP\ra\CP^{\rm F}_{\pm}$ 
indeed transform $T_{\pm\pm}[\vf]$ into $T_{\pm\pm}^{\rm F}[\vf^{\rm F}]$
\cite{PP}. It follows in particular that the maps $W_\pm$
intertwine the actions of conformal symmetry generated 
by $T_{\pm\pm}^{\rm F}$ and $T_{\pm\pm}$ respectively. A check
of this statement may be based on the observation that
$f_i^{\pm}$ $i=1,2$ transform under the conformal transformations 
generated by $T_{\pm\pm}^{\rm F}$ as tensors of weight $\frac{1}{2}$:
\begin{equation}
\{T_{\pm\pm}^{\rm F},f_i^\pm(x_\pm)\}\;=\;e^{inx_\pm}
(\pa_\pm -in\fr{1}{2})f_i^\pm(x_\pm).
\end{equation}
Closely related is the fact that
the map $S:\CP^{\rm F}\ra\CP^{\rm F}$ commutes with the action
of conformal symmetry as generated by $T_{\pm\pm}^{\rm F}$. Very similar
considerations for the case of Liouville theory on the strip have
first appeared in \cite{GN0}.

\section{Quantization of the free field theory}

Given the possibility to map classical Liouville theory to a free 
field theory it is of course natural to approach quantization
of Liouville theory by first quantizing the free field theory
and then trying to reconstruct the Liouville field operators
in terms of operators in the free field theory.

So let us introduce the free field $\phi^{\rm F}(\si,t)$ 
with canonical commutation relations
\begin{equation}
{[}\phi^{\rm F}(\si),\pa_{t}\phi^{\rm F}(\si){]}=2\pi i\de(\si-\si')
\end{equation} 
at time $t=0$. Modes are introduced via 
\begin{equation}
\phi^{\rm F}(\si,t)\;=\;\sq^{\rm F}+2\spp t+\sum_{n\neq 0}
\frac{1}{n}\bigl(\sa_n^{\rm F}e^{-inx_+}+\ba_n^{\rm F}e^{-inx_-}\bigr),
\end{equation}
and the Hilbert-space will be defined as
\begin{equation}
\CH^{\rm F}\;\equiv \;L^2(\BR)\ot\CF\ot\CF,
\end{equation}
where $\CF\ot\CF$
is the Fock-space generated by action of the modes $\sa_n^{\rm F}$, 
$\ba_n^{\rm F}$, 
$n<0$ on the Fock-vacuum $\Om$
that satisfies $\sa_n^{\rm F}\Om=0$ and $\ba_n^{\rm F}\bar{\Om}=0$,
$n>0$. We will work in a representation where $P$ is diagonal.

\subsection{Conformal symmetry}

Conformal symmetry is realized on $\CH^{\rm F}$ by means of
\begin{equation}\label{freefield}
\ST_{++}(x_+)\;=\;\sum_{n\in\BZ}\SL_n^{\rm F}e^{-inx_+}\qquad
\ST_{--}(x_-)\;=\;\sum_{n\in\BZ}\bL_n^{\rm F}e^{-inx_-},
\end{equation}
where the expressions for $\SL_n^{\rm F}$, $\bL_n^{\rm F}$ are 
obtained from the formulae for $\SL_n^{\rm F}(P)$, $\bL_n^{\rm F}(P)$
(cf. eqn. (122)) by replacing $\sa_n\ra\sa_n^{\rm F}$, 
$\sbb_n\ra\sbb_n^{\rm F}$ and $P\ra \spp$.

This means that $\CH^{\rm F}$ decomposes as the direct integral of
Fock-space representations of the Virasoro algebra:
\begin{equation}\label{decFP}
\CH^{\rm F}\;\simeq\; \int\limits_{-\infty}^{\infty}dP
\;\CF_P\ot\bar{\CF}_P,
\end{equation}
where $\CF_P$ (resp. $\bar{\CF}_P$) denote the Virasoro representations
defined on $\CF$ by means of the generators $\SL_n^{\rm F}$ 
(resp. $\bL_n^{\rm F}$) defined in \rf{freefield}.

\subsection{Building blocks}

The basic building blocks of all constructions will be the following 
operators:

\noindent {\sc Normal ordered exponentials: }
\begin{equation}\begin{aligned}
\SE^{\al}(x_+)\;\equiv\;&
e^{\al \sq^{\rm F}}
\exp\Biggl(\sum_{n< 0}
\frac{2\al}{n}\sa_n^{\rm F}e^{-inx_+}\Biggr)e^{2\al x_+ \spp} 
\exp\Biggl(\sum_{n> 0}
\frac{2\al}{n}\sa_n^{\rm F}e^{-inx_+}\Biggr)e^{\al \sq^{\rm F}}\\
\bar{\SE}^{\al}(x_-)\;\equiv\;&
e^{\al \sq^{\rm F}}
\exp\Biggl(\sum_{n< 0}
\frac{2\al}{n}\ba_n^{\rm F}e^{-inx_-}\Biggr)e^{2\al x_- \spp} 
\exp\Biggl(\sum_{n> 0}
\frac{2\al}{n}\ba_n^{\rm F}e^{-inx_-}\Biggr)e^{\al \sq^{\rm F}}.
\end{aligned} \end{equation}
{\sc Screening charges: }
\begin{equation}
\SQ(x_+)\;\equiv\;\int\limits_0^{2\pi}dy_+\;\SE^b(x_++y_+),\qquad
\bar{\SQ}(x_-)\;\equiv\;\int\limits_0^{2\pi}dy_+\;\bar{\SE}^b(x_-+y_-).
\end{equation}
The normal ordered exponentials can be understood as 
true (unbounded) operators if $\Im(x_\pm)>0$ (negative euclidean time),  
as operator valued distributions for 
real values of $x_\pm$. This being understood we will often call the 
normal ordered exponentials and functions thereof ``operators'' in the
following. 

The screening charges on the 
contrary represent densely defined unbounded operators even for 
$x_\pm\in\BR$. They can be seen to have a canonical self-adjoint extension
due to their property of positivity. This allows to take 
arbitrary powers of these operators.

The behavior of these operators under conformal transformations can be 
summarized by 
\begin{equation}\begin{aligned}\label{E_cov}
\bigl[\SL_n,\SE^{\al}(x_+)\bigr]= & e^{inx_+}
\bigl(-i\pa_++n\De_\al)\bigr)\;\SE^{\al}(x_+),\\
\bigl[\bL_n,\bar{\SE}^{\al}(x_-)\bigr]=& e^{inx_-}
\bigl(-i\pa_-+n\De_\al)\bigr)\;\bar{\SE}^{\al}(x_-),
\end{aligned}\end{equation}
whereas the screening charges transform as 
\begin{equation}\label{Qcov}
\bigl[\SL_n,\SQ(x_+)\bigr]=-i e^{inx_+}
\pa_+\SQ(x_+),\qquad
\bigl[\bL_n,\bar{\SQ}(x_-)\bigr]=-ie^{inx_-}
\pa_-\bar{\SQ}(x_-),
\end{equation}

\subsection{Quantum counterparts of $f_i^{\pm}$}

Let us define the following set of operators:
\begin{equation}\begin{aligned}
\sf_1(x_+)=& \;\SE^{-\frac{b}{2}}(x_+),\\
\bar{\sf}_1(x_-)=& \;\bar{\SE}^{-\frac{b}{2}}(x_-),
\end{aligned}\qquad
\begin{aligned}
\sf_2(x_+)=& \;\frac{e^{-2\pi b(\spp+i\frac{b}{2})}}{\sin(\pi b(Q-2i\spp))}
\SQ(x_+)\sf_1(x_+),\\ 
\bar{\sf}_2(x_-)=& \;\bar{\sf}_1(x_-)\bar{\SQ}(x_-)
\frac{e^{-2\pi b(\spp-i\frac{b}{2})}}{2\sin(\pi b(Q+2i\spp))}. 
\end{aligned}
\end{equation}
$\sf_i$, $\bar{\sf}_i$ 
should be considered as the quantum counterparts of $f_i^+$,
$f_i^-$ respectively. This correspondence is quite obvious when comparing
the expressions (176) to the expressions for $f_i^+$,
$f_i^-$ that follow from eqns. (157), (159) and (160)
(The shifts of $\spp$ in the $\spp$-dependent pre-factors could
be absorbed by choosing a different ordering). 

It is encouraging to note that the operators $\sf_i$ $i=1,2$ satisfy 
a second order differential equation of a very similar form as
their classical counterparts: 
\begin{equation}\label{qlin}
 \pa_+^2  \;\sf_i \;= \;-b^2:\ST\;\sf_i:,\quad i=1,2,
\end{equation}
where the normal ordering of the expression on the right hand side 
is defined as follows:
\begin{equation}\label{normord}\begin{aligned}
{} :\ST\;\sf_i:\;\,& = \;\sum_{n<0}\SL_n e^{-in x_+}\;
\sf_i+\sum_{n>0}
\;\sf_1\;e^{-in x_+}\SL_n \\
& \qquad\qquad\qquad\qquad +\frac{1}{2}\Bigl(\SL_0\sf_i^{+} 
+\sf_i^{+}\SL_0\Bigr)
-\Bigl(\frac{c-1}{24}-\frac{b^2}{16}\Bigr)\sf_i.
\end{aligned}\end{equation}
The operators $\bar{\sf}_i$ $i=1,2$ satisfy a second order differential
equation that is obtained by obvious replacements.

\subsection{Quantum counterpart of $e^{-\frac{1}{2}\vf}$}

Let us consider the following operator: 
\begin{equation}\label{Vansatz}
\SV(\si,t)\;=\;\sf_1(x_+)\,e^{b\sq}\,
\bar{\sf}_1(x_-)+\mu_{\rm e}\,\sf_2(x_+)
\,e^{-b\sq}\,\bar{\sf}_2(x_-).
\end{equation}
It can be shown to satisfy the following properties:
\begin{itemize}
\item[a)] {\sc Conformal covariance}
\begin{equation*}\begin{aligned}
\bigl[\SL_n,\SV(\si,t)\bigr]=&e^{+inx_+}
\bigl(-i\pa_{+}+n\De_b)\bigr)\;\SV(\si,t),\\ 
\bigl[\bL_n,\SV(\si,t)\bigr]=&e^{-inx_-}
\bigl(-i\pa_{-}+n\De_b)\bigr)\;\SV(\si,t)
\end{aligned}\end{equation*}
\item[b)] {\sc Equation of motion}
\begin{equation*}\begin{aligned}
 \pa_+^2\;\SV\;= & \;-b^2:\ST(x_+)\;
\SV:,\\
 \pa_-^2\;\SV\;= & \;-b^2:\bar{\ST}(x_-)\;
\SV:,
\end{aligned}\end{equation*}
\item[c)] {\sc Locality}
\begin{equation*}
\bigl\bra\psi_2|\,\bigl[ \;\SV(f)\;,\;
\SV(f')\;\bigr]\,|\psi_1\bigr\ket\;=\;0,
\end{equation*}
\item[d)]{\sc Positivity}
\begin{equation*}
\bra\psi|\,\SV(f)\,|\psi\ket \;>\; 0,
\end{equation*}
\end{itemize}
where $|\psi\ket$, $|\psi_1\ket$, $|\psi_2\ket$
are taken from a dense subset of $\CH^{\rm F}$
and the {\it smeared} operator 
$\SV(f)$ is defined as
\begin{equation}\label{sm}
\SV(f)
\;\equiv\; \int\limits_{S^1}d\si \;\,f(\si)\,
\SV(\si).
\end{equation}
Properties a) and b) are obvious. Locality is proved by 
using the exchange relations of the operators $\sf_i$, $\bar{\sf}_i$
\cite{GN2}, and property d)
is easy to prove by means of the reflection operator 
introduced in the next subsection.

\subsection{Reflection operator}

We had found in our discussion of the classical integrability of the 
Liouville equation that the free field with unrestricted zero mode $p$
parameterizes the Liouville field in an ambiguous way: There exists
a transformation $S$ (defined in \rf{S-classdef})
of free field configurations $\vf^{\rm F}(t,\si)$ that leaves
the Liouville field unchanged. When using the free field 
phase space $\CP^{\rm F}$ as a parameterization of the
space of solutions to the Liouville equation 
one should therefore identify any two configurations 
of the free field that are related by $S$: $\vf_1^{\rm F}\sim
\vf_2^{\rm F}$ iff $S[\vf_1^{\rm F}]=\vf_2^{\rm F}$.

In the quantization scheme that we have discussed so far, we have
started from a realization of the 
zero mode $\spp$ with spectrum being the entire real line. 
The quantum analogue of $S$ 
should be an operator $\SS$ on $\CH^{\rm F}$ that maps $\spp$ to $-\spp$
but leaves $[e^{-b\phi}]_b$ unchanged,
\begin{equation}\label{refl-inv}
\SS^{-1}\cdot[e^{-b\phi}]_b\cdot \SS= [e^{-b\phi}]_b.
\end{equation}
This operator can be considered
as expressing the ambiguity in the parameterization of Liouville states
by free field variables on the quantum level. The true 
Liouville Hilbert $\CH$ space should then be identified with the 
subspace in $\CH^{\rm F}$ of $\SS$-invariant vectors. For the 
identification of $\CH$ with a subspace in $\CH^{\rm F}$ to be compatible
with conformal symmetry one evidently needs that $\SS$ commutes with
the Virasoro generators, 
\begin{equation}\label{R-intertw}
\SS \cdot\SL_n^{\rm F}(\spp)\;=\;\SL_n^{\rm F}(-\spp)\cdot\SS,\qquad
\SS \cdot\bL_n^{\rm F}(\spp)\;=\;\bL_n^{\rm F}(-\spp)\cdot\SS.
\end{equation}
Such an operator can be constructed as follows: Since the 
Fock-space representations $\CF_P$ and $\CF_{-P}$ are unitarily equivalent
(cf. Subsection \ref{refl_op}) 
one has a unique operator $\SS(P):\CF\ot\CF
\ra\CF\ot\CF$ that satisfies \rf{R-intertw}
and $\SS(P)\Om=\Om$. The sought-for operator $\SS$ must therefore be
of the form 
\begin{equation}\label{SS_def}
\SS\;=\;
\SP \cdot R(\spp)\SS(\spp),
\end{equation}
where $\SP$ denotes the parity operation in $L^2(\BR)\ot\CF\ot\CF$,
$\SP\psi(P)=\psi(-P)$.
One may then show that there exists a unique choice of the function
$R(P)$ that was introduced in \rf{SS_def} such that \rf{refl-inv} 
is satisfied as well. This choice is given as 
\begin{equation}\label{R-explicit}
R(P)=-\bigl(\mu_{\rm e}\Ga^2(b^2)\bigr)
^{-\frac{2iP}{b}}\frac{\Ga(1+2ibP)\Ga(1+2ib^{-1}P)}
{\Ga(1-2ibP)\Ga(1-2ib^{-1}P)},
\end{equation}
which coincides with the reflection amplitude $R(P)$
obtained from the DOZZ-proposal if we identify the constant 
$\mu_{\rm e}$ introduced in the definition \rf{Vansatz} of $\SV$ 
with the constant $\mu$ that appears in the DOZZ-formula via
$\mu_{\rm e}=\mu \sin(\pi b^2)$.

By means of $\SS$ we may now identify the Liouville Hilbert space $\CH$
as 
\begin{equation}
\CH\;=\;\{|\psi\ket\in\CH^{\rm F}; (1-\SS)\Psi=0\}.
\end{equation}

\section{General exponential operators}\label{genexps}

Let us start by reconsidering the classical expression for $e^{\la\vf}$ which
may be written in terms of $f^{\pm}_s$ as (cf. \rf{liouf})
\begin{equation}\label{exp-class}
e^{\la\vf(\si,t)}\;=\; 
 \bigr(f_1^+f_1^-+ 2\pi\mu_c f_2^+f_2^-\bigr)^{-2\la}.
\end{equation}
Let us note that at least for $\la>0$ one has a useful 
representation of \rf{exp-class} as sum of 
{\it imaginary} powers of $f_i^+f_i^{-}$:
\begin{equation}\label{binom}
e^{\la\vf(\si,t)}\; =  \; \frac{i}{2\pi}\int\limits_{i\BR}ds
\;(2\pi\mu_c)^{s}\;
\frac{\Ga(s+2\la)\Ga(-s)}{\Ga(2\la)}\,\bigl(f_2^+f_2^-\bigr)^{s}
\,\bigl(f_1^+f_1^-\bigr)^{-2\la-s}.
\end{equation}
This expansion is related to the binomial expansion by writing the 
integral as sum over residues. However, it has the advantage to be 
valid both for $f_1^+f_1^->f_2^+f_2^-$ and $f_1^+f_1^-<f_2^+f_2^-$,
which is important for quantization: 

The essential point is captured
by the following example: Consider the operator $\sv_\la$ 
of multiplication by the function $h_{\la}(q)\equiv
(e^{\frac{q}{2}}+e^{-\frac{q}{2}})^{-2\la}$ on $L^2(\BR)$ in the case that
$\Re(\la)>0$. $\sv_\la$ is a nice, bounded operator on $L^2(\BR)$ since
$h_{\la}(q)$ behaves asymptotically as $h_{\la}(q)\sim e^{-\la |q|}$.
Now let us try to represent this operator by using the usual binomial
expansion:
\begin{equation}\label{g-exp1}
\sv_{\la}\;=\;e^{\la \sq}\sum_{n=0}^{\infty}
\frac{(-)^n\Ga(n+2\la)}{\Ga(2\la)\Ga(n+1)}e^{n\sq}
\end{equation}
which clearly can represent the operator on wave-functions with 
support on the negative half-axis only. The domain in
which \rf{g-exp1} serves to represent the operator $\sv_\la$ is 
not even dense in $L^2(\BR)$! This problem does not occur if one uses 
the integral version \rf{binom} of the binomial expansion. 
By means of a shift of the contour of integration one may 
even write it as an expansion over the unitary operators $e^{itb\sq}$.

\subsection{Definition}

We would like to define $[e^{2\al\phi}]_b^{}(\si)$ as something like 
$[e^{-b\phi}]_b^{}\equiv \SV$ raised to the power $-2\al$. This is of 
course a nontrivial thing to do in the case of local operators.
One may, however, recall that the usual exponential function 
is the unique solution to the functional equation 
$\exp(x)\exp(y)=\exp(x+y)$ within the class of continuous functions.
This motivates the following definition:
\begin{defn}\label{Val_def}
Let $\SV_{\al}(\si)$ be a family of operators that has the properties
\begin{itemize}
\item[i)] There exists a complex number $\rho$ such that 
\begin{equation*}
\bigl(\SV_{\al}\star\SV_{\be}\bigr)(\si)\;\equiv\;
\lim_{\si'\ra\si}\bigl|\si'-\si\bigr|^{\rho}\SV_{\al}(\si')\SV_{\be}(\si)
\end{equation*}
exists and satisfies
\begin{equation*} 
\bigl(\SV_{\al}\star\SV_{\be}\bigr)(\si)
\;=\;\SV_{\al+\be}(\si).
\end{equation*}
\item[ii)] Each two operators $\SV_{\al}(\si)$ and $\SV_{\be}(\si)$ are
mutually local:
\begin{equation*}
\bigl[ \;\SV_{\al}(\si') \;,\; \SV_{\be}(\si)\;\bigr]=0.
\end{equation*}
\item[iii)] $\SV_{\al}(\si)$ reduces to $\SV(\si)$ for $\al=-\frac{b}{2}$.
\end{itemize} 
We then call $\SV_{\al}(\si)$ a renormalized power function of $\SV$, 
in symbols $\SV_{\al}(\si)\equiv \SV^{-2\al}(\si)$.
\end{defn}

It turns out that there there exists a family of operators 
$\SV_{\al}(\si)$ that fulfills these properties. This family 
$\SV_{\al}(\si)$ can be shown to be unique at least for 
irrational $b\in(0,1)$. It will be constructed
by an expression similar to \rf{binom} out of building blocks 
that can be considered as quantum analogs of $
(f_2^+)^{is}(f_1^+)^{-2\la-is}$ and $
(f_2^-)^{is}(f_1^-)^{-2\la-is}$, which will be introduced in the 
next subsection.

\subsection{Covariant chiral operators} \label{chirvert}

We will define operators $\sf_s^{\al}(x_+)$, $\bar{\sf}_s^{\al}(x_-)$
in the spirit of Definition \ref{Val_def} such that 
\newcommand{\bsf}{\bar{\sf}}
\begin{equation}\label{sf_def}
\begin{aligned}
\bigl(\sf_s^{\al}\star \sf_1\bigr)(x_+)=&\;\sf_s^{\al-\frac{b}{2}}(x_+),\\
\bigl(\sf_2\star\sf_s^{\al}\bigr)(x_+)=& \;\sf_{s+1}^{\al-\frac{b}{2}}(x_+),
\end{aligned} 
\qquad
\begin{aligned}
\bigl(\bsf_1\star\bsf_s^{\al}\bigr)(x_-)=& \;\bsf_s^{\al-\frac{b}{2}}(x_-),\\
\bigl(\bsf_s^{\al}\star \bsf_2\bigr)(x_-)=&\;\bsf_{s+1}^{\al-\frac{b}{2}}(x_-).
\end{aligned} 
\end{equation}
The operators $\sf_s^{\al}(x_+)$, $\bar{\sf}_s^{\al}(x_-)$ can be 
represented explicitly as follows:
\begin{equation}\label{sfdefs}\begin{aligned}
\sf_s^{\al}(x_+)\;\equiv & \;e^{-2\pi bs(\spp+i\frac{b}{2})}
\frac{S_b(Q-2i\spp)}{S_b(Q-2i\spp+bs)}\bigl(\SQ(x_+)\bigr)^s\SE^{\al}(x_+),\\
\bsf_s^{\al}(x_-)\;\equiv & \;
\bar{\SE}^{\al}(x_-)\bigl(\bar{\SQ}(x_-)\bigr)^s
\frac{S_b(Q+2i\spp)}{S_b(Q+2i\spp+bs)}e^{-2\pi bs(\spp-i\frac{b}{2})},
\end{aligned} 
\end{equation}
where the special function $S_b(x)$ is defined as 
\begin{equation}\label{sbdef}
S_b(x)=\Ga_b(x)/\Ga_b(Q-x).
\end{equation}
Let us observe that the 
definition \rf{sfdefs} makes sense for complex values of $s$
due to the positivity of $\SQ$.
The operators $\sf_s^{\al}(x_+)$, $\bar{\sf}_s^{\al}(x_-)$
transform under conformal symmetry the same way as the operators 
$\SE^{\al}(x_+)$, $\bar{\SE}^{\al}(x_-)$, cf. \rf{E_cov}.

It is technically often more convenient to work with the 
operators $\sg_s^{\al}(x_+)=\SE^{\al}(x_+)\bigl(\SQ(x_+)\bigr)^s$, 
$\bar{\sg}_s^{\al}(x_-)=\bar{\SE}^{\al}(x_-)
\bigl(\bar{\SQ}(x_-)\bigr)^s$, which are related to 
$\sf_s^{\al}(x_+)$, $\bar{\sf}_s^{\al}(x_-)$ by multiplication with a 
$\spp$-dependent factor. The following two results are the main technical
ingredients to our proof of the DOZZ-proposal, the details of which will
appear in \cite{TO}.

\noindent {\sc Normalization} $\frac{\quad}{}$
The 
matrix elements of operators 
$\sg_s^{\al}$ between primary states for the Virasoro algebra 
are given as 
\begin{equation}
\bra P'|\sg_s^{\al}(x_+)|P\ket\;=\;\de\bigl(P_s^{\al}-P'\bigr)
\,e^{ix_+(\De(P')-\De(P))}
\,G_{s}^{\al}(P),
\end{equation}
where $P_s^{\al}\equiv P-i\al-ibs$ and 
\begin{equation}\label{G-norm}\begin{aligned}
G_{s}^{\al}(P)\;=\;& 
\Bigl(\Ga(1+b^2)b^{-1-b^2}\Bigr)^s e^{2\pi bsP} e^{-\pi i b^2s^2}\\
& \ti 
\frac{\Ga_b(Q-2iP-2\al-sb)\Ga_b(Q+2iP+sb)}
{\Ga_b(Q-2iP-2\al-2sb)\Ga_b(Q+2iP)}
\frac{\Ga_b(Q-2\al-sb)\Ga_b(Q+sb)}{\Ga_b(Q-2\al)\Ga_b(Q)}.
\end{aligned}\end{equation}
The corresponding formula for $\bar{\sg}_s^{\al}(x_-)$ is obtained
by simply replacing $\sg_s^{\al}\ra \bar{\sg}_s^{\al}$ and $x_+\ra x_-$.

\noindent {\sc Braid relations} $\frac{\quad}{}$
The operators 
$\sg_s^{\al}$ satisfy braid relations of the form
\begin{equation}\label{braidrelg}
\sg_{s_2}^{\al_2}(\si_2)\sg_{s_1}^{\al_1}(\si_1)
\;=\;\frac{1}{4}\int\limits_{\BT}dt_2dt_1\;\,
\sg_{t_1}^{\al_1}(\si_1)\sg_{t_2}^{\al_2}(\si_2)\;
B_{\ep}(\al_2,\al_1|\spp)^{s_2s_1}_{t_2t_1},
\end{equation}
where $\BT\equiv -\frac{Q}{2}+i\BR$ and $\ep\equiv \sgn(\si_2-\si_1)$. 
The distributional kernel $
B_{\ep}(\al_2,\al_1|\spp)^{s_2s_1}_{t_2t_1}$ 
has support for $t_2+t_1=s_2+s_1$
only. 

\subsection{Powers of $\SV$}

Let us now consider the operator
\begin{equation}\label{Valdef}
\SV_{\al}(t,\si)\;\equiv \;
\int\limits_{i\BR}ds \;\mu_{\rm e}^s \;B_b(\al,s)
\;\sf_s^{\al}(x_+) \,\ST^{\al}_s \;
\bar{\sf}_s^{\al}(x_-),
\end{equation}
where $\ST^{\al}_s\equiv e^{-2(\al+bs)\sq}$ and 
the $b$-binomial coefficients $B_b(\al,s)$ are given by
\begin{equation}
B_b(\al,s)\;=\;
\frac{S_b(-bs)S_b(2\al+bs)}{S_b(2\al)}.
\end{equation}
$\SV_{\al}(t,\si)$ is clearly a primary field for the left- and right Virasoro 
algebras generated by $\SL_n$, $\bL_n$.
The main result of \cite{TO} will be that $\SV_{\al}(\si)\equiv 
\SV_{\al}(t=0,\si)$ is indeed a renormalized power of $\SV$ in the
sense of Definition \ref{Val_def}. The least trivial part of this statement
is of course the verification of mutual locality, which becomes
possible thanks to the existence of the braid relations \rf{braidrelg}.
We will outline in the following Part \ref{boots} how this 
verification can be done 
using the explicit calculation of the kernel  $
B_{\ep}(\al_2,\al_1|\spp)^{s_2s_1}_{t_2t_1}$ in 
Section \ref{braid_cco} below.

The matrix elements of $\SV_{\al}$ can be assembled from \rf{G-norm}, 
\rf{sf_def} and \rf{Valdef}. By identifying $\mu_{\rm e}=\mu\sin(\pi b^2)$
one finds that
\begin{equation}
\bra P_2|\SV_{\al}(0)|P_1\ket\;=\;
C\bigl(\fr{Q}{2}-iP_2,\al,\fr{Q}{2}+iP_1\bigr),
\end{equation}
where $C(\al_3,\al_2,\al_1)$ is the DOZZ three point function. 
It follows that the operators $\SV_{\al}$ are indeed identical to
those discussed in Part \ref{pathint}.

Let us also note 
that $\SS\cdot\SV_{\al}(x)\cdot\SS^{-1}=\SV_{\al}(x)$
as a consequence of the reflection property of the DOZZ three point function.
This means in particular that $\SV_{\al}(x)$ leaves the Liouville 
Hilbert-space $\CH$ invariant.

\subsection{Related work}

The operator approach to Liouville theory based on the quantization
of the classical map to a free field has a long tradition going back to
\cite{BCT1}\cite{BCT2},\cite{GN1}\cite{GN2} and \cite{OW}. We should therefore
discuss the relations between the treatment given here to the
results of these works.

The approach taken in this paper 
is close in spirit to the approach of Gervais and
Neveu, which was originally developed for Liouville theory on a 
strip instead of the cylinder. Subsequent work aimed at the 
development of this approach for Liouville theory on the cylinder
includes \cite{LS}\cite{GS3} and references therein. 
These works are concerned with the construction of field operators
in a space of states called the ``elliptic sector'' which in our notation
would be generated by states $|P\ket$ with purely {\it imaginary} 
values of $P$. One needs to note, however, that the elliptic sector
unavoidably contains non-unitary representations of the Virasoro
algebra \cite{LS} and that exponential field operators must have 
unusual hermiticity properties if such a spectrum is assumed 
\cite{LS}\cite{BP}\cite{GS3}. Unfortunately we could not find 
a treatment of the so-called hyperbolic sector (generated by
$|P\ket$ with $P\in\BR$) in the spirit of Gervais and
Neveu. Moreover, we do not know how to employ the 
results of \cite{GS3} for the construction of general exponential operators
in the hyperbolic sector that is studied in this paper.
 
The approaches of \cite{BCT1}\cite{BCT2}
and \cite{OW} on the other hand
were in fact devoted to the hyperbolic sector. 
However, it was not obvious to us how to
define and calculate the
matrix elements of general exponential operators from 
the results of these references. The points where it is possible to
compare the results of \cite{BCT1}\cite{BCT2}
and \cite{OW} to the consequences of the DOZZ-proposal
as discussed in Part \ref{pathint} are therefore somewhat 
limited.\footnote{These include comparison of certain structure 
functions related 
to the cases for the three point functions that are calculable
in terms of Dotsenko-Fateev integrals 
(cf. part \ref{pathint}). A detailed comparison
of such data as obtained from \cite{BCT1}\cite{BCT2}, \cite{GN1}\cite{GN2}
and \cite{OW} was carried out in \cite{GS2},
where mutual consistency of these
results was found.} 
We would like to mention, however, that 
the necessity to restrict to ``half'' of the free field space of states
in the scheme of \cite{BCT1}\cite{BCT2} was also observed in \cite{BCGT},
but the relation with the reflection operator $\SR(P)$ was not discussed 
there.

\section{Appendix: Braiding of covariant chiral operators} \label{braid_cco}

We would like to outline the derivation of the key 
fact that the operators $\sg_s^{\al}(x_+)$,
$\bar{\sg}_s^{\al}(x_-)$ satisfy braid relations of the form \rf{braidrelg}.
More details will be given in \cite{TO}. 
In the following we will 
mainly consider the case $\al_i\in\frac{Q}{2}+i\BR$, $s_i\in-\frac{Q}{2}+i\BR$,
$t_i\in-\frac{Q}{2}+i\BR$, $i=1,2$.
The braid relations in the general case can be obtained by analytic
continuation.  

\subsection{Reformulation in terms of Weyl-type operators}
 
The starting point of our calculation will be the following nice trick 
introduced in \cite{GS1}. We will only 
consider the case $\si_2<\si_1\leftrightarrow \ep=-1$ 
explicitly in the following,
the other case being completely analogous.
Let us split $\SQ(\si)=\SQ_I^c+\SQ_I^{}$, 
$\SQ(\si')=\SQ_I^c+\SQ_I'$, where 
\begin{equation}
\SQ_I^c\;\equiv\;\int\limits_{\si'}^{\si+2\pi}d\vf\;\SE^b(\vf),\quad
\SQ_I^{}\;\equiv\;\int\limits_{\si}^{\si'}d\vf\;\SE^b(\vf),\quad
\SQ'_I\;\equiv\;\int\limits_{\si+2\pi}^{\si'+2\pi}d\vf\;\SE^b(\vf).
\end{equation}
It follows from the exchange relations 
\begin{equation}
\SE^{\al_2}(\si_2)\SE^{\al_1}(\si_1)\;=\;e^{-2\pi i\al_2\al_1\ep(\si_2-\si_1)}
\SE^{\al_1}(\si_1)\SE^{\al_2}(\si_2),
\end{equation}
where $\ep(\si)=1$ if $\si>0$ and $\ep(\si)=-1$ if $\si<0$, 
that \rf{braidrelg} is equivalent to the following identity:
\begin{equation}\label{braidq}\begin{aligned}
e^{4\pi i \al_2\al_1}\bigl( & e^{-2\pi i b\al_1} 
\SQ_I^c  +e^{2\pi i b\al_1}\SQ_I^{}\bigr)^{s_2}
\bigl(\SQ_I^c+\SQ_I'\bigr)^{s_1}=\\
 & =\;\int\limits_{\BT}dt_2dt_1\;
\bigl(e^{-2\pi i b\al_2}\SQ_I^c+e^{-6\pi i b\al_2}\SQ_I'\bigr)^{t_1}
\bigl(\SQ_I^c+\SQ_I^{}\bigr)^{t_2}\; 
B_{-}^{}(\al_2,\al_1|\spp)^{s_2s_1}_{t_2t_1}.
\end{aligned}\end{equation}
The operators $\SQ_I^{}$, $\SQ_I'$, $\SQ_I^c$ satisfy the following algebra:
\begin{equation}\begin{aligned}
\SQ_I^c\SQ_I^{}\;=& \;q^{-2}\SQ_I^{}\SQ_I^c,\\
\SQ_I^c\SQ_I'\;=& \;q^{+2}\SQ_I'\SQ_I^c,
\end{aligned}\qquad \SQ_I^{}\SQ_I'\;=\;q^4\SQ_I'\SQ_I^{},
\end{equation}
where $q=e^{\pi i b^2}$. It follows that these operators can be 
realized as 
\begin{equation}
\SQ_I^c\;=\;e^{2b\sx}e^{-\pi i bt}, \qquad
\SQ_I^{}\;=\;e^{b\sx}e^{-2\pi b \spp}e^{b\sx}e^{\pi i bt},\qquad
 \SQ_I'\;=\;e^{b\sx}e^{2\pi b \spp}e^{b\sx}e^{\pi i bt},
\end{equation}
where $\sx$ and $\spp$ satisfy the 
commutation relations
$[\sx,\spp]=\frac{i}{2}$ and $t\in i\BR$ commutes with $\sx$ and $\spp$.

\subsection{Ordering the operators}

It turns out to be possible to translate \rf{braidq} into a relation 
that only contains commuting quantities: This becomes possible thanks 
to identities such as\footnote{Some care is needed in applying 
these identities in the present case: They represent relations
between unitary operators as long as $\al_k\in i\BR$ and $s_k\in i\BR$,
$k=1,2$. For the case we are presently interested in
($\al_i\in\frac{Q}{2}+i\BR$, $s_i\in-\frac{Q}{2}+i\BR$,
$t_i\in-\frac{Q}{2}+i\BR$, $i=1,2$) one needs to interpret these
relations by analytic continuation from the unitary case.} 
\begin{equation*}
\bigl(e^{-2\pi i b\al_1} \SQ_I^c  +e^{2\pi i b\al_1}\SQ_I^{}\bigr)^{s_2}
\;=\; e^{2bs_2\sx}e^{-\pi bs_2(\spp-is_2\frac{b}{2})}
\frac{S_b\bigl(\frac{Q}{2}+2\al_1+i\spp+s_2b+t\bigr)}
{S_b\bigl(\frac{Q}{2}+2\al_1+i\spp+t\bigr)},
\end{equation*}
which follow quite easily from the functional equation $S_b(x+b)=2\sin(\pi bx)
S_b(x)$. These identities allow one to collect all $\sx$-dependent factors
to the left of the terms containing $\spp$. It follows that \rf{braidq}
is equivalent to\footnote{In the presently considered case 
$\al_i\in\frac{Q}{2}+i\BR$, $s_i\in-\frac{Q}{2}+i\BR$,
$t_i\in-\frac{Q}{2}+i\BR$, $i=1,2$ one needs to interpret \rf{braidq2} 
as relation between tempered distributions defined by replacing
$\pm i\spp$ by $\pm i(\spp\pm i0)$, cf. the previous footnote.}
\begin{equation}\label{braidq2}\begin{aligned}
 {} e^{\frac{\pi i}{2}b^2(s_2^2-s_1^2)}e^{\pi i b^2s_1s_2}
e^{\pi b (s_1-s_2)\spp}\; & \frac{S_b\bigl(\frac{Q}{2}+i
\spp+2\al_1+(s_2+s_1)b+t\bigr)S_b\bigl(\frac{Q}{2}-i\spp+t\bigr)}
{S_b\bigl(\frac{Q}{2}+i\spp+2\al_1+bs_1+t\bigr)
S_b\bigl(\frac{Q}{2}-i\spp-bs_1+t\bigr)}=\\
\quad =\int\limits_{\BT}dt_2dt_1\;e^{-4\pi i (\al_1+bt_1)\al_2}
 & e^{\frac{\pi i}{2}b^2(t_2^2-t_1^2)}e^{-\pi i b^2t_1t_2}
e^{\pi b (t_1-t_2)\spp}
B_{-}^{}(\al_2,\al_1|\spp)^{s_2s_1}_{t_2t_1}
\\[-3ex]  \hspace{4cm}\ti & \frac{S_b\bigl(\frac{Q}{2}+i
\spp+2\al_2+(t_2+t_1)b-t\bigr)S_b\bigl(\frac{Q}{2}-i\spp-t\bigr)}
{S_b\bigl(\frac{Q}{2}+i\spp+2\al_2+bt_2-t\bigr)
S_b\bigl(\frac{Q}{2}-i\spp-bt_2-t\bigr)}.
\end{aligned}\end{equation}
It is furthermore convenient to take the Fourier-transformation
w.r.t. the variable $t$ of this identity. Let us introduce the 
following notation: 
\begin{equation}\begin{aligned}
\Phi_{\la}(A,B,C;y)\;=&\;\int_{\BR}d\tau\; 
e^{2\pi i \tau y}\,\tilde{\Phi}_{\la}(A,B,C;\tau),\\
\tilde{\Phi}_{\la}(A,B,C;\tau)\;\equiv & \;
\frac{S_b\bigl(\frac{Q}{2}+i(\tau+A)\bigr)
S_b\bigl(\frac{Q}{2}+i(\tau+B)\bigr)}
{S_b\bigl(Q+i(\tau-C+\la+i0)\bigr)S_b\bigl(Q+i(\tau-C-\la+i0)\bigr)},
\end{aligned}\end{equation}
where $A,B,C$ are real parameters. The identity obtained 
by Fourier-transformation of \rf{braidq2} takes the form
\begin{equation}\label{braidq3}\begin{aligned}
 {} e^{4\pi i (\al_1+bs)\al_2} & 
e^{\pi ib^2(s^2_{}-s_1^2)}e^{2\pi b (s_1-s)\spp}
\Phi_{\la}(A_1,B_1;C_1;y)=\\
=\;&  \int\limits_{\BT}dt_2dt_1\;e^{4\pi i bt_2\al_2}
e^{\pi ib^2t_2^2}
e^{-2\pi b t_2\spp}
B_{-}^{}(\al_2,\al_1|\spp)^{s_2s_1}_{t_2t_1}
\,\Phi_{\la}(A_2,B_2;C_2;-y),
\end{aligned}\end{equation}
with parameters $A_1,B_1,C_1$ and $A_2,B_2,C_2$ that can be 
easily read off from \rf{braidq2}, and $s=s_1+s_2=t_1+t_2$.

\subsection{Exploiting the completeness of the $\Phi_{\la}$}

\cite[Theorem 4]{PT2} is equivalent to the statement that
for any choice of $A,B,C\in\BR$ 
the set $\{\Phi_{\la}(A,B;C;y);\la\in \BR^+\}$
forms a basis for $L^2(\BR)$ with normalization given by
\[ 
\int_{\BR}dy \;\bigl(\Phi_{\la}(A,B;C;y)\bigr)^*\Phi_{\la'}(A,B;C;y)=
|S_b(Q+2i\la)|^{-2}\de(\la-\la').
\]
This means that in the case 
$\al_i\in\frac{Q}{2}+i\BR$, $s_i\in-\frac{Q}{2}+i\BR$,
$t_i\in-\frac{Q}{2}+i\BR$, $i=1,2$, the sought-for relation 
\rf{braidq2} is nothing but a representation for the
unitary transformation relating two such bases
$\{\Phi_{\la}(A_1,B_1;C_1;y);\la\in \BR^+\}$ and 
$\{\Phi_{\la}(A_2,B_2;C_2;-y);\la\in \BR^+\}$, with kernel 
$B_{\ep}(\al_2,\al_1|\spp)^{s_2s_1}_{t_2t_1}$ given in terms
of the overlap
\[ \int_\BR dy\;\bigl
(\Phi_{\la}(A_1,B_1;C_1;y) \bigr)^*\Phi_{\la'}(A_2,B_2;C_2;-y).\]
The rest of the calculation is straightforward. The result can be 
written as
\begin{equation}\label{braidcoeff1}\begin{aligned}
B_{\ep}(\al_2,\al_1|P)^{s_2s_1}_{t_2t_1}\;=\;
\de_{st}\;\frac{e^{\ep A}e^B}{|S_b(2C)|^2}
\int_{i\BR}ds\;\prod_{i=1}^4\frac{S_b(s+R_i)}{S_b(s+S_i)},
\end{aligned}\end{equation}
where $\de_{st}\equiv\de(s_1+s_2-t_1+t_2)$, 
$A$, $B$, and $C$ are given as 
\begin{equation}\begin{aligned}
A\;=\;& \pi i b^2(s_1^2+t_2^2-s^2)-2\pi bP(s_1+t_2-s)-2\pi i b\bigl(
(s-t_2)\al_2+(s-s_1)\al_1\bigr),\\
B\;=\;& 2\pi i b\bigl(
(s-t_2)\al_2-(s-s_1)\al_1\bigr),\\
C\;=\;& -iP-\al_2-bt_2,
\end{aligned}\end{equation}
and the coefficients $R_i$ and $S_i$, $i=1,\ldots,4$ are defined by
\begin{equation}  
\begin{aligned} R_1=&\fr{Q}{2}-2\al_2-iP-b(t_1+t_2), \\
         R_2=& \fr{Q}{2}+2\al_1+iP+b(s_1+s_2),\\
         R_3=& \fr{Q}{2}+iP,\\
        R_4=& \fr{Q}{2}-iP,
\end{aligned}\qquad
\begin{aligned} 
       S_1=&\fr{3Q}{2}-2\al_2-iP-bt_2, \\
         S_2=& \fr{Q}{2}+2\al_1+iP+bs_1,\\
         S_3=& \fr{3Q}{2}+iP+bt_2,\\
        S_4=& \fr{Q}{2}-iP-bs_1. 
\end{aligned}
\end{equation}

\newpage

\part{Chiral bootstrap}\vspace{0.5cm}\label{boots}

In this part we would like to explain how Liouville theory
fits into a ``chiral bootstrap'' formalism similar to the 
one developed by Moore and Seiberg \cite{MS} for rational 
conformal field theories (see
\cite{FFK} for a very similar formalism). 
Such a formalism is useful to gain further insight into the mathematics that
is behind the consistency of
Liouville theory, like fusion of unitary Virasoro representations, 
and the connection with quantum group representation theory \cite{PT1}.
It furthermore provides a convenient framework for completing the
proof of locality and crossing symmetry of operators $\SV_{\al}$
that are characterized by the DOZZ three-point function. 
Let us finally note that 
having such a formulation is important when trying to construct
Liouville theory with conformally invariant boundary conditions 
by means of a formalism that resembles the one introduced in 
\cite{BPPZ}\cite{FFFS} 
for rational conformal field theories in the presence of boundaries. 

\section{Formal chiral vertex operators} \label{CVOsec}

\subsection{Definition}

Chiral vertex
operators $\SV_{\BA}^{\rm N}(z)$, $\BA\equiv
\bigl({}_{\al_3}^{}\!{}^{\al_2}_{}\!{}_{\al_1}^{}\bigr)$
may be introduced
as maps from the Verma module 
$\CV_{\al_1}$ (cf. section 6 for notation) 
to $\CV_{\al_3}$ with
the properties
\begin{equation}\label{CVOdef}
\begin{aligned}
{\rm (i)} &\quad \bigl[
L_n,\SV_{\BA}^{\rm N}(z)\bigr]
\;=\; z^n(z\pa_z +\De_{\al_2}(n+1))\SV_{\BA}^{\rm N}(z),\\
{\rm (ii)} & \quad 
\SV_{\BA}^{\rm N}(z)
v_{\al_1}
\;=\;z^{\De_{\al_3}-\De_{\al_2}-\De_{\al_1}}\bigl({\rm N}(\al_3,\al_2,\al_1)
v_{\al_3}+\CO(z)\bigr),
\end{aligned}
\end{equation}
where $v_{\al}$ denotes the highest weight vector in $\CV_{\al}$.
We will denote $\SV_{\BA}^{}(z)\equiv \SV_{\BA}^{N\equiv 1}(z)$. 
Knowing $\SV_{\BA}^{}(z)$
is good enough since $\SV_{\BA}^{\rm N}(z)={\rm N}(\al_3,\al_2,\al_1)
\SV_{\BA}^{}(z)$.
\begin{rem}
The variable $z$ is to be considered as a formal variable for the moment: 
$\SV_{\BA}^{}(z)$ is considered as a generating device for a 
collection of maps $\SV_{\BA}^{}(k):\CV_{\al_1}[n]\ra\CV_{\al_3}[n+k]$ via
\begin{equation}
\SV_{\BA}^{}(z)\;=\;z^{\De_{\al_3}-\De_{\al_2}-\De_{\al_1}}
\sum_{k\in\BZ}\;z^k\;\SV_{\BA}^{}(k).
\end{equation}
Similar remarks will apply to the generalizations of the operators 
$\SV_{\BA}^{}(z)$ that we will introduce in the rest of this section.
This should be kept in mind as 
we will not present a formalism that makes the ``formal'' 
interpretation of chiral vertex operators mathematically precise.
\end{rem}
It is often convenient to generalize 
the $\SV_{\BA}^{}(z)$ by introducing
a family of operators $\SV_{\BA}^{}(\xi|z)$
called {\it descendants} of $\SV_{\BA}^{}(z)$ 
that are indexed by an element $\xi\in\CV_{\al_2}$.
The operators $\SV_{\BA}^{}(\xi|z)$ are defined in terms of $\SV_{\BA}^{}(z)$  
by means of $\SV_{\BA}^{}(v_{\al_2}|z)
\equiv \SV_{\BA}^{}(z)$, $\SV_{\BA}^{}(L_{-1}\zeta|z)
=   \pa_z\SV_{\BA}^{}(\zeta|z)$ and 
\begin{equation}\label{CVOdesc}\begin{aligned}
 {}& \SV_{\BA}^{}(L_{-n}\zeta|z)
\;=\; \frac{1}{(n-2)!}:(\pa_z^{n-2}T(z))\SV_{\BA}^{}(\zeta|z):,\\
& :T(w)\SV_{\BA}^{}(\zeta|z):\;
\equiv \;  \sum_{n<-1}w^{-n-2}L_n\SV_{\BA}^{}(\zeta|z)+\SV_{\BA}^{}(\zeta|z)
\sum_{n>-2}w^{-n-2}L_n.
\end{aligned}\end{equation}
The operators $\SV_{\BA}^{}(\zeta|z)$, $\zeta\in\CV_{\al_2}$ satisfy
the commutation relations 
\begin{equation}\label{desctrf}
{[}L_n,\SV_{\BA}^{}(\zeta|z){]}=\sum_{k=-1}^{l(n)}z^{n-k}
\biggl( \begin{matrix}n+1\\ k+1\end{matrix} \biggr)\SV_{\BA}^{}(L_k\zeta|z), 
\end{equation}
where $l(n)=n$ if $n>-2$ and $l(n)=\infty$ otherwise.
It is not difficult to show that chiral vertex operators are in fact
uniquely characterized by the properties \rf{CVOdef}: Transformation 
property \rf{desctrf} and definition \rf{CVOdesc} of the operators 
$\SV_{\BA}^{}(\xi|z)$ are closely related to the conformal Ward identities
in the case of the three-punctured Riemann sphere. It follows that
the matrix elements of $\SV_{\BA}^{}(\xi|z)$
can be expressed in terms of the trilinear
form $\rho^{\al_3,\al_2,\al_1}_{\infty,z,0}$ as follows 
(cf. Part \ref{pathint}, Section \ref{Verma}
for the notation):
\begin{equation}
\bigl(\xi_{3}^{}\, ,\, \SV_{\BA}^{}(\xi_2|z)\xi_{_1}\bigr)_{{\al_3}}\;=\;
\rho^{\al_3,\al_2,\al_1}_{\infty,z,0}(\xi_{3}^{},\xi_{2}^{},\xi_{1}^{}),
\end{equation}
which is equivalent to the following expression for 
the image of $\xi_{1}^{}\in\CV_{\al_1}$ under $\SV_{\BA}^{}(\xi_2^{}|z)$:
\begin{equation}
\SV_{\BA}^{}(\xi_2^{}|z)\xi_{1}^{}\;=\;
\sum_{\nu\in\CT}\;z^{\De(v_{\nu})-\De(\xi_{2})
-\De(\xi_{1})}
\;v_{\nu,\al_3}^{t}\;\rho(v_{\nu}^{},\xi_{2}^{},\xi_{1}^{}).
\end{equation}

\subsection{Generalized chiral vertex operators}

It is often useful to further 
generalize the notion of chiral vertex operators \cite{MS}:
One considers them as describing the ``fusion'' of two 
``in-going'' representations $\CV_{\al_2}$ and $\CV_{\al_1}$ 
associated to points $z_2$ and $z_1$ on the Riemann sphere
into a third representation $\CV_{\al_3}$. 
A generalized vertex operator 
$\SV_{}^{\al_3;}{}^{\al_2}_{z_2}{}^{\al_1}_{z_1}$, 
will then represent  
a map $\CV_{\al_2}\ot\CV_{\al_1}\ra\CV_{\al_3}$ which is defined by 
\newcommand{\fmult}[2]{\,{}^{}_{#1}\!\ot^{}_{\,#2}\!}
\begin{equation}\label{genCVO}
\SV_{}^{\al_3;}{}^{\al_2}_{z_2}{}^{\al_1}_{z_1}(\xi_2\ot\xi_1)\;=\;
\sum_{\nu\in\CT}\;v_{\nu,\al_3}^{t}\;
\rho_{\infty,z_{2},z_{1}}^{\al_3,\al_2,\al_1}(v_{\nu}^{},\xi_2^{},\xi_1^{}).
\end{equation}
These objects are related to the operators $\SV_{\BA}^{}(\xi|z)$ 
as follows:
\begin{equation}
\SV_{}^{\al_3;}{}^{\al_2}_{z_2}{}^{\al_1}_{z_1}(\xi_2\ot\xi_1)\;=\;
\SV_{\al_2\al_1}^{\;\,\al_3}(\xi_2|z_2)
\SV_{\al_10}^{\;\,\al_1}(\xi_1|z_1).
\end{equation}

Introducing the operators $\SV_{}^{\al_3;}{}^{\al_2}_{z_2}{}^{\al_1}_{z_1}$ 
is useful to make a profound analogy
between conformal field theory 
and representation theory more manifest: Let us define an action of the
Virasoro algebra on the tensor product $\CV_{\al_2}\ot\CV_{\al_1}$
by means of the co-product
\begin{equation}
\label{coprod}
\De_{z_2,z_1}^{}(L_n)\equiv 
\sum_{k=-1}^{l(n)} \biggl( \begin{matrix}n+1\\ k+1\end{matrix} \biggr)
\bigl(z_2^{n-k}L_k\ot\id+z_1^{n-k}\id\ot L_k\bigl). 
\end{equation}
The definition of $\SV_{}^{\al_3;}{}^{\al_2}_{z_2}{}^{\al_1}_{z_1}$ 
in terms of the conformal Ward identities
(via $\rho$ in \rf{genCVO}) is then equivalent to 
the following intertwining property:
\begin{equation}\label{CVOintw}
L_n \;\SV_{}^{\al_3;}{}^{\al_2}_{z_2}{}^{\al_1}_{z_1}(\xi_2\ot\xi_1)
\;=\;  
\SV_{}^{\al_3;}{}^{\al_2}_{z_2}{}^{\al_1}_{z_1}(\De_{z_2,z_1}(L_n)\cdot
\xi_2\ot\xi_1).
\end{equation}
The operators $\SV_{}^{\al_3;}{}^{\al_2}_{z_2}{}^{\al_1}_{z_1}$ 
are analogous to the
Clebsch-Gordon maps that describe the projection of the tensor product 
of two representations onto a third one. They satisfy the following 
simple relation w.r.t. the exchange of the two tensor
factors in $\CV_{\al_2}\ot\CV_{\al_1}$: 
\begin{equation}\label{elembraid}\begin{aligned}
 \SV_{}^{\al_3;}{}^{\al_2}_{z_2}{}^{\al_1}_{z_1}(\xi_2\ot\xi_1)\; =\;
O_{\ep}  \bigl({}_{\al_3}^{}\!{}^{\al_2}_{}\!{}_{\al_1}^{}\bigr)
\SV_{}^{\al_3;}{}^{\al_1}_{z_1}{}^{\al_2}_{z_2}(\xi_1\ot\xi_2)
,
\end{aligned}\end{equation}
where $O_{\ep}({}_{\al_{3}}^{}\!{}^{\al_2}_{}\!{}_{\al_1}^{}\bigr)
\equiv e^{\ep\pi i (\De_{\al_3}-\De_{\al_2}-\De_{\al_1})}$
and $\ep=1$ if $\arg(z_2-z_1)\in(0,\pi]$ and
$\ep=-1$ if $\arg(z_2-z_1)\in(-\pi,0]$.
\begin{rem}
In order to make the interpretation of the generalized chiral vertex
operators as generalizations of Clebsch-Gordon maps
more precise, one would need to show that a certain set 
$\{\CV_{\al};\al\in\BF\}$ of Virasoro
representations is indeed closed under the tensor product
operation defined by means of the co-product $\De_{z_2,z_1}^{}$,
i.e. that there exists a canonical decomposition of the representation
defined on $\CV_{\al_2}\ot\CV_{\al_1}$ by means of $\De_{z_2,z_1}^{}$
into irreducible representations $\CV_{\al}$ with $\al\in\BF$.
We will outline an approach to do so (based on ``Connes-fusion'') in 
Section \ref{unfus}.
\end{rem}

\subsection{Composition of chiral vertex operators}

By forming compositions of chiral vertex operators it is possible
to construct maps $\CV_{\al_{n-1}}\ot\dots\ot\CV_{\al_1}\ra\CV_{\al_{n}}$. 
Let us consider the example $n=4$:
There are two natural types 
of compositions of chiral vertex operators:
On the one hand (``s-channel''):
\begin{equation}
\SV_{}^{\al_4;}{}^{\al_3}_{z_3}{}^{\al_s}_{z_1}\bigl(\xi_3^{}\ot
\SV_{}^{\al_s;}{}^{\al_2}_{z_{21}}{}^{\al_1}_{\;0}(\xi_2^{}\ot\xi_1^{})\bigr),
\end{equation}
where $z_{21}=z_2-z_1$, and on the other hand (``t-channel''):
\begin{equation}
\SV_{}^{\al_4;}{}^{\al_t}_{z_2}{}^{\al_1}_{z_1}\bigl(
\SV_{}^{\al_t;}{}^{\al_3}_{z_{32}}{}^{\al_2}_{\;0}
(\xi_3^{}\ot\xi_2^{})\ot\xi_1)
\bigr),
\end{equation} where $z_{32}=z_3-z_2$.
The corresponding
conformal blocks are then recovered as the matrix elements of compositions
of generalized chiral vertex operators, for example: 
\newcommand{\RA}{\al}
\begin{equation}\begin{aligned}
\CfBls{\al_s}{\RA_3}{\RA_2}{\RA_4}{\RA_1}(\sz)\;=&\;
\bigl\bra v\,,\,
\SV_{}^{\al_4;}{}^{\al_3}_{z_3}{}^{\al_s}_{z_1}\bigl(v\ot
\SV_{}^{\al_s;}{}^{\al_2}_{z_{21}}{}^{\al_1}_{\;0}(v\ot v)\bigr)
\bigr\ket_{{\al_4}}^{},\\
\CfBlt{\al_t}{\RA_3}{\RA_2}{\RA_4}{\RA_1}(\sz)
\;=& \;\bigl\bra v\,,\,
\SV_{}^{\al_4;}{}^{\al_t}_{z_2}{}^{\al_1}_{z_1}\bigl(
\SV_{}^{\al_t;}{}^{\al_3}_{z_{32}}{}^{\al_2}_{\;0}
(v\ot v)\ot v)
\bigr)
\bigr\ket_{{\al_4}}^{},
\end{aligned}\end{equation}
where $\sz=(z_3,z_2,z_1)$.

\section{Fusion and braiding}

So far we have considered chiral vertex operators 
and conformal blocks in the sense of formal power series only.
It appears to be difficult to get 
information on the analytical properties of these 
objects from the above definition in terms
of representation theory of the Virasoro algebra.
It will therefore be important to have an alternative 
realization that allows one to prove convergence of 
the power series defining the conformal blocks and to study 
their analytic properties (monodromies, braiding...). 

\subsection{Free field realization} \label{freeCVO}

A useful realization of chiral vertex operators is furnished by the 
operators $\sg_s^{\al}(x_+)$ that were introduced in Part \ref{gervnev},
Subsection \ref{chirvert}: Let us recall that these were well-defined
as operators for negative euclidean time $\tau=it$. The corresponding
objects on the punctured Riemann sphere are obtained by introducing
the variable $z=e^{ix_+}$ and letting $\sg_s^{\al}(z)\equiv
z^{-\De_{\al}}\sg_s^{\al}(x_+)$. 

We need to explain how to go from such 
operators on $\CH^F$ to operators between Verma modules.
To simplify slightly one may assume  
$\al+bs\in i\BR$, from which one may get the 
general case by analytic continuation. We will furthermore
assume having chosen 
a Gelfand triple $\CT\subset\CH^{\rm F}\subset\CT^{\dagger}$
(cf. the discussion in Part \ref{pathint}, Subsection 
\ref{micmac}). The direct integral representation
\begin{equation}\label{dirintH}
\CH^{\rm F} \;\simeq\; \int\limits_{\BS\cup\bar{\BS}}^{\oplus}d\al\;
\CV_{\al}\ot\CV_{\al}
\end{equation}
allows to identify $\CV_{\al}\simeq\CV_{\al}\ot v$ 
with a subspace $\CT_{\al}^{\dagger}
\subset\CT^{\dagger}$.

Due to the smooth dependence of 
$G_{s}^{\al}(P)$ on $P$ it is for $|z|<1$ possible to define
$ \sg_s^{\al_2}(z)|P_1,\xi\ket$ as an element of $\CT^{\dagger}$. 
Since \[ 
\spp\;\sg_s^{\al_2}(z)|P_1,\xi\ket=
\bigl(P_1-i(\al_2+bs)\bigr)\sg_s^{\al_2}(z)|P_1,\xi\ket,
\] 
we may identify  $ \sg_s^{\al_2}(z)|P_1,\xi\ket$ 
as an element of $\CT_{\al_1+\al_2+bs}^{\dagger}$. 
Let us denote the resulting operator $\CT_{\al_1}^{\dagger}\ra
\CT_{\al_1+\al_2+bs}^{\dagger}$ by $\sg_s^{\al_2}(P_1|z)$. We have
\begin{equation*}\begin{aligned}
\SL_n(P_1-i(\al_2+bs))\sg_s^{\al_2}(P_1|z)|P_1,\xi\ket &- 
\sg_s^{\al_2}(P_1|z)\SL_n(P_1)|P_1,\xi\ket=\\
& =\;z^n(z\pa_z+\De_{\al_2}(n+1))
\sg_s^{\al_2}(P_1|z)|P_1,\xi\ket,
\end{aligned}\end{equation*}
so that the identification \rf{dirintH} induces 
\begin{equation}\label{chnorm}
\sg_s^{\al_2}(P_1|z)\;\simeq\;\SV^{\rm G}
\bigl({}_{\al_3}^{}\!{}^{\al_2}_{}\!{}_{\al_1}^{}\bigr)(z)\;\equiv\;
{\rm G}(\al_3,\al_2,\al_1)\SV
\bigl({}_{\al_3}^{}\!{}^{\al_2}_{}\!{}_{\al_1}^{}\bigr)(z), 
\end{equation}
where $\al_1=\fr{Q}{2}+iP_1$, $\al_3=\al_1+\al_2+bs$
and ${\rm G}(\al_3,\al_2,\al_1)\equiv G^{\al_2}_{s}(P_1)$.  
It now follows from the corresponding properties of the $\sg_s^{\al_2}(z)$
that the chiral vertex operators $\SV_{\BA}(z)$ indeed make sense
as operators for $|z|<1$ and as operator-valued distributions for 
$|z|=1$.

\subsection{Generalized braiding}

It is then straightforward to translate the 
braid relations \rf{braidrelg}
for the covariant chiral 
operators $\sg_s^{\al_2}(\si)$ into the corresponding
relations for the chiral vertex operators 
$\SV^{\rm G}
\bigl({}_{\al_3}^{}\!{}^{\al_2}_{}\!{}_{\al_1}^{}\bigr)(z)$.
They take the following form:
\begin{equation}\begin{aligned}\label{braidrel}
\SV^{\rm G}
\bigl({}_{\al_4}^{}\!{}^{\al_3}_{}\!{}_{\al_s}^{}\bigr)
(\si_2) & \SV^{\rm G}
\bigl({}_{\al_s}^{}\!{}^{\al_2}_{}\!{}_{\al_1}^{}\bigr)(\si_1)
\;=\;\\
=\;& \int\limits_{\BS}d\al_u\;\,
\Braidg{\al_{s}}{\al_{u}}{\al_3}{\al_2}{\al_4}{\al_1}\;
\SV^{\rm G}
\bigl({}_{\al_4}^{}\!{}^{\al_2}_{}\!{}_{\al_u}^{}\bigr)(\si_1)
\SV^{\rm G}
\bigl({}_{\al_u}^{}\!{}^{\al_3}_{}\!{}_{\al_1}^{}\bigr)(\si_2)\;.
\end{aligned}\end{equation}
It will be useful to write the braiding coefficients 
$B$ as follows:
\begin{equation}\label{braidcoeffs}\begin{aligned}
\Braidg{\al_{s}}{\al_{u}}{\al_3}{\al_2}{\al_4}{\al_1}\;=\;
{e^{\ep A}e^B}
\frac{S_b(\al_2+\bar{\al}_u-\bal_4)
S_b(\al_s+\bal_4-\al_3)}
     {S_b(\al_2+\al_{s}-\al_1)
S_b(\bal_{u}+\al_1-\al_3)}
\SJS{\al_1}{\al_2}{\bar{\al}_4}{\bar{\al}_3}{\al_{s}}{\bar{\al}_{u}},
\end{aligned}\end{equation}
where $\bar{\al}\equiv Q-\al$, $A$ and $B$ are given as
\begin{equation}\begin{aligned}
A\;=\;& \pi i b^2(s_1^2+t_2^2-s^2)-2\pi bP(s_1+t_2-s)-2\pi i b\bigl(
(s-t_2)\al_2+(s-s_1)\al_1\bigr)\\
V\;=\;& 2\pi i b\bigl(
(s-t_2)\al_2-(s-s_1)\al_1\bigr),
\end{aligned}\end{equation}
and $\{\dots\}_b$ are the so-called b-Racah-Wigner symbols \cite{PT2},
\begin{equation}\label{RWexplicit} \begin{aligned}
{}\SJS{\al_1}{\al_2}{\al_3}{\bar{\al}_4}{\al_s}{\al_{t}} =
& \quad\frac{S_b(\al_2+\al_s-\al_{1})
S_b(\al_{t}+\al_1-\al_4)}
     {S_b(\al_2+\al_{t}-\al_3)
S_b(\al_s+\al_3-\al_4)}\cdot\\
&\quad\cdot|S_b(2\al_{t})|^2
\int\limits_{-i\infty}^{i\infty}ds \;\;
\frac{S_b(U_1+s)S_b(U_2+s)S_b(U_3+s)S_b(U_4+s)}
{S_b(V_1+s)S_b(V_2+s)S_b(V_3+s)S_b(V_4+s)},
\end{aligned} 
\end{equation}
with coefficients $U_i$ and $V_i$, $i=1,\ldots,4$ given by
\begin{equation}  
\begin{aligned} U_1=& \al_s+\al_1-\al_2 \\
        U_2=& Q+\al_s-\al_2-\al_1 \\
        U_3=& \al_s+\al_3-\al_4 \\
        U_4=& Q+\al_s-\al_3-\al_4
\end{aligned}\qquad
\begin{aligned} 
        V_1=& 2Q+\al_s-\al_{t}-\al_2-\al_4 \\
        V_2=& Q+\al_s+\al_{t}-\al_4-\al_2 \\
        V_3=& 2\al_s \\
        V_4=& Q.
\end{aligned}
\end{equation}
Taking into account the change of normalization \rf{chnorm}
it is straightforward to obtain the corresponding relations 
for the generalized chiral vertex operators 
$\SV_{}^{\al_3;}{}^{\al_2}_{z_2}{}^{\al_1}_{z_1}$ from \rf{braidrel}.
In particular it is possible to show by 
direct calculation  that \rf{braidrel}
simplifies for $\al_1\ra 0$ to 
\begin{equation}
\SV
\bigl({}_{\al_4}^{}\!{}^{\al_3}_{}\!{}_{\al_2}^{}\bigr)
(\si_2)  \SV
\bigl({}_{\al_2}^{}\!{}^{\al_2}_{}\!{}_{0}^{}\bigr)(\si_1)
\;=
\; O_{\ep}({}_{\al_{4}}^{}\!{}^{\al_3}_{}\!{}_{\al_2}^{}\bigr)\SV
\bigl({}_{\al_4}^{}\!{}^{\al_2}_{}\!{}_{\al_3}^{}\bigr)(\si_1)
\SV
\bigl({}_{\al_3}^{}\!{}^{\al_3}_{}\!{}_{0}^{}\bigr)(\si_2)\;,
\end{equation}
showing that the generalized braid relations \rf{braidrel}  
are indeed consistent with the elementary one \rf{elembraid}. 
One thereby concludes that the Liouville conformal blocks form a 
representation of the braid group.

\subsection{Fusion}

The generalized braid relations \rf{braidrel} also allow one to construct 
a kind of associativity relation between s- and t-channel 
compositions of chiral
vertex operators that is analogous to the Moore-Seiberg ``fusion move'': 
It is a relation of the following form:
\begin{equation}\begin{aligned}\label{fusrel}
 \SV_{}^{\al_4;}{}^{\al_3}_{z_3}{}^{\al_s}_{z_1}\bigl(\xi_3^{}\ot & 
\SV_{}^{\al_s;}{}^{\al_2}_{z_{21}}{}^{\al_1}_{\;0}(\xi_2^{}\ot\xi_1^{})\bigr)
\;=\;\\
=\;& \int\limits_{\BS}d\al_t\;\,
\Fus{\al_{s}}{\al_{t}}{\al_3}{\al_2}{\al_4}{\al_1}\;
\SV_{}^{\al_4;}{}^{\al_t}_{z_2}{}^{\al_1}_{z_1}\bigl(
\SV_{}^{\al_t;}{}^{\al_3}_{z_{32}}{}^{\al_2}_{\;0}
(\xi_3^{}\ot\xi_2^{})\ot\xi_1)
\bigr).
\end{aligned}\end{equation}
Such relations can be read as expressing some sort of associativity of the 
``fusion products'' of representations defined by means of 
$\De_{z_2,z_1}^{}$, with fusion coefficients $F_{\al_{s}\al_{t}}$ 
playing the role of the Racah-Wigner coefficients from angular momentum 
theory. The associativity relation \rf{fusrel} is essentially equivalent to
relations for the corresponding conformal blocks like
\begin{equation}\label{fus_cfbl}
\CfBls{\al_s}{\RA_3}{\RA_2}{\RA_4}{\RA_1}(\sz)\;=\;
\int\limits_{\BS}d\al_t\;\,
\Fus{\al_{s}}{\al_{t}}{\al_3}{\al_2}{\al_4}{\al_1}\;
\CfBlt{\al_t}{\RA_3}{\RA_2}{\RA_4}{\RA_1}(\sz).
\end{equation}

Relations of the form \rf{fusrel} can indeed be constructed in terms of the 
braid relations \rf{braidrel} and \rf{elembraid}. 
This is done by {\it defining} fusion as the result of the sequence of 
braid-moves that may be schematically indicated
by $3(21)\ra 3(12)\ra 1(32)\ra (32)1$, cf. \cite{MS}. 
The first and the last of these moves 
are represented by \rf{elembraid}, the middle one by \rf{braidrel}. 
By means of the explicit expressions
for $O_{\ep}$ and $B_{\al_s\al_u}^{\ep}$ it is possible to verify
that the resulting relation \rf{fusrel} does not depend on the sign
choices made
to define the generalized chiral vertex operators involved in \rf{fusrel}.

Taking into account the explicit expression for $G(\al_3,\al_2,\al_1)$
that follows from \rf{G-norm}, it is straightforward to deduce the
following explicit expression for $F_{\al_{s}\al_{t}}$ from \rf{braidcoeffs}:
\newcommand{\RN}{{\rm N}}
\begin{equation}
\label{Fexplicit1}
\Fus{\al_{s}}{\al_{t}}{\al_3}{\al_2}{\al_4}{\al_1}\;\, \equiv\;\,
\frac{\RN(\al_4,\al_3,\al_s)
\RN(\al_s,\al_2,\al_1)}
{\RN(\al_4,\al_{t},\al_1)\RN(\al_{t},\al_3,\al_2)}\;
\SJS{\al_2}{\al_1}{\bal_4}{\bal_3}{\al_{s}}{\bal_{t}},
 \end{equation}
with $\RN(\al_3,\al_2,\al_1)$ being defined by the expression
\begin{equation}\label{RNexplicit}\begin{aligned}
\RN & (\al_3,\al_2,\al_1)=\\[-1ex]
= & \; \frac{\Ga_b(2\al_1)\Ga_b(2\al_2)\Ga_b(2Q-2\al_3)}
{\Ga_b(2Q-\al_1-\al_2-\al_3)
\Ga_b(Q-\al_1-\al_2+\al_3)\Ga_b(\al_1+\al_3-\al_2)\Ga_b(\al_2+\al_3-\al_1)}.
\end{aligned}\end{equation}
We will discuss the relation between the b-Racah-Wigner coefficients
that appear in \rf{Fexplicit1} and the representation
theory of $\CU_{q}(\fsl(2,\BR))$ in Section \ref{q-stuff}.

\subsection{Consistency conditions} \label{conscond}

Different ways to construct conformal blocks 
by composing generalized chiral vertex operators 
can be associated with the tree-level Feynman graphs in a $\vf^3$ 
theory, where all lines are ``colored'' with labels $\al\in\BS$ of 
Virasoro Verma modules. The corresponding graphs with 
colors only on the external lines parameterize {\it sets} of 
conformal blocks which have elements distinguished 
by the coloring of the internal lines.
One may view the elementary
braiding transformation \rf{elembraid} and the fusion relation
\rf{fusrel} as elementary moves that allow one to relate 
sets of conformal blocks which have graphs 
with the same number and coloring
of the external lines. The relation between two such graphs
can generically
be decomposed into elementary braid- and fusion transformations
in more than one way, but the resulting relation between 
the two sets of conformal blocks must be identical. 
This leads to a bunch of identities that fusion and braid coefficients
have to satisfy \cite{MS}\cite{FFK}. These identities include:\\
{\sc Pentagon:} $\frac{\quad}{}$

\begin{equation}\begin{aligned}\label{pentagon}
\int\limits_{\BS} d\de_1 \;\Fus{\be_1}{\de_1}{\al_3}{\al_2}{\be_2}{\al_1}
 \Fus{\be_2}{\ga_2}{\al_4}{\de_1}{\al_{5}}{\al_1} &
\Fus{\de_1}{\ga_1}{\al_4}{\al_3}{\ga_2}{\al_2}
=\\
& =\Fus{\be_2}{\ga_1}{\al_4}{\al_3}{\al_5}{\be_1}
\Fus{\be_1}{\ga_2}{\ga_1}{\al_2}{\al_5}{\al_1},\end{aligned}
\end{equation}
{\sc Hexagon:} $\frac{\quad}{}$
\begin{equation*}\begin{aligned}
\int\limits_{\BS}d\al_{32}\;
\Fus{\al_{21}}{\al_{32}}{\al_3}{\al_2}{\al_4}{\al_1}
& O_{\pm}({}_{\al_{32}}^{}\!{}^{\al_4}_{}\!{}_{\al_1}^{}\bigr)
\Fus{\al_{32}}{\al_{31}}{\al_1}{\al_3}{\al_4}{\al_2}
\;=\;\\
& =\; O_{\pm}({}_{\al_{21}}^{}\!{}^{\al_2}_{}\!{}_{\al_1}^{}\bigr)
\Fus{\al_{21}}{\al_{31}}{\al_3}{\al_1}{\al_4}{\al_2}
O_{\pm}({}_{\al_{31}}^{}\!{}^{\al_3}_{}\!{}_{\al_1}^{}\bigr)
\end{aligned}\end{equation*}
It can be shown \cite{MS} that
the pentagon and hexagon identities suffice to ensure 
consistency of fusion and braiding in general.

\section{Relation to tensor category of quantum group representations}
\label{q-stuff}

We had expressed the fusion- and braid coefficients 
of Liouville theory in terms of some object that
we called ``b-Racah-Wigner coefficients''. 
These objects were first constructed as ``re-coupling'' coefficients that
describe the relation between two canonical ways to 
reduce the triple tensor products of certain quantum group 
representations into irreducible representations \cite{PT2}. 
Their role within quantum group representation theory
is therefore analogous to the role of the fusion coefficients
for the representation theory of the Virasoro algebra. 
We will see that fusion- and b-Racah-Wigner coefficients become
{\it identical} by a suitable choice of normalization of the 
chiral vertex operators. 
We therefore observe a deep relationship 
between representation theory of the Virasoro algebra 
and quantum group representation theory, for which a more direct 
explanation remains to be found.

\subsection{A class of representations of $\CU_q(\fsl(2,\BR))$}

Let us recall the definition of the quantum group in question:
$\usl$ is a Hopf-algebra with 
\begin{equation}
\begin{aligned}
{}& \text{generators:}\quad  E,\quad F,\quad K,\quad K^{-1}\\
& \text{relations:}\quad 
KE=qEK\qquad\quad KF=q^{-1}FK\qquad[E,F]=-\frac{K^2-K^{-2}}{q-q^{-1}}\\
& \text{co-product:}\quad\De(K)=K\ot K\qquad 
\begin{aligned}\De(E)=&E\ot K+K^{-1}\ot E\\
\De(F)=&F\ot K+K^{-1}\ot F
\end{aligned}\\
& \text{q-Casimir:}\quad C=FE-(q-q^{-1})^{-2}\bigl(qK^2+q^{-1}K^{-2}-2\bigr)
\end{aligned}\end{equation}
We will use the notation
$U_q(\fsl(2,\BR))$ for 
$\usl$ supplemented by the following star-structure
\begin{equation}\label{star-struct}
\text{star-structure:}\quad 
K^*=K \qquad E^*=E \qquad F^*=F,
\end{equation}
which defines the hermiticity assignments for what is called 
a unitary representation of $U_q(\fsl(2,\BR))$.

The following set of representations $\CP_{\al}$ by unbounded operators
on the Hilbert-space $L^2(\BR)$ was considered in \cite{PT1}\cite{PT2}:
\begin{equation}\label{uslgens}
\begin{aligned}
\SE_{\al}=
\SU^{+1}\frac{e^{\pi i b (Q-\al)}\SV- 
e^{-\pi i b (Q-\al)}\SV^{-1}}{e^{\pi i b^2}-
e^{-\pi i b^2}}\\
\SF_{\al}=
\SU^{-1}\frac{e^{-\pi i b (Q-\al)}\SV- 
e^{\pi i b (Q-\al)}\SV^{-1}}{e^{\pi i b^2}-
e^{-\pi i b^2}}
\end{aligned}\qquad\qquad \SK_{\al}=\SV,
\end{equation}
where $\SU$ is the operator of multiplication by $e^{2\pi b x}$, and 
$\SV$ acts on $f(x)$ as $\SV f(x)=f(x+i\frac{b}{2})$. 
These representations appear in the list of ``well-behaved''
unitary irreducible representations of $\CU_q(\fsl(2,\BR))$ that
was obtained in \cite{S1} when $\al\in\frac{Q}{2}+i\BR$. However, 
these representations 
do not reproduce representations of the
classical Lie algebra $\fsl(2,\BR)$  in the limit $b\ra 0$, 
which is why we will call this series of representations 
the ``strange series'' of $\CU_q(\fsl(2,\BR))$.

\subsection{Why to consider the strange series?}
  
The proposal \cite{PT1} to consider the
representations introduced in the 
previous subsection was motivated by 
previous proposals concerning the appearance of objects from the
representation theory of quantum groups in Liouville theory, going back
to \cite{FT,B1,G1} (see \cite{CGR,GS1} for more recent developments).
It was shown in particular in \cite{CGR,GS1} that chiral vertex operators
which correspond to triples $(\al_3,\al_2,\al_1)$ subject to a certain 
integrality constraint have fusion coefficients that can be expressed
in terms of q-6j symbols associated to the quantized universal enveloping
algebra $\usl$. 
It is then natural to expect that fusion and braiding
transformations of general chiral vertex operators 
can also be expressed in terms of q-6j symbols associated to
$\usl$.

The problem is to choose the right set of representations of $\usl$. 
Let us present an a-posteriori line of argument that leads to the 
proposal of \cite{PT1}. This proposal was originally
made for the ``weak-coupling'' case of real $b$. 
The first question is: What hermiticity relations should the generators
of $\usl$ satisfy in the representation (choice of star-structure for
$\usl$)? The results of \cite{CGR,GS1} indicate that the deformation
parameter $q$ is related to the coupling constant $b$ of Liouville theory via
\begin{equation}\label{q-b}
q=e^{\pi i b^2}.
\end{equation}  
It is shown in \cite{MMNNSU} that the only star-structure on $\usl$
that is compatible with \rf{q-b} is \rf{star-struct}.
This motivates us
to look for unitary representations of $U_q(\fsl(2,\BR))$. A classification
result for such representations can be found in \cite{S1}. 
But which of these representations should we choose? 

The representation introduced in \rf{uslgens} is distinguished by
having a certain remarkable
self-duality property under the replacement $b\ra b^{-1}$: 
These representations have the property that replacing 
$b\ra b^{-1}$ in the expressions for the generators $\SE_{\al}$, 
$\SF_{\al}$, $\SE_{\al}$
above yields operators $\tilde{\SE}_{\al}$  
$\tilde{\SF}_{\al}$, $\tilde{\SK}_{\al}$ 
that generate a representation of the ``dual'' algebra 
$\CU_{\tilde{q}}(\fsl(2,\BR))$ that commutes with the 
$\CU_q(\fsl(2,\BR))$-representation generated by   
$\SE_{\al}$, 
$\SF_{\al}$, $\SK_{\al}$
\footnote{The precise 
sense of this term is subtle in this context involving unbounded
operators. In fact, these two sets of operators do not commute in
the strict sense (commutativity of spectral projections). The statement
rather is that there exists a natural domain $\CP_{\al}$ 
such that the representations of $\CU_q(\fsl(2,\BR))$ and 
$\CU_{\tilde{q}}(\fsl(2,\BR))$ that are generated by   
$\SE_{\al}$, 
$\SF_{\al}$, $\SK_{\al}$ and 
$\tilde{\SE}_{\al}$  
$\tilde{\SF}_{\al}$, $\tilde{\SK}_{\al}$ respectively commute on $\CP_{\al}$.
The domain $\CP_{\al}$ may be considered as an analogue of the Schwartz-space
for the representations of the two Hopf-algebras}.

\begin{rem}
An alternative point of view on this self-duality phenomenon is to regard
the strange series representations 
as representations of the {\it modular double} 
of $\CU_q(\fsl(2,\BR))$
introduced by Faddeev\cite{F}, see \cite[Section 1]{KLS} 
for a nice discussion of this concept.
\end{rem}

It is this self-duality 
that may be taken as a hint that this class of representations
is well-suited for making contact with the DOZZ-proposal on the one hand
(where we had observed such a self-duality earlier), and with the results
of \cite{CGR,GS1} on the other hand. In these latter references it was
in particular found that the fusion coefficients for
the special class of chiral 
vertex operators considered therein show a factorization into
q-6j symbols for $\usl$ times $\tilde{q}$-6j symbols for 
$\CU_{\tilde{q}}(\fsl_2))$ where $\tilde{q}=e^{\pi i/b^2}$,
expressing a form of $b\ra b^{-1}$ duality for that 
particular class of operators.


\subsection{Generalized Clebsch-Gordan coefficients}

We will now briefly review the results of \cite{PT2} which is devoted to
the construction of the b-Racah-Wigner coefficients of the strange series
representations. 

The first main result of \cite{PT2} is the closure of the  
strange series representations under tensor 
products:
\begin{equation}\label{CGdecoQG}
\CP_{\al_2}\ot \CP_{\al_1}\;\,\simeq\;\, \int\limits_{\BS}^{\oplus}
\!d\al\;\, \CP_{\al,}\qquad \BS\equiv \frac{Q}{2}+i\BR^+.
\end{equation}
This result forms the basis for an associated 
calculus of Clebsch-Gordan and Racah-Wigner coefficients
that strongly resembles standard angular momentum theory:

To begin with, one needs to note that 
the projection operators $\SC(\al_3|\al_2,\al_1):\CP_{\al_2}\ot\CP_{\al_1}\ra
\CP_{\al_3}$ may be explicitly represented by an integral transform
\begin{equation}
\SC(\al_3|\al_2,\al_1): f(x_2,x_1)\longrightarrow F[f](\al_3|x_3)
\equiv \int_{\BR}dx_2dx_1 \;\CGC{\al_3}{x_3}{\al_2}{x_2}{\al_1}{x_1}\;
f(x_2,x_1).
\end{equation}
Explicit expressions for the 
distributional kernel $[\ldots]$ (the "Clebsch-Gordan coefficients")
can be found in \cite{PT2}. 
The kernel $[\ldots]$ can be shown to satisfy orthogonality and completeness
relations of the form:
\begin{equation}\begin{aligned}
{} & \int\limits_{\BR}dx_1dx_2\;\,
\CGC{\al_3}{x_3}{\al_2}{x_2}{\al_1}{x_1}^*
\CGC{\be_3}{y_3}{\al_2}{x_2}{\al_1}{x_1}
=|S_b(2\al_3)|^{-2}\de(\al_3-\be_3)\de(x_3-y_3).\\
&  \int\limits_{\BS}d\al_3 \; |S_b(2\al_3)|^2
\int\limits_{\BR}dx_3 \;\,
\CGC{\al_3}{x_3}{\al_2}{x_2}{\al_1}{x_1}^*
\CGC{\al_3}{x_3}{\al_2}{y_2}{\al_1}{y_1}
=\de(x_2-y_2)\de(x_1-y_1). 
\end{aligned}\end{equation}

\subsection{Generalized Racah-Wigner coefficients}

Triple tensor products $\CP_{\al_3}\ot
\CP_{\al_2}\ot\CP_{\al_1}$ carry a representation $\pi_{321}$ of 
$\CU_{q}(\fsl(2,\BR))$ given by 
$(\id\ot\De)\circ\De=(\De\ot\id)\circ\De$.
The projections affecting the decomposition of this representation into
irreducibles can be constructed by iterating Clebsch-Gordan maps.
One thereby obtains two canonical bases
in the sense of generalized eigenfunctions for $\CP_{\al_3}\ot
\CP_{\al_2}\ot\CP_{\al_1}$ given by the sets of distributions 
($\fx=(x_4,\ldots,x_1)$)
\begin{equation}\label{q-blocks}
\begin{aligned}
\CpBls{\al_{s}}{\al_3}{\al_2}{\al_4}{\al_1}(\fx)=& 
\int_{\BR}dx_{s}\;\,
\CGC{\al_4}{x_4}{\al_3}{x_3}{\al_{s}}{x_{s}}\,
\CGC{\al_{s}}{x_{s}}{\al_2}{x_2}{\al_1}{x_1}
\quad\al_4,\al_{s}\in\BS,x_4\in\BR\\
\CpBlt{\al_{t}}{\al_3}{\al_2}{\al_4}{\al_1}(\fx)=& 
\int_{\BR}dx_{t}\;\,
\CGC{\al_4}{x_4}{\al_{t}}{x_{t}}{\al_1}{x_1}\,
\CGC{\al_{t}}{x_{t}}{\al_3}{x_3}{\al_2}{x_2}.
\quad\al_4,\al_{t}\in\BS,x_4\in\BR\end{aligned}.
\end{equation}
It is possible to show that the resulting spectral decompositions for the 
operators $\pi_{321}(C)$ and $\pi_{321}(K)$ do not depend on the 
order in which the Clebsch-Gordan decompositions were performed:
\begin{equation}
\CP_{a_3}\ot\CP_{a_2}\ot \CP_{a_1}\simeq 
\int\limits_{\BS}^{\oplus}d\al\int\limits_{\BR}^{\oplus}dk
\;\, \CH_{\al,k}.
\end{equation}
It then follows from completeness of the bases $\FB_{321}^s$ and $\FB_{321}^t$
and orthogonality of the eigenspaces $ \CH_{\al,k}$ that 
the bases $\Phi^s$ and $\Phi^t$ must be related by a transformation of the
form
\begin{equation}\label{Racahdef}
\CpBls{\al_{s}}{\al_3}{\al_2}{\al_4}{\al_1}(\fx)\;=\;
\int\limits_{\BS}d\al_{t}\;\,
\SJS{\al_1}{\al_2}{\al_3}{\al_4}{\al_{s}}{\al_{t}}\;\,
\CpBlt{\al_{t}}{\al_3}{\al_2}{\al_4}{\al_1}(\fx)
\end{equation}
thereby defining the b-Racah-Wigner symbols $\SJS{\cdot}{\cdot}
{\cdot}{\cdot}{\cdot}{\cdot}$. Their explicit expression was found in
\cite{PT2} to be given by equation \rf{RWexplicit}.

\subsection{Equivalence to Liouville fusion coefficients}

Let us reconsider the expression \rf{Fexplicit1} that we had found for
the fusion coefficients. It is natural to introduce 
\begin{equation}
\SC_{}^{\al_3;}{}^{\al_2}_{z_2}{}^{\al_1}_{z_1}(\xi_2\ot\xi_1)\;=\;
\RN^{-1}(\al_3,\al_2,\al_1)
\SV_{}^{\al_3;}{}^{\al_2}_{z_2}{}^{\al_1}_{z_1}(\xi_2\ot\xi_1),
\end{equation} 
since the chiral vertex operators $\SC$ will then satisfy 
a fusion relation like \rf{fusrel} that involves only the 
b-Racah-Wigner symbols:
\begin{equation}\begin{aligned}\label{fusrel+}
 \SC_{}^{\al_4;}{}^{\al_3}_{z_3}{}^{\al_s}_{z_1}\bigl(\xi_3^{}\ot & 
\SC_{}^{\al_s;}{}^{\al_2}_{z_{21}}{}^{\al_1}_{\;0}(\xi_2^{}\ot\xi_1^{})\bigr)
\;=\;\\
=\;& \int\limits_{\BS}d\al_t\;\,
\SJS{\al_2}{\al_1}{\bal_4}{\bal_3}{\al_{s}}{\bal_{t}}\;
\SC_{}^{\al_4;}{}^{\al_t}_{z_2}{}^{\al_1}_{z_1}\bigl(
\SC_{}^{\al_t;}{}^{\al_3}_{z_{32}}{}^{\al_2}_{\;0}
(\xi_3^{}\ot\xi_2^{})\ot\xi_1)
\bigr).
\end{aligned}\end{equation}
It remains to observe that the  
b-Racah-Wigner symbols satisfy a 
symmetry relation of the following form:
\begin{equation}\label{tetrsymm1}
\SJS{\al_2}{\al_1}{\bal_4}{\bal_3}{\al_{s}}{\bal_{t}}\;=\;
\SJS{\al_1}{\al_2}{\al_3}{\al_4}{\al_{s}}{\al_{t}}.
\end{equation}
The proof of this relation involves the 
behavior of the kernel $[\dots]$ under complex conjugation,
reflection identities that express the equivalence of representations
$\CP_{\al}$ and $\CP_{Q-\al}$ \cite{PT2}, together with some simple
symmetry properties of the integral representation 
\rf{RWexplicit}. Details will be given elsewhere. 

By inserting \rf{tetrsymm1} into \rf{fusrel+} one has finally 
brought the fusion transformations of conformal blocks
into a form that makes 
the analogy with the definition \rf{Racahdef} of
the b-Racah-Wigner symbols perfect.

\subsection{Locality and crossing symmetry}

The relation between fusion coefficients and b-Racah-Wigner symbols
is quite useful: Having established the existence of
fusion transformations of the form \rf{fusrel} allows one to
translate the condition of crossing symmetry into 
a relation involving the three point functions $C(\al_3,\al_2,\al_1)$
together with the fusion coefficients:
\begin{equation}\label{orth1}\begin{aligned}
\int\limits_{\BS}d\al_{s} \; C(\al_4,\al_3,\al_{s})  
C(\bal_{s},\al_2,\al_1) &
\Fus{\al_{s}}{\be_{t}}{\al_3}{\al_2}{\al_4}{\al_1}
\Fus{\bal_{s}}{\bal_{t}}{\bal_3}{\bal_2}{\bal_4}{\bal_1}
=\\
 & \quad=\de_{\BS}(\al_{t},\be_{t})C(\al_4,\al_{t},\al_1)
C(\bal_{t},\al_3,\al_2),
\end{aligned}\end{equation}
where $\de_{\BS}(\al_{t},\be_{t})=\de(P-P')$ if $\al_t=\frac{Q}{2}+iP$,
$\be_t=\frac{Q}{2}+iP'$. We have used that
$F_{\al_s\al_t}\bigl[{}_{\al_4}^{\al_3}{}_{\al_1}^{\al_2}\bigr]
=F_{\bal_s\bal_t}\bigl[{}_{\bal_4}^{\bal_3}{}_{\bal_1}^{\bal_2}\bigr]$,
which trivially follows from the facts that the conformal blocks depend
on the conformal dimensions only and $\De_{\al}=\De_{\bal}$.

But on the other hand one may observe that the construction 
of the b-Racah-Wigner symbols 
in terms of representation theory of 
$\CU_q(\fsl(2,\BR))$ implies the following 
unitarity relation \cite{PT2} :
\begin{equation}\label{orth2}
\int\limits_{\BS}d\al_{s}\;\,M_b(\al_{s}) 
\;\, \SJS{\al_1}{\al_2}{\al_3}{\al_4}{\al_{s}}{\al_{t}}
\SJS{\bal_1}{\bal_2}{\bal_3}{\bal_4}{\bal_{s}}{\bal_{t}'}\;=\;
M_b(\al_{t})\;\,\de_{\BS}(\al_{t}^{},\al_{t}'),
\end{equation}
where $\bal_i=Q-\al_i$, $\al_t,\be_t\in\BS$, and the measure $M_b(\al)$ is
given by
\begin{equation}
M_b(\al)\;=\;\left|S_b(2\al)\right|^2
=\;-4\sin(\pi b(2\al-Q))\sin(\pi b^{-1}(2\al-Q)).
\end{equation}

It now suffices to observe \rf{orth2} ensures validity
of \rf{orth1} for all $C(\al_3,\al_2,\al_1)$ of the form
\begin{equation}\label{CN-rel}
C(\al_3,\al_2,\al_1)\;=\;N_b \;\prod_{i=1}^3 \;\kappa(\al_i)
\RN^{-1}(\al_3,\al_2,\al_1)\RN^{-1}(\bal_3,\bal_2,\bal_1),
\end{equation}
with arbitrary function $\kappa(\al)$ and constant $N_b$.
The DOZZ-formula for $C(\al_3,\al_2,\al_1)$ is easily found to be 
of the form \rf{CN-rel} for suitable choice of $\kappa(\al)$, $N_b$.

This establishes crossing symmetry for the four-point functions
of operators $\SV_{\al}$ that are characterized by the 
DOZZ-formula for $C(\al_3,\al_2,\al_1)$. The proof of locality
is almost identical.

\begin{rem}
Up to now we have mostly assumed that the relevant representations 
$\CV_{\al}$ all correspond to the unitary representations 
in the spectrum, $\al\in\BS\equiv \frac{Q}{2}+i\BR$. 
However, fusion-coefficients $F$ and normalization coefficients
$N(\al_3,\al_2,\al_1)$ all possess a meromorphic continuation to
generic complex values of the representation labels. Due to the
analytic properties of the conformal blocks w.r.t. the representation
labels (cf. Part \ref{pathint}, Subsection \ref{analcfbl})
one may study the analytic continuation of the fusion relations
\rf{fusrel} in a way that is very similar to our discussion in 
Part \ref{pathint}, Subsection \ref{mercont+}. 
\end{rem}

\subsection{Remark on the strong coupling regime}\label{strc_rem}

Our derivation of braiding and fusion transformations of the conformal
blocks in Liouville theory was based on the results of \cite{PT2}.
Strictly speaking, it is therefore not directly applicable
to the regime of strong coupling ($b=e^{i\vf}$, $\vf\in[0,\frac{\pi}{2})$).
However, it seems to us that the arguments of \cite{PT2}
can be modified to become applicable to the 
strong coupling case as well. Moreover, if one considers the relations
on the level of conformal blocks, cf. e.g. \rf{fus_cfbl}, 
one may note that all the appearing objects can be analytically continued
w.r.t. the parameter $b$ from the weak- to the strong coupling 
regime. We consider it therefore as very likely that all of our discussion
applies to the strong coupling case with hardly any change.

Let us note, however, that there exists an alternative proposal for 
fusion and braid relations of chiral vertex operators at special values
of $b$ \cite{G2} (see also \cite{GR1}\cite{GR2}). This proposal 
is based on the construction of a solution to the 
consistency conditions (cf. Subsection \ref{conscond})
that fusion and braiding coefficients must satisfy, which involves only
a {\it discrete} set of representation labels corresponding 
to unitary representations of the Virasoro algebra. 

So far it seems difficult to decide whether this alternative proposal 
is actually realized. On the one hand it is not clear whether a 
given solution of the consistency conditions discussed in 
Subsection \ref{conscond} must in fact be realized by the 
chiral vertex operators that are uniquely defined in terms of the 
representation theory of the Virasoro algebra. To the author's 
knowledge there does not exist a calculation like the one given in 
Section \ref{braid_cco} which would prove that the chiral 
vertex operators for irreducible representation of the Virasoro algebra
satisfy the braid relations proposed in
\cite{G2}\cite{GR1}\cite{GR2}. 

On the other hand, it is also not obvious that the proposal of \cite{G2}
\cite{GR1}\cite{GR2} contradicts our results. So far we do not see 
any reason why there should not exist two different representations
of fusion- and braid relations, at least for special values of $b$
and under suitable restrictions on the representation labels. 

\section{Unitary fusion?} \label{unfus}

As an outlook, we would like to discuss a notion of ``fusion''
for unitary representations of the Virasoro algebra that should allow 
to make the deeper mathematical reasons for the consistency
of Liouville theory more transparent.

Viewing the chiral vertex operators as Clebsch-Gordan maps for a modified
tensor product (``fusion product'')
of Virasoro representations naturally leads to the
question of compatibility of the fusion product with unitarity of the
representation. More precisely: Is it possible to relate the above 
notion of fusion product to a concept of fusion which manifestly 
creates a unitary ``fused'' representation as product of two given
unitary representations of the Virasoro algebra? In more physical
terms this is of course closely related to the natural question
whether the set of vertex operators creating {\it unitary} representations
of the Virasoro algebra is closed under operator product expansion. 
A concept of fusion (``Connes-fusion'') that fulfills such a task is 
known for the WZNW-models associated to the group $SU(N)$ \cite{Wa}.
We will try to outline how a similar treatment should look like 
in the case of unitary Virasoro representations. But before let us indicate 
how Connes-fusion is related to the more standard picture of fusion 
that goes back to \cite{BPZ}:

\subsection{Heuristics: Fusion on the unit circle}

Let us briefly recall (from e.g. \cite{BPZ}\cite{MS}) the complex analytic
picture of fusion and the associated notion of chiral vertex operators:
A chiral vertex operator $\SV_{}^{\al_3;}{}^{\al_2}_{z_2}{}^{\al_1}_{z_1}$
is thought 
of as being associated with a Riemann sphere with three marked points, one of
which was chosen to coincide with infinity. The representations
$\CV_{\al_3}$, $\CV_{\al_2}$ and $\CV_{\al_1}$ are assigned to the 
marked points at $\infty$, $z_2$ and $z_1$ respectively. A 
Virasoro generator $T[v]=\sum_{n\in\BZ}v_nL_{n}$ that ``goes out to'' $\infty$
would then be represented by the contour integral $\int_{\CC_{\infty}}dz
T(z)v(z)$, $v(z)=\sum_{n\in\BZ}z^{n+1}v_n$, where $\CC_{\infty}$ is a small
circle around the point $\infty$. Its action on the 
``ingoing'' representation $\CV_{\al_2}$ and $\CV_{\al_1}$ is obtained by
deforming the contour $\CC_{\infty}$ into two small circles around 
the points $z_2$ and $z_1$ respectively. This prescription directly leads  
to the formula \rf{coprod} for the co-product. 

This prescription encodes the analyticity of $T(z)$ within euclidean
vacuum expectation values. It is clear that the euclidean picture
is not well-suited for questions of unitarity: The euclidean time
evolution is not unitary. One should therefore try to change 
(``Wick-rotate'') to a Minkowskian picture.  
Let us observe that in the case of the cylinder
the possibility of deforming the contour $\CC$ that enters the 
definition of $T[v]=\int_{\CC}dzT(z)v(z)$ would
correspond to conservation of the 
charge $T[v]\equiv\int_{0}^{2\pi}d\si v(\si)T(\si)$, $\pa_{t}T[v]=0$
in a minkowskian framework. 

In order to change to a Minkowskian picture in the case 
of the sphere with three marked points, one needs a global notion 
of time on that geometry. A convenient choice
is given by the mapping 
\begin{equation}
r(z)\;=\; d\ln(z-z_2)+(1-d)\ln(z-z_1).
\end{equation}
The global
(euclidean) time-coordinate is given by $t=\Re(r)$. Lines of constant $t$
may on the Riemann sphere be visualized as electrostatic
``equipotential lines'' for the potential around the charges at points $z_2$
and $z_1$.
$r(z)$ maps the complex plane to a diagram that is known as Mandelstam 
diagram in light-cone string theory:
\medskip\begin{center}
\setlength{\unitlength}{0.0005in}
\begingroup\makeatletter\ifx\SetFigFont\undefined%
\gdef\SetFigFont#1#2#3#4#5{%
  \reset@font\fontsize{#1}{#2pt}%
  \fontfamily{#3}\fontseries{#4}\fontshape{#5}%
  \selectfont}%
\fi\endgroup%
{\renewcommand{\dashlinestretch}{30}
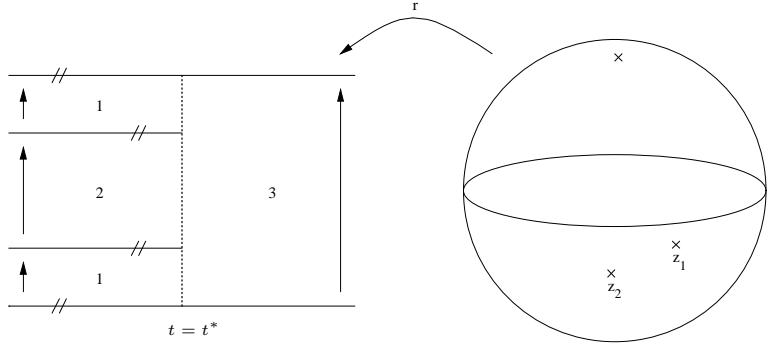
\begin{picture}(7895,3545)(0,-10)
\put(6312,1583){\ellipse{3150}{3150}}
\put(6312,1583){\ellipse{3150}{750}}
\path(12,2783)(3612,2783)
\drawline(12,383)(12,383)
\path(12,383)(3612,383)
\path(12,983)(1812,983)
\path(12,2183)(1812,2183)
\path(162,2333)(162,2633)
\blacken\path(192.000,2513.000)(162.000,2633.000)(132.000,2513.000)(192.000,2513.000)
\path(162,1133)(162,2033)
\blacken\path(192.000,1913.000)(162.000,2033.000)(132.000,1913.000)(192.000,1913.000)
\path(162,533)(162,833)
\blacken\path(192.000,713.000)(162.000,833.000)(132.000,713.000)(192.000,713.000)
\path(3462,533)(3462,2633)
\blacken\path(3492.000,2513.000)(3462.000,2633.000)(3432.000,2513.000)(3492.000,2513.000)
\dottedline{45}(1812,2783)(1812,383)
\path(462,308)(537,458)
\path(537,308)(612,458)
\path(462,2708)(537,2858)
\path(537,2708)(612,2858)
\path(1287,908)(1362,1058)
\path(1362,908)(1437,1058)
\path(1287,2108)(1362,2258)
\path(1362,2108)(1437,2258)
\path(6912,1058)(6987,983)
\path(6912,983)(6987,1058)
\path(6237,683)(6312,758)
\path(6312,683)(6237,758)
\path(6312,2933)(6387,3008)
\path(6387,2933)(6312,3008)
\path(5037,3008)        (4959.383,3059.182)
        (4886.649,3106.024)
        (4754.625,3187.237)
        (4638.512,3252.738)
        (4535.891,3303.624)
        (4444.347,3340.994)
        (4361.461,3365.948)
        (4212.000,3383.000)

\path(4212,3383)        (4074.385,3362.112)
        (3998.600,3336.199)
        (3915.233,3298.509)
        (3822.086,3247.942)
        (3716.963,3183.401)
        (3597.667,3103.787)
        (3462.000,3008.000)

\blacken\path(3541.796,3102.513)(3462.000,3008.000)(3576.885,3053.842)(3541.796,3102.513)
\put(912,608){\makebox(0,0)[lb]{\smash{{{\SetFigFont{6}{7.2}{\rmdefault}{\mddefault}{\updefault}1}}}}}
\put(912,1508){\makebox(0,0)[lb]{\smash{{{\SetFigFont{6}{7.2}{\rmdefault}{\mddefault}{\updefault}2}}}}}
\put(912,2408){\makebox(0,0)[lb]{\smash{{{\SetFigFont{6}{7.2}{\rmdefault}{\mddefault}{\updefault}1}}}}}
\put(2712,1508){\makebox(0,0)[lb]{\smash{{{\SetFigFont{6}{7.2}{\rmdefault}{\mddefault}{\updefault}3}}}}}
\put(1662,83){\makebox(0,0)[lb]{\smash{{{\SetFigFont{6}{7.2}{\rmdefault}{\mddefault}{\updefault}$t=t^*$}}}}}
\put(6912,833){\makebox(0,0)[lb]{\smash{{{\SetFigFont{6}{7.2}{\rmdefault}{\mddefault}{\updefault}z}}}}}
\put(6987,758){\makebox(0,0)[lb]{\smash{{{\SetFigFont{5}{6.0}{\rmdefault}{\mddefault}{\updefault}1}}}}}
\put(4212,3458){\makebox(0,0)[lb]{\smash{{{\SetFigFont{6}{7.2}{\rmdefault}{\mddefault}{\updefault}r}}}}}
\put(6237,533){\makebox(0,0)[lb]{\smash{{{\SetFigFont{6}{7.2}{\rmdefault}{\mddefault}{\updefault}z}}}}}
\put(6312,458){\makebox(0,0)[lb]{\smash{{{\SetFigFont{5}{6.0}{\rmdefault}{\mddefault}{\updefault}2}}}}}
\end{picture}
}

\end{center}\medskip
The markings on the lines in the left part of the diagram are supposed to 
indicate identifications. The regions marked by 1 and 2 therefore
represent (part of) ``ingoing'' semi-infinite cylinders, whereas 
the region 3 represents an ``outgoing'' semi-infinite cylinder. 
These cylinders
are joined at time $t=t^*$, where $t^*$ is a function of $z_1$, $z_2$ and
$d$ that can be figured out from the definition of $r(z)$. The time-slice
at $t=t^*$ looks like the unit circle $S^1$ divided into two intervals $I$ and
$I^c$, where $S(I)$ ($I$ with end-points identified) represents the ending
of the semi-infinite cylinder 2, $S(I^c)$ the ending of cylinder 1. 

By means of $r(z)$ one therefore maps the previous contour-deformation
picture for the definition of the co-product into a picture with 
obvious Minkowskian counterpart: On each of the cylinders one may
use charge conservation to shift $T[v]$ to the left or right, so 
everything boils down to the splitting at $t=t^*$. This
splitting should clearly be purely geometrical:
If $T[f]=\int_{S^1}d\si f(\si)T(\si)$
is the generator of an infinitesimal transformation, it should
split into $T[f\Theta_I]$ and $T[f\Theta_{I^c}]$, where 
$\Theta_I$ denotes the characteristic
function of the interval $I$. 

It is not a trivial task to turn this heuristic idea into a
rigorous definition of fusion that makes the issue of
unitary more transparent: First, it is not clear which scalar
product to put on $\CV_{\al_2}\ot\CV_{\al_1}$ such that 
the action defined by the geometric splitting of the unit circle
into two subintervals is well-defined and unitary.
If one would just take the canonical scalar product on the
tensor product of representations, one would get the following 
problem: $f\Theta_I$ and $f\Theta_{I^c}$, 
when considered as functions on the 
circles $S(I)$ and $S(I^c)$, will generically have jumps at the points
corresponding to the end-points of $I$. 
This leads to the problem that the 
vector $T[f\Theta_I]v_2\ot T[f\Theta_{I^c}]v_1$ will 
generically not be normalizable when the standard norm in 
$\CV_{\al_2}\ot\CV_{\al_1}$ is taken.

It seems to us that these problems are just 
what is overcome by the so-called ``Connes-fusion'', more precisely its
more explicit version
that was developed in \cite{Wa} in the case of loop groups.
We would next
like to give an idea of such a formulation for the case of unitary 
representations of the Virasoro algebra.

\subsection{Connes-fusion for unitary Virasoro-representations?}

First, it is technically better to consider the 
group of diffeomorphisms of the 
unit circle and its
(projective) unitary 
representations instead of its Lie algebra (the Virasoro) 
algebra and their representations. It is known \cite{GW} that for
$c>1$ the Virasoro algebra representation $\CV_{\al}$ indeed exponentiates
to a projective unitary representation of ${\rm Diff}(S^1)$ if $\De>0$, 
i.e. for all unitarizable representations of the Virasoro algebra.

It should be possible to ``restrict'' ${\rm Diff}(S^1)$ to an interval
$I$ and its complement $I^c$ in a suitable sense, 
e.g. by considering the subgroups 
${\rm Diff}(I)$ and ${\rm Diff}(I^c)$ generated by
elements of the algebra like 
$\int_{S^1}d\si T(\si)f(\si)$ and $\int_{S^1}d\si T(\si)g(\si)$,
where $f$, $g$ have support only in $I$, $I^c$ respectively.  

Central objects are then the intertwining operators $\SV_{\al}^{}$ 
that ``create'' 
the representation $\CV_{\al}$ from the vacuum. Here one would consider
in particular maps 
$\SV_{\al}^{}\in{\rm Hom}_{{\rm Diff}(I)}(\CV_{0},\CV_{\al})$ that 
intertwine the respective actions of ${\rm Diff}(I)$ according to
\begin{equation}\label{interI}
\pi_{\al}^{}(g)\SV_{\al}^{}\;=\;\SV_{\al}^{}\pi_{0}^{}(g), \qquad g\in {\rm Diff}(I),
\end{equation}
as well as their counterparts for $I^c$. Such maps should be 
given by smeared chiral vertex operators:
\begin{equation}
\SV_{\al}^{}(\xi|f)\;\equiv\;
\int_{S^1}d\si \;f(\si)\;
\SV\bigl({}_{\al}^{}\!{}^{\al}_{}\!{}_{0}^{}\bigr)(\xi|\si).
\end{equation}
where $f(\si)$ has support only in $I^c$. 
It is important that $f$ vanishes in $I$ in order for
the ordinary intertwining property \rf{desctrf}
to translate into \rf{interI}. 

Let $\CO_{\al_2}^{I}$, $\CO_{\al_1}^{I^c}$ be the spaces 
of operators ${\rm Hom}_{{\rm Diff}(I)}(\CV_{0},\CV_{\al})$
and ${\rm Hom}_{{\rm Diff}(I^c)}(\CV_{0},\CV_{\al})$ 
respectively. One has a correspondence between 
elements $\SV_{\al}^{}\in
{\rm Hom}_{{\rm Diff}(I)}(\CV_{0},\CV_{\al})$ and the 
state $\SV_{\al}^{}v_0$ that they create when acting on the vacuum.
In view of that correspondence one may define fusion for elements 
of $\CO_{\al_2}^{I}$, $\CO_{\al_1}^{I^c}$ instead of 
$\CV_{\al_2}$, $\CV_{\al_1}$.

First define a scalar product on $\CO_{\al_2}^{I}\ot\CO_{\al_1}^{I^c}$
via the four-point conformal block
\begin{equation}\label{Wassscprd}\begin{aligned}
\bigl( \SV_{\al_2}^{}(\zeta_2^{}|g_2^{}) & 
\ot\SV_{\al_1}^{}(\xi_2^{}|f_2^{})\;,\; 
\SV_{\al_2}^{}(\zeta_1^{}|g_1^{})\ot\SV_{\al_1}^{}(\xi_1^{}|f_1^{})
\bigr)^{}_{\CO_{\al_2}^{I}\boxtimes\CO_{\al_1}^{I^c}}\;\equiv\\
& \equiv\;\bigl\bra v_0\, ,\,
\SV_{\al_2}^{\dagger}(\zeta_2^{}|g_2^{})\SV_{\al_2}^{}(\zeta_1^{}|g_1^{})
\, \SV_{\al_1}^{\dagger}(\xi_2^{}|f_2^{})\SV_{\al_1}^{}(\xi_1^{}|f_1^{})
\, v_0\bigr\ket_{\CV_0},
\end{aligned}\end{equation}
where $g_i$, $f_i$, $i=1,2$ 
have support in $I^c$, $I$ respectively, and
$\SV_{\al}^{\dagger}(\zeta|f)$ is the adjoint of
$\SV_{\al}^{\dagger}(\zeta|f)$ that may be expressed as 
\begin{equation}
\SV_{\al}^{\dagger}(\xi|f)\;\equiv\;
\int_{S^1}d\si \; f^*(\si)\;
\SV\bigl({}_{0}^{}\!{}^{\al}_{}\!{}_{\al}^{}\bigr)(\xi^*|\si).
\end{equation}
The fusion $\CO_{\al_2}^{I}\boxtimes\CO_{\al_1}^{I^c}$ is defined as the
completion of $\CO_{\al_2}^{I}\ot\CO_{\al_1}^{I^c}$ w.r.t. the scalar product
introduced in \rf{Wassscprd}.

$\CO_{\al_2}^{I}\ot\CO_{\al_1}^{I^c}$ carries a natural action of 
${\rm Diff}(I)\ti{\rm Diff}(I^c)$. It is easy to see that 
this action is unitary w.r.t. the scalar product \rf{Wassscprd}. 
The crucial question now is whether the action of 
${\rm Diff}(I)\ti{\rm Diff}(I^c)$ uniquely extends to a unitary action of 
${\rm Diff}(S^1)$ on $\CO_{\al_2}^{I}\boxtimes\CO_{\al_1}^{I^c}$. This would
be the desired representation of ${\rm Diff}(S^1)$ obtained as fusion
of representations $\CV_{\al_2}$ and $\CV_{\al_1}$ associated to the
intervals $I$ and $I^c$ respectively. The unique extension of the 
${\rm Diff}(I)\ti{\rm Diff}(I^c)$-action looks intuitively plausible:
As ${\rm Diff}(I)$ and ${\rm Diff}(I^c)$ contain diffeomorphisms that are
nonzero arbitrarily close to the end points of the respective intervals,
it seems unlikely that there is much freedom in ``defining the action
at the touching points''.

Let us note that the proof of 
the corresponding unique extension property for
loop group representations given in \cite{Wa} makes essential 
use of the braiding relations for 
chiral vertex operators. In our case this would
be the relations of the form
\begin{equation}\label{0-braid}\begin{aligned}
\bigl\bra v_0\, ,\, 
\SV_{\al_2}^{\dagger} & (\zeta_2^{}|g_2^{})\SV_{\al_2}^{}(\zeta_1^{}|g_1^{})
\, \SV_{\al_1}^{\dagger}(\xi_2^{}|f_2^{})\SV_{\al_1}^{}(\xi_1^{}|f_1^{})
\, v_0\bigr\ket_{\CV_0}\;=\;\\
=\; &\int\limits_{\BS}d\al\;
O_{\ep}({}_{\al}^{}\!{}^{\al_2}_{}\!{}_{\al_1}^{}\bigr)
\Braid{0}{\al}{\al_2}{\al_1}{\al_2}{\al_1}{-\ep}
\;\bigl\bra v_0\, ,\,
\SV\bigl({}_{0}^{}\,{}^{\al_1}_{}\!{}_{\al_1}^{}\bigr)(\xi_2^*|f_2^*)
\SV\bigl({}_{\al_1}^{}\!{}^{\al_2}_{}{}_{\al}^{}\bigr)(\zeta_2^*|g_2^*)
\\[-1ex]
& \hspace{5.1cm}
\SV\bigl({}_{\al}^{}{}^{\al_2}_{}\!{}_{\al_1}^{}\bigr)(\zeta_1^{}|g_1^{})
\SV\bigl({}_{\al_1}^{}\!{}^{\al_1}_{}\,{}_{0}^{}\bigr)(\xi_1^{}|f_1^{})
\, v_0\bigr\ket_{\CV_0}.
\end{aligned}\end{equation}
that we had found previously. These relations would imply that
\begin{equation}\label{Complfus}\begin{aligned}
\bigl|\!\bigl|\SV_{\al_2}^{}(\zeta_1^{}|g_1^{}) & \ot
\SV_{\al_1}^{}(\xi_1^{}|f_1^{})
\bigr|\!\bigr|^2_{\CO_{\al_2}^{I}\boxtimes\CO_{\al_1}^{I^c}}\;=\;\\
=&\;
\int\limits_{\BS}d\mu_{\al_2\al_1}(\al)\;
\bigl|\!\bigl|
\SV\bigl({}_{\al}^{}{}^{\al_2}_{}\!{}_{\al_1}^{}\bigr)(\zeta_1^{}|g_1^{})
\SV\bigl({}_{\al_1}^{}\!{}^{\al_1}_{}\,{}_{0}^{}\bigr)(\xi_1^{}|f_1^{})
\, v_0\bigr|\!\bigr|^2_{\CV_{\al}},
\end{aligned}\end{equation}
where $d\mu_{\al_2\al_1}(\al_3)$ can be worked out from the explicit expression
for the braiding coefficients given in \rf{braidcoeffs} as
\begin{equation}\label{Nexplicit}\begin{aligned}
{} & d\mu_{\al_2,\al_1}(\al_3) \;= 
\;\frac{\Ga_b(2Q)}{\Ga_b(Q)}\;d\al_3\;|S_b(2\al_3)|^2\;\ti \\
&\quad \ti\left|\frac
{\Ga_b(\al_1+\al_2+\al_3-Q)
\Ga_b(\al_1+\al_2-\al_3)\Ga_b(\al_1+\al_3-\al_2)\Ga_b(\al_2+\al_3-\al_1)}
{\Ga_b(2\al_1)\Ga_b(2\al_2)\Ga_b(2\al_3)}\right|^2.
\end{aligned}\end{equation}
Relation \rf{Complfus} can be read as an expression for the unitary equivalence
\begin{equation}\label{Cofudeco}
\CO_{\al_2}^{I}\boxtimes\CO_{\al_1}^{I^c}\;\simeq\;
\int\limits_{\BS}^{\oplus}d\mu_{\al_2\al_1}(\al)\;\CV_{\al}.
\end{equation}
The unique extension of the ${\rm Diff}(I)\ti{\rm Diff}(I^c)$-action to an 
action of ${\rm Diff}(S^1)$ should then follow as in \cite{Wa} from 
similar statements concerning the respective actions on the irreducible
representations $\CV_{\al}$ that appear in \rf{Cofudeco}. 

Having established the braid relations \rf{0-braid} motivates 
our hope that a treatment along such lines is within reach.
We find it particularly satisfactory to observe that the change 
of normalization of the chiral vertex operators that 
established the relation between fusion coefficients 
and b-Racah-Wigner symbols is precisely such that the 
fusion density $d\mu_{\al_2\al_1}(\al)$ would become proportional to the
canonical measure $d\al |S_b(2\al)|^2$  if we had used the 
chiral vertex operators 
$\SC_{}^{\al_3;}{}^{\al_2}_{z_2}{}^{\al_1}_{z_1}$ 
instead of the $\SV_{}^{\al_3;}{}^{\al_2}_{z_2}{}^{\al_1}_{z_1}$. 
This means that the normalization of chiral vertex operators 
that makes the relation to quantum group representation theory manifest
is simultaneously a natural one from the point of view of 
Connes-Wassermann fusion. We believe that these ideas should
make up for a rather beautiful story when being worked out.

\newcommand{\CMP}[3]{{\it Comm. Math. Phys. }{\bf #1} (#2) #3}
\newcommand{\LMP}[3]{{\it Lett. Math. Phys. }{\bf #1} (#2) #3}
\newcommand{\IMP}[3]{{\it Int. J. Mod. Phys. }{\bf A#1} (#2) #3}
\newcommand{\NP}[3]{{\it Nucl. Phys. }{\bf B#1} (#2) #3}
\newcommand{\PL}[3]{{\it Phys. Lett. }{\bf B#1} (#2) #3}
\newcommand{\MPL}[3]{{\it Mod. Phys. Lett. }{\bf A#1} (#2) #3}
\newcommand{\PRL}[3]{{\it Phys. Rev. Lett. }{\bf #1} (#2) #3}
\newcommand{\PRD}[3]{{\it Phys. Rev.}{\bf D#1} (#2) #3}
\newcommand{\AP}[3]{{\it Ann. Phys. (N.Y.) }{\bf #1} (#2) #3}
\newcommand{\LMJ}[3]{{\it Leningrad Math. J. }{\bf #1} (#2) #3}
\newcommand{\FAA}[3]{{\it Funct. Anal. Appl. }{\bf #1} (#2) #3}
\newcommand{\PTPS}[3]{{\it Progr. Theor. Phys. Suppl. }{\bf #1} (#2) #3}
\newcommand{\LMN}[3]{{\it Lecture Notes in Mathematics }{\bf #1} (#2) #2}

\end{document}